\documentclass[
  reprint,
  final,
  amsmath,
  amssymb,
  aps,
  floatfix,
]{revtex4-2}

\pdfoutput=1

\usepackage{graphicx}
\usepackage{ifpdf}

\newcommand{\figfile}[2][]{
  \ifpdf
    \IfFileExists{#2.pdf}
      {\includegraphics[#1]{#2.pdf}}
      {\includegraphics[#1]{#2}}
  \else
    \IfFileExists{#2.eps}
      {\includegraphics[#1]{#2.eps}}
      {\includegraphics[#1]{#2}}
  \fi
}

\setkeys{Gin}{keepaspectratio}

\ifpdf
  \usepackage{epstopdf}
\fi
\ifpdf
\usepackage{epstopdf}
\fi

\usepackage{dcolumn}
\usepackage{bm}
\usepackage{appendix}
\usepackage{soul}
\usepackage[utf8]{inputenc}
\usepackage[T1]{fontenc}
\usepackage{lmodern}
\usepackage{geometry}
\usepackage{amsmath, amssymb, amsthm}
\usepackage{microtype}
\usepackage{hyperref}
\usepackage{natbib}
\usepackage{comment}
\newcommand{\bq}{\begin{equation}}
\newcommand{\eq}{\end{equation}}

\newcommand{\bea}{\begin{eqnarray}}
\newcommand{\eea}{\end{eqnarray}}

\newcommand{\ba}{\begin{eqnarray}}
\newcommand{\ea}{\end{eqnarray}}

\begin{document}
\preprint{APS/123-QED}

\title{Physics of the Propagated Action Potential
}
\author{Nikola K. Jurisic}
\email{nikola.k.jurisic@gmail.com}
\affiliation{Independent researcher (Retired from UCLA Stein Eye Institute), 12013 Navy St,
Los Angeles, CA 90066, USA}
\author{Fred Cooper} 
\email{cooper@santafe.edu}
\affiliation{The Santa Fe Institute, 1399 Hyde Park Road, Santa Fe, NM 87501, USA}
\affiliation{Theoretical Division and Center for Nonlinear Studies,
 Los Alamos National Laboratory,
 Los Alamos, NM 87545}

\date{\today}
\begin{abstract}
We analyze the Rosenthal--Bezanilla experimental action potential waveforms using a charge-conserving phase-space formulation of the cable equation for a steadily propagated action potential. Each measured waveform determines the axial Ohmic current $J_z$, its geometry-converted divergence $J_m$, the capacitive current, and the total ionic current without requiring voltage-clamp currents as independent inputs. The current balance is expressed branchwise through $V$, $\Phi(V)=dV/dt$, and $d\Phi(V)/dV$; laboratory time remains recoverable from $dt=dV/\Phi(V)$. At points where $\Phi(V)\neq0$, the total ionic current vanishes when $d\Phi/dV=k$, producing one zero crossing on each branch. Both crossings are present in the Rosenthal--Bezanilla data. The reconstructed currents remain continuous through the action-potential peak, while their state-dependent decomposition changes, supporting a continuous $M\rightarrow H$ transition between sodium-channel regimes. The second crossing occurs shortly after the peak on the recovery branch, after which the total ionic current is outward. Its effective reversal potential, channel-completion kinetics, and temperature dependence support a sodium-dominated interpretation. Quasilinear phase-space regions yield effective conductances, reversal potentials, and characteristic time rates. Channel completion follows a modified Avrami relation with nearly temperature-independent exponents and Arrhenius time rates. The fitted processes share an approximately invariant dimensionless prefactor, numerically close to the fine-structure constant, whose physical origin remains undetermined. This framework provides a unified charge-conserving, clock-free relational description of propagated action potentials and reveals current and channel-state structure not resolved by stationary voltage-clamp measurements.
\end{abstract}

\maketitle
\section{Introduction}
\label{sec:introduction}

\subsection{Classical formulation, propagation, and invariance}

\quad
The classical Hodgkin--Huxley (HH) description of the propagated action
potential \cite{Hodgkin1952a,Hodgkin1952b} combines the cable equation
with empirical ionic-current relations reconstructed from voltage-clamp
experiments in which spatial propagation is suppressed. This framework
has provided the foundation of membrane electrophysiology for more than
seventy years and remains remarkably successful in describing excitable
cells. In its conventional use, ionic-current relations obtained under
voltage clamp are inserted into the cable equation to calculate a
propagating waveform.

\quad
Steady propagation, however, contains an electrodynamic structure that
is absent from a spatially uniform voltage-clamp experiment. A
propagated waveform establishes an axial electric field and, through
Ohm's law, an axial Ohmic current $J_z$ within the axoplasm. Because the
waveform varies along the axon, $J_z$ is generally nonuniform. Its
spatial divergence supplies or removes charge locally and must be
balanced continuously by membrane charging and ionic charge transfer.
Axial conduction, capacitive current, and ionic current are therefore
not independent additions to the propagated solution; they are linked
by local charge conservation.

\begin{figure*}
\centering

\includegraphics[width=2.08\columnwidth]{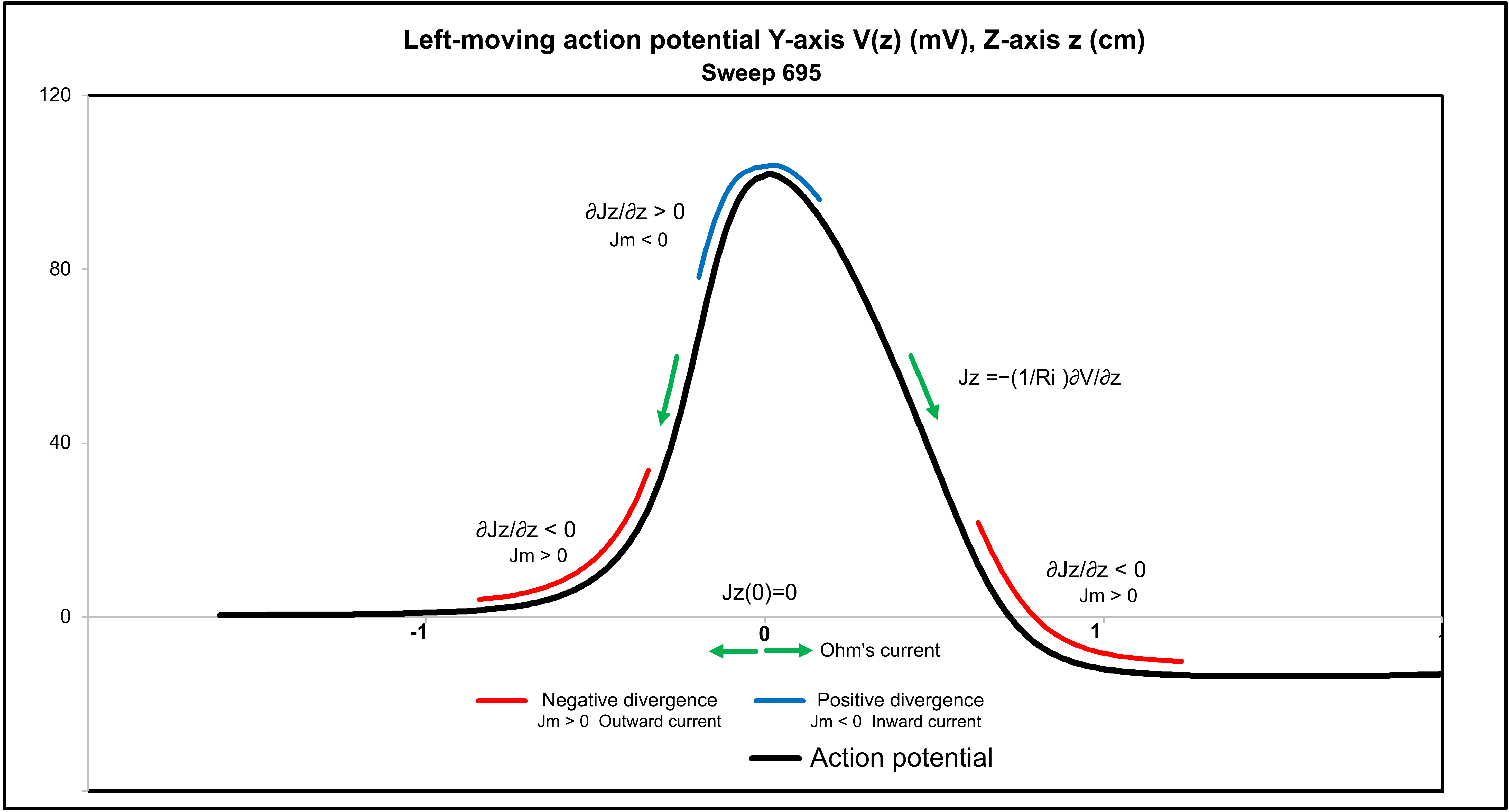}
\caption{\label{fig:ActionPotL}
Action-potential waveform illustrating the electrodynamic structure of
propagation. The spatial variation of the potential defines the axial
Ohmic current $J_z\propto-\partial V/\partial z$. Because the waveform
slope varies along the axon, $J_z$ is nonuniform. Its
geometry-converted divergence,
$J_m=-(R/2)\,\partial J_z/\partial z$, is expressed on the membrane-area
basis and represents the local axial supply or removal of charge. Local
charge conservation requires
$J_m=C_m\partial V/\partial t+J_I$, where the two contributions on the
right describe membrane charging and ionic charge transfer,
respectively.
}
\end{figure*}

\quad
This structure is illustrated schematically in
Fig.~\ref{fig:ActionPotL}, drawn from the Rosenthal--Bezanilla data
\cite{RB}. The measured waveform determines $J_z$, its
geometry-converted axial-current divergence $J_m$, the capacitive
current, and the total ionic current. The propagated waveform therefore
contains dynamical information that is not retained when propagation is
suppressed.

\quad
A further observation motivating the present work is that the shape of
a steadily propagated action potential and the current--voltage
relations reconstructed from it are preserved under Galilean
transformation of the propagating waveform
\cite{Jurisic1987,Jurisic1988}. This invariance concerns the structural
relations along the trajectory, not the disappearance of temporal
ordering. It motivates a formulation in which the propagated trajectory,
rather than externally reconstructed voltage-clamp currents, is the
starting point of the analysis.

\quad
The objective of the present work is to formulate the charge-conserving
cable equation in the phase plane $(V,\Phi)$, with
$\Phi=dV/dt$, and to reconstruct the associated currents directly from
the measured propagated waveform. This representation reveals
structural properties of propagation before any decomposition into
sodium, potassium, or other ionic components is introduced.

\subsection{Generalized phase-space form of the charge-conserving cable equation}

\quad
Here ``phase space'' denotes the dynamical state plane $(V,\Phi)$ and
not the canonical Hamiltonian coordinates $(q,p)$. For a steadily
propagated waveform,
\begin{equation}
\Phi(V)\equiv\frac{dV}{dt},
\label{eq:PhiIntro}
\end{equation}
is a single-valued function of membrane potential along each monotonic
branch. The measured waveform $V(t)$ can therefore be represented by
the phase-space curve $\Phi(V)$. This transformation introduces no new
dynamical assumption; it expresses the same propagated solution through
correlations between the potential and its rate of change.

\quad
The spatial divergence of the axial Ohmic current, converted by
cylindrical geometry to the membrane-area normalization used in the
local charge balance, is
\begin{equation}
J_m(V)
=
\frac{C_m}{k}\,
\Phi(V)\frac{d\Phi(V)}{dV},
\label{eq:JmIntro}
\end{equation}
where $C_m$ is the membrane capacitance and $k$ is the propagation
parameter determined by axonal geometry, axoplasmic resistivity, and
propagation velocity. The quantity $J_m$ is not an independently
transported radial current. It is the geometry-converted
axial-current divergence defined more fully in Appendix~A.

\quad
Local charge conservation gives
\begin{equation}
J_m(V)=J_C(V)+J_I(V),
\label{eq:ChargeIntro}
\end{equation}
with
\begin{equation}
J_C(V)=C_m\Phi(V).
\label{eq:JCIntro}
\end{equation}
Here $J_C$, $J_I$, and $J_m$ are expressed on the same membrane-area
basis, but they have different physical origins. The total ionic
current is consequently
\begin{equation}
J_I(V)
=
C_m\Phi(V)
\left[
\frac{1}{k}\frac{d\Phi(V)}{dV}-1
\right].
\label{eq:JIIntro}
\end{equation}
Only $J_z$ obeys Ohm's law directly; $J_m$ is determined by its spatial
variation.

\quad
Equation~\eqref{eq:JIIntro} is the central phase-space cable relation
used in this work. Once the propagated waveform is measured,
$\Phi(V)$ and $d\Phi(V)/dV$ determine the capacitive current, the total
ionic current, and $J_m$ without requiring voltage-clamp currents as
independent inputs. The contribution proportional to
$\Phi\,d\Phi/dV$ reflects axial-current divergence, whereas the
contribution proportional to $-\Phi$ is associated with membrane
charging or discharging. Their changing balance determines the sign
and zero crossings of the reconstructed total ionic current.

\subsection{Structural predictions of propagation}

\quad
At points for which $\Phi(V)\neq0$, Eq.~\eqref{eq:JIIntro} predicts that
the total ionic current vanishes when
\begin{equation}
\frac{d\Phi(V)}{dV}=k.
\label{eq:ZeroCIntro}
\end{equation}
This condition is determined by the measured trajectory and the
propagation parameter, not by an assumed decomposition into individual
ionic components. Waveform turning points, where $\Phi=0$, require
separate limiting treatment and are not identified with these
branchwise zero crossings.

\quad
The Rosenthal--Bezanilla trajectories satisfy
Eq.~\eqref{eq:ZeroCIntro} once on each branch of the propagated
action potential \cite{RB}. The first crossing occurs near the end of
the action-potential foot. The second occurs shortly after the peak on
the recovery branch; beyond it, the total ionic current is outward. At
both crossings,
\[
J_m=C_m\Phi.
\]
The outward recovery current beyond the second crossing is therefore a
structural consequence of steady propagation and local charge
conservation rather than an auxiliary component inserted to reproduce
the falling phase.

\quad
The phase-space reconstruction also displays extended quasilinear
regions in which the ionic current varies approximately linearly with
membrane potential. Their slopes and intercepts provide effective
conductances and reversal potentials. Distinct quasilinear branches
during depolarization and recovery show that the ionic-current
organization changes across the action-potential peak even though the
experimentally reconstructed currents remain continuous.

\quad
These predictions precede any channel-kinetic model. The two
branchwise zero crossings, the outward recovery current, and the
quasilinear regimes follow from steady propagation, Ohmic conduction in
the axoplasm, and local charge conservation. Basic electrodynamics
determines the admissible current balance of propagation; Darwinian
evolution has produced ion-channel systems capable of satisfying that
balance while generating a biologically useful action potential.

\subsection{Recovery currents and the sodium-channel transition}

\quad
The quasilinear recovery region can be resolved into two outward
components. One, denoted $J_N(V)$, has the conductance and
effective-reversal-potential structure associated with potassium
activation. The second, denoted $J_H(V)$, forms a distinct branch with
effective reversal potential $V_H$. Its relation to the preceding
sodium activation current $J_M(V)$, the near-complete sodium fraction
at the peak, and its kinetic and temperature dependence support a
sodium-dominated interpretation. This ionic assignment remains
inferential in the absence of selective ion-substitution or blocking
experiments.

\quad
The transition from $J_M$ to $J_H$ occurs at the action-potential peak.
The total reconstructed currents remain continuous, whereas the
state-dependent decomposition changes from the sodium activation branch
to the recovery branch. In the present interpretation, the peak
therefore marks a continuous $M\rightarrow H$ symmetry transition
between sodium-channel regimes having different conductances,
effective reversal potentials, time rates, and completion variables.
Potassium activation begins later in recovery and does not supply the
initial branch leaving the peak.

\quad
The capacitive contribution also has an unexpectedly direct role in the
reconstructed recovery current. On the recovery branch,
$\Phi(V)<0$, so $C_m\Phi(V)<0$, and
\[
J_I(V)=J_m(V)-C_m\Phi(V)
\]
shows that capacitive discharge shifts the required ionic current in
the outward direction. The capacitor does not itself drive sodium ions
outward, and its charge $q_C=C_mV$ remains a single-valued function of
$V$. Nevertheless, the branch dependence of both $J_m$ and
$C_m\Phi$ contributes to the separation of the ionic-current
trajectories in the $(V,J_I)$ plane.

\quad
This branch dependence produces electrical hysteresis without requiring
a history-dependent membrane capacitance. Appendix~B develops a reduced
representation in terms of branch-dependent effective-equilibrium
potentials and polarization-related state coordinates. The designation
\emph{ferroelectric-like} refers to the cooperative and
history-dependent structure of the reconstructed trajectory; it does
not identify the axonal membrane or sodium channel as a crystalline
ferroelectric material, nor does it assume that a thermodynamic
polarization charge has already been reconstructed.

\subsection{Modified Avrami representation of channel completion}

\quad
The phase-space cable equation determines the total ionic current from
the propagated waveform, but decomposition into sodium, potassium, and
polarization components requires a representation of the corresponding
completion processes. Channel activation, deactivation, and
polarization are described here by a modified Avrami (mAvrami)
relation. The Avrami equation
\cite{Avrami1,Avrami2,Avrami3}, originally developed for
crystallization kinetics, is the only phenomenological kinetic law
introduced in the channel-completion analysis. Its use for axonal
conductance has an early precedent in the work of Cope \cite{Cope}.

\quad
For a process $X$ defined by an inception time $t_{iX}$, the completion
fraction is
\begin{equation}
\frac{X_o(t)}{X}
=
1-
\exp\left[
-\alpha_X
\left\{
\mu_X\left(t-t_{iX}\right)
\right\}^{\theta_X}
\right],
\label{eq:mAvramiIntro}
\end{equation}
and
\begin{equation}
-\ln\left[
1-\frac{X_o(t)}{X}
\right]
=
\alpha_X
\left\{
\mu_X\left(t-t_{iX}\right)
\right\}^{\theta_X}
\label{eq:mAvramiExponentIntro}
\end{equation}
is the exponent of the mAvrami law. Here $\mu_X$ is the characteristic
time rate, $\theta_X$ is the completion exponent, and $\alpha_X$ is a
dimensionless relational parameter. The $H$ recovery process is
parameterized analogously using its closure time $t_{cH}$ rather than
an inception time.

\quad
The mAvrami representation treats the reconstructed channel kinetics as
collective completion processes rather than as independent transitions
assigned to empirical gating variables. It is not intended as a
microscopic model of channel-protein structure. Its purpose is to
separate temperature-dependent time rates from dimensionless
completion parameters and to permit comparisons among sodium
activation, sodium recovery, potassium activation, and polarization
processes.

\subsection{Temperature dependence and universal scaling}

\quad
For each reconstructed process, the characteristic time rate follows an
Arrhenius relation,
\begin{equation}
\mu_X(T)
=
\mu_{0X}
\exp\left(
-\frac{E_{aX}}{k_B T}
\right),
\label{eq:ArrheniusIntro}
\end{equation}
where $\mu_{0X}$ is the pre-exponential rate and $E_{aX}$ is the
activation energy. The distinct activation energies obtained for
sodium activation, pre-peak inactivation, sodium recovery, and
potassium activation show that these are kinetically distinct
processes rather than portions of a single temperature-scaled
waveform.

\quad
In contrast, the fitted completion exponents remain nearly
temperature independent, and the dimensionless coefficients
$\alpha_X$ cluster near a common value,
\begin{equation}
\alpha=0.007297\ldots,
\label{eq:AlphaIntro}
\end{equation}
numerically close to the fine-structure constant. Using this common
scale reduces the fitting degeneracy without degrading the
reconstruction. The present analysis does not derive the coefficient
from microscopic theory; its identification with the fine-structure
constant therefore remains an empirical proposition. The observed
stability across processes and temperatures motivates the search for a
more fundamental derivation but does not by itself establish one.

\subsection{Relational formulation and scope of the paper}

\quad
Along each monotonic branch of the propagated action potential,
membrane potential and laboratory time are in one-to-one
correspondence. Equation~\eqref{eq:PhiIntro} gives
\begin{equation}
dt=\frac{dV}{\Phi(V)}.
\label{eq:dtRelIntro}
\end{equation}
The mAvrami completion law can therefore be written exactly in terms of
membrane potential and the measured phase-space velocity:
\begin{equation}
\frac{X_o(V)}{X}
=
1-
\exp\left[
-\alpha_X
\left\{
\mu_X
\int_{V_{iX}}^{V}
\frac{d\widetilde V}{\Phi(\widetilde V)}
\right\}^{\theta_X}
\right],
\label{eq:mAvramiRelIntro}
\end{equation}
where $V_{iX}$ corresponds to the inception time $t_{iX}$; the
$H$ recovery branch is treated analogously using the potential
corresponding to its closure time $t_{cH}$.

\quad
The temporal and phase-space forms describe the same completion
process. The completion fractions and dimensionless parameters are
unchanged; only the variable used to traverse the trajectory is
different. Laboratory time is therefore not removed from the physical
process but can be reconstructed from the measured correlations. The
formulation is relational rather than timeless. Its comparison with
Rovelli \cite{Rovelli1} is interpretive rather than foundational: the
phase-space relations are derived independently from steady
propagation, Ohm's law, and charge conservation, but are consistent
with the view that physical laws may be expressed through correlations
among observables rather than evolution with respect to a privileged
external clock.

\quad
The paper develops this framework in three stages. Part~II derives the
charge-conserving phase-space cable equation and reconstructs the axial
Ohmic current $J_z$, the geometry-converted axial-current divergence
$J_m$, the capacitive current, and the total ionic current from the
Rosenthal--Bezanilla waveforms. Part~III resolves the total ionic
current into sodium activation, sodium recovery, potassium activation,
and polarization components and develops the mAvrami representation,
temperature dependence, activation energies, symmetry transition, and
dimensionless scaling. The appendices clarify the current terminology,
develop the hysteretic representation, document the fitting procedure,
and give the exact relational reformulation of the completion kinetics.

\section{\label{sec:level1}Phase-space cable equation}

\quad
We derive the charge-conserving cable equation for a steadily
propagating action potential directly in current--voltage phase space,
providing a compact state-based description analogous to an equation
of state. For a traveling wave of fixed shape, the measured trace
$V(t)$ determines the branchwise function
$\Phi(V)=dV/dt$, and the current balance can be expressed through
correlations among $V$, $\Phi(V)$, and $d\Phi(V)/dV$. Laboratory time
is therefore not an independent variable in the current equations,
although temporal direction and elapsed time remain recoverable along
each monotonic branch from the sign of $\Phi(V)$ and the relation
$dt=dV/\Phi(V)$. This representation is natural for propagation, where
spatial coupling and charge conservation constrain the admissible
current--voltage structure.

\subsection{Charge-conserving cable equation and current definitions}

\quad
In the laboratory frame, steady propagation is governed by the
charge-conserving cable equation
(see, e.g., Ref.~\cite{Jurisic1987}),
\bq
\label{cable}
C_m\frac{dV}{dt}+J_I(t)
=
-\frac{R}{2}\frac{dJ_z}{dz},
\eq
where $C_m$ is the membrane capacitance per unit area, $R$ is the axon
radius, and $J_I$ is the total ionic current expressed per unit
membrane area. The axial Ohmic current $J_z$, defined per unit axonal
cross-sectional area, obeys Ohm's law,
\bq
\label{Ohm}
J_z
=
-\frac{1}{R_i}\frac{dV}{dz},
\eq
where $R_i$ is the axoplasmic resistivity.

\quad
For a pulse propagating at constant velocity $v$, the cable equation
can be written in terms of the measured waveform $V(t)$ as
\bq
\label{cableV}
C_m\frac{dV(t)}{dt}+J_I(t)
=
\frac{R}{2v^2R_i}\frac{d^2V(t)}{dt^2},
\eq
where $v$ is the propagation velocity. The right-hand side is commonly
designated the ``membrane current.'' Its physical origin in the
propagated solution is more precisely expressed by defining
\bq
\label{cableJm}
J_m
\equiv
-\frac{R}{2}\frac{dJ_z}{dz}.
\eq
Thus, $J_m$ is the geometry-converted axial-current divergence. Its
definition follows directly from Ohm's law, its spatial derivative, and
axonal geometry and is independent of the membrane capacitance $C_m$.

\quad
Equation~\eqref{cableV} then gives the exact identity
\begin{equation}
J_I
=
J_m-C_m\frac{dV}{dt}.
\end{equation}
For the propagated solution, the total ionic current is therefore
reconstructed by subtracting the capacitive current from $J_m$. A more
detailed discussion of the terminology and physical distinctions among
these quantities is given in Appendix~A.

\subsection{Phase-space representation}

\begin{figure*}
\includegraphics[width=2.1\columnwidth]{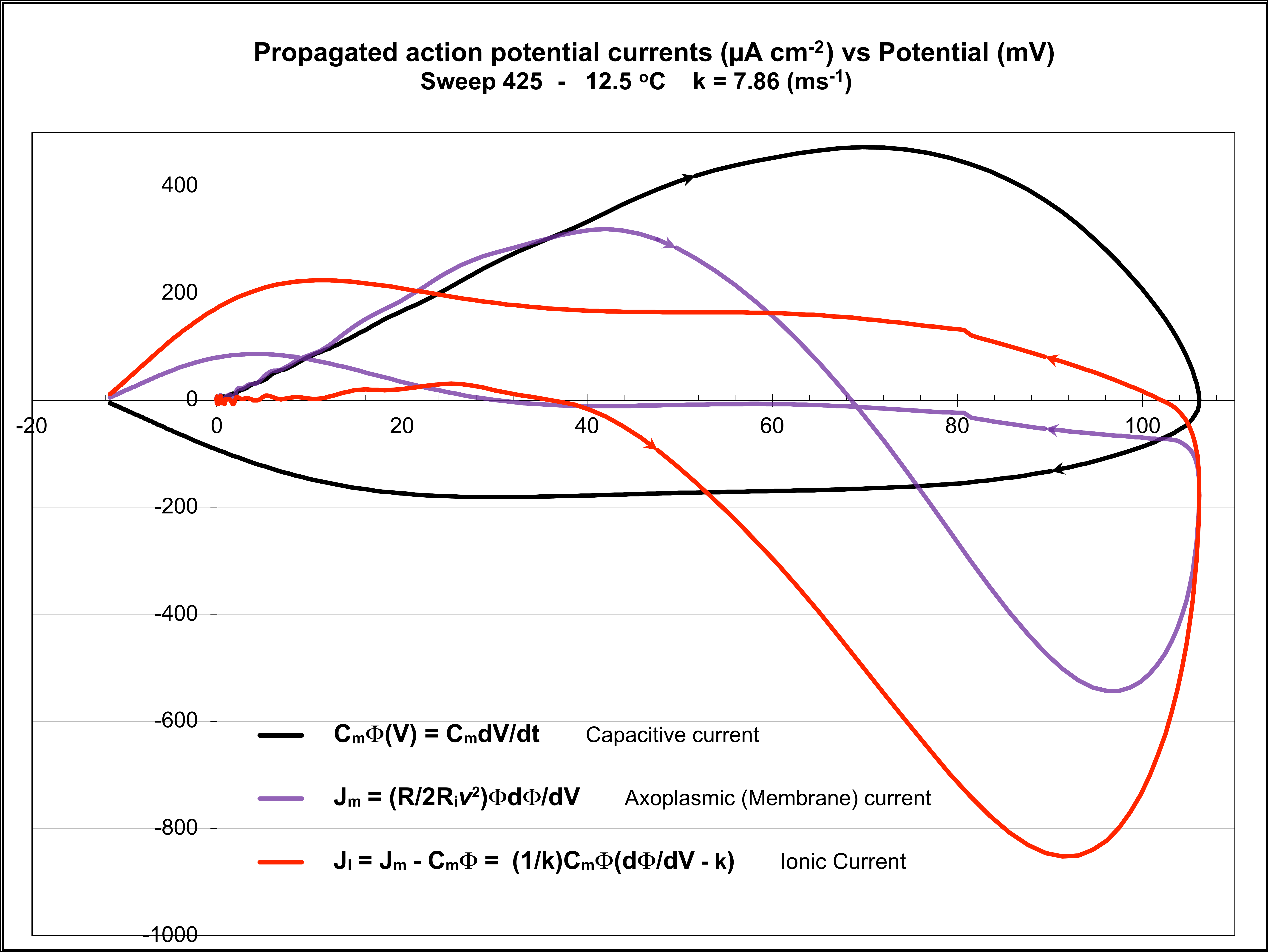}
\caption{\label{fig:CurrentsSweep425L}
Currents reconstructed from the phase-space trajectory for Sweep~425.
At points where $\Phi(V)\neq0$, the total ionic current vanishes
when $d\Phi/dV=k$, where $\Phi(V)=dV/dt$ and $k$ is the propagation
parameter. This condition is satisfied once on the rising branch and
once on the recovery branch, producing two ionic-current zero
crossings, marked at $V\approx35.6$~mV and
$V\approx102$~mV. The capacitive current, $J_m$, and the total ionic
current exhibit approximately quasilinear behaviour at the foot and
tail of the action potential and in shorter segments surrounding the
two zero crossings. These regions later provide effective conductances
and reversal potentials associated with distinct conducting states
(see Figs.~\ref{fig:RisingEdgeSweep425L} and
\ref{fig:RecoveryCurrentsSweep425L}). Arrows indicate the direction of
time along the trajectory.
}
\end{figure*}

\begin{figure*}
\includegraphics[width=2.08\columnwidth]
{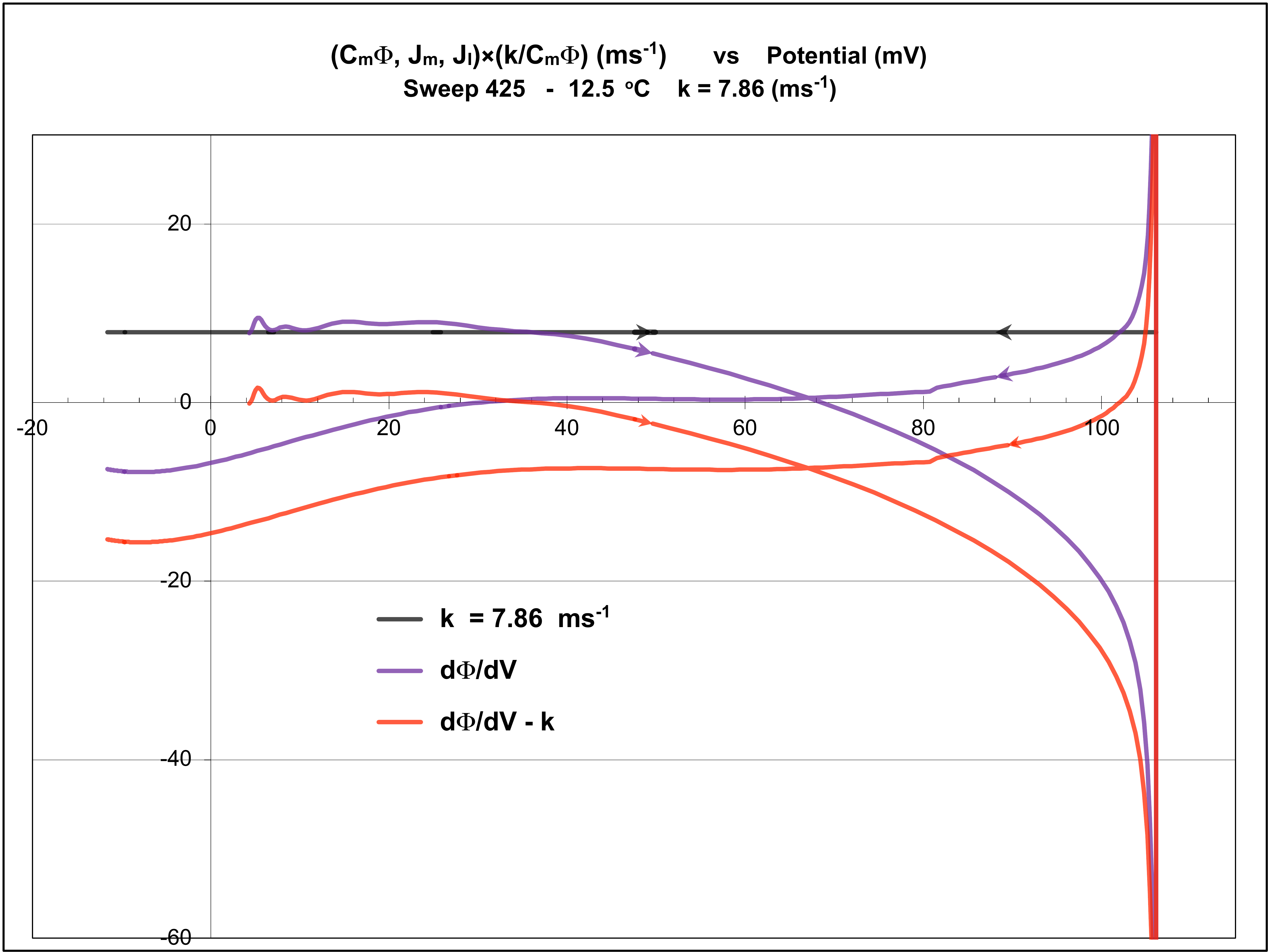}
\caption{\label{fig:Currents-kCmF-V-Sweep425L}
Normalized phase-space trajectories
$k(C_m\Phi,J_m,J_I)/(C_m\Phi)$ versus $V$.
Away from turning points, the condition $J_I=0$ corresponds to
$d\Phi/dV=k$, highlighting the two branchwise ionic-current zero
crossings satisfying Eq.~\eqref{cableVa}. The first occurs at
$V\approx35.6$~mV and the second at $V\approx102$~mV.
The normalized $J_I$ and $J_m$ ratios are discontinuous at the
action-potential peak because the normalizing quantity $C_m\Phi$
vanishes there; the underlying reconstructed currents remain
continuous.
}
\end{figure*}

\quad
Along each monotonic branch of the steadily propagating
action-potential trajectory, time can be represented locally as a
function of membrane potential. Defining
\[
\Phi(V)\equiv\frac{dV}{dt},
\qquad
\frac{d}{dt}
=
\Phi(V)\frac{d}{dV},
\]
and substituting into Eq.~\eqref{cableV} gives the phase-space cable
equation
\bq
\label{cableVa}
J_I(V)
=
C_m\Phi(V)
\left[
\frac{1}{k}\frac{d\Phi(V)}{dV}-1
\right],
\eq
where
\bq
\label{propagationkL}
k
=
\frac{2C_mR_iv^2}{R}
\eq
is the propagation parameter that combines the geometric and material
properties of the axon with the propagation velocity.

\quad
Away from the waveform turning points at which $\Phi(V)=0$, a key
consequence of Eq.~\eqref{cableVa} is that the total ionic current
vanishes when
\begin{subequations}
\label{cableVd_all}
\bq
\label{cableVd}
\frac{d\Phi(V)}{dV}
=
k,
\eq
that is, when $J_m$ equals the capacitive current,
\bq
\label{cableVe}
J_m
=
C_m\Phi.
\eq
\end{subequations}

\quad
The quantities entering the charge balance are displayed in
Fig.~\ref{fig:CurrentsSweep425L}. The capacitive current is
$C_m\Phi(V)$, and the axial Ohmic current is
\bq
\label{OhmV}
J_z(V)
=
-\frac{1}{vR_i}\Phi(V).
\eq
The corresponding geometry-converted axial-current divergence is
\bq
\label{MembraneV}
J_m(V)
=
\frac{C_m}{k}
\Phi(V)\frac{d\Phi(V)}{dV}.
\eq
The total ionic current is therefore
\bq
\label{IonicCV}
J_I(V)
=
C_m\Phi(V)
\left[
\frac{1}{k}\frac{d\Phi(V)}{dV}-1
\right].
\eq

\quad
The axial Ohmic current $J_z$ is defined per unit axonal
cross-sectional area. The capacitive current, $J_m$, and the total ionic
current are expressed on the membrane-area basis. All four quantities
are determined by correlations among $V$, $\Phi(V)$, and
$d\Phi(V)/dV$. Laboratory time does not appear as an independent
variable in these relations. Along each monotonic branch, however,
temporal direction is retained in the sign of $\Phi(V)$, and elapsed
time can be reconstructed from
\[
dt=\frac{dV}{\Phi(V)}.
\]
The formulation is therefore correlational rather than timeless: the
propagated trajectory determines both the current structure and its
branchwise temporal parametrization, up to an arbitrary time origin.

\quad
The experimental trajectories contain one branchwise ionic-current
zero crossing on each side of the action-potential peak. Their measured
locations and physical interpretation are examined below.

\quad
The identity
\[
\frac{d}{dt}
=
\Phi(V)\frac{d}{dV}
\]
is mathematically a change of variables. The physical content of the
present formulation arises when this transformation is combined with
charge conservation, steady propagation, and Ohm's law to produce a
closed relation among the measured trajectory, the axial Ohmic current,
$J_m$, the capacitive current, and the total ionic current. Unlike a
re-expression of preassigned voltage-clamp kinetics, the propagated
waveform itself determines the current balance and imposes structural
conditions, including the two branchwise ionic-current zero crossings.
This structure is therefore a consequence of charge conservation in a
propagating system and does not arise merely from changing variables in
a preexisting kinetic description.

\quad
The identification of the zero crossings is robust with respect to the
reconstruction procedure. Although evaluation of $J_I(V)$ requires
differentiation of the measured $V(t)$, the crossings correspond to
sign changes that are insensitive to moderate smoothing and persist
across the full temperature range of the Rosenthal--Bezanilla data
\cite{RB}. In particular, the recovery-branch crossing is a stable
feature of the reconstructed curves and does not depend on the details
of numerical differentiation.

\quad
The interpretation of the recovery-branch crossing begins with the
current balance imposed by the phase-space cable equation. Because
$J_z$ and $J_m$ are determined from the measured waveform, the sign of
the total ionic current is fixed by the difference between $J_m$ and
the capacitive current. The outward direction of $J_I(V)$ beyond the
recovery-branch crossing is therefore not an adjustable feature. Its
interpretation as a sodium-dominated recovery current requires the
additional continuity, reversal-potential, and channel-completion
analysis developed in Part~III.

\subsection{Structural implication for voltage-clamp descriptions}

\quad
In the Hodgkin--Huxley construction, ionic-current relations are
obtained from stationary voltage-clamp protocols and then inserted into
the cable equation to reconstruct propagation. This procedure
successfully reproduces the gross waveform, but stationary
voltage-clamp measurements do not independently isolate every
conducting state that may occur during propagation. Equation
\eqref{cableVa} instead determines $J_I(V)$ self-consistently from the
propagated waveform and can therefore reveal current structures not
resolved as independent components under stationary clamp conditions.
The recovery-branch zero crossing is one such feature.

\quad
The two zero crossings follow directly from the phase-space structure
of the propagated solution and do not depend on the functional form
later used to represent channel completion. The distinction between the
effective reversal potentials $V_M$ and $V_H$ emerges from the
quasilinear component analysis developed below and in Part~III.

\subsection{Two zero-current crossings from data}

\quad
For Sweep~170, the first crossing occurs at
\begin{subequations}
\bq
\label{cableVb}
\left.
\frac{d\Phi}{dV}
\right|_{V\approx38\,\mathrm{mV}}
\approx
4.12\,\mathrm{ms}^{-1},
\eq
\end{subequations}
and for Sweep~425 at
\begin{subequations}
\bq
\label{cableVc}
\left.
\frac{d\Phi}{dV}
\right|_{V\approx35.6\,\mathrm{mV}}
\approx
7.81\,\mathrm{ms}^{-1}.
\eq
\end{subequations}
Using the experimental parameters in Eq.~\eqref{propagationkL} gives
$k=4.1\,\mathrm{ms}^{-1}$ and $k=7.86\,\mathrm{ms}^{-1}$ for
Sweeps~170 and~425, respectively, consistent with the observed
crossings.

\quad
The second, recovery-branch crossing occurs at
\begin{subequations}
\bq
\label{cableVb2}
\left.
\frac{d\Phi}{dV}
\right|_{V\approx105.5\,\mathrm{mV}}
\approx
4.1\,\mathrm{ms}^{-1}
\qquad
\text{(Sweep~170)},
\eq
\bq
\label{cableVpeakb}
\left.
\frac{d\Phi}{dV}
\right|_{V\approx102\,\mathrm{mV}}
\approx
7.89\,\mathrm{ms}^{-1}
\qquad
\text{(Sweep~425)}.
\eq
\end{subequations}

\begin{figure*}
\includegraphics[width=2.1\columnwidth]{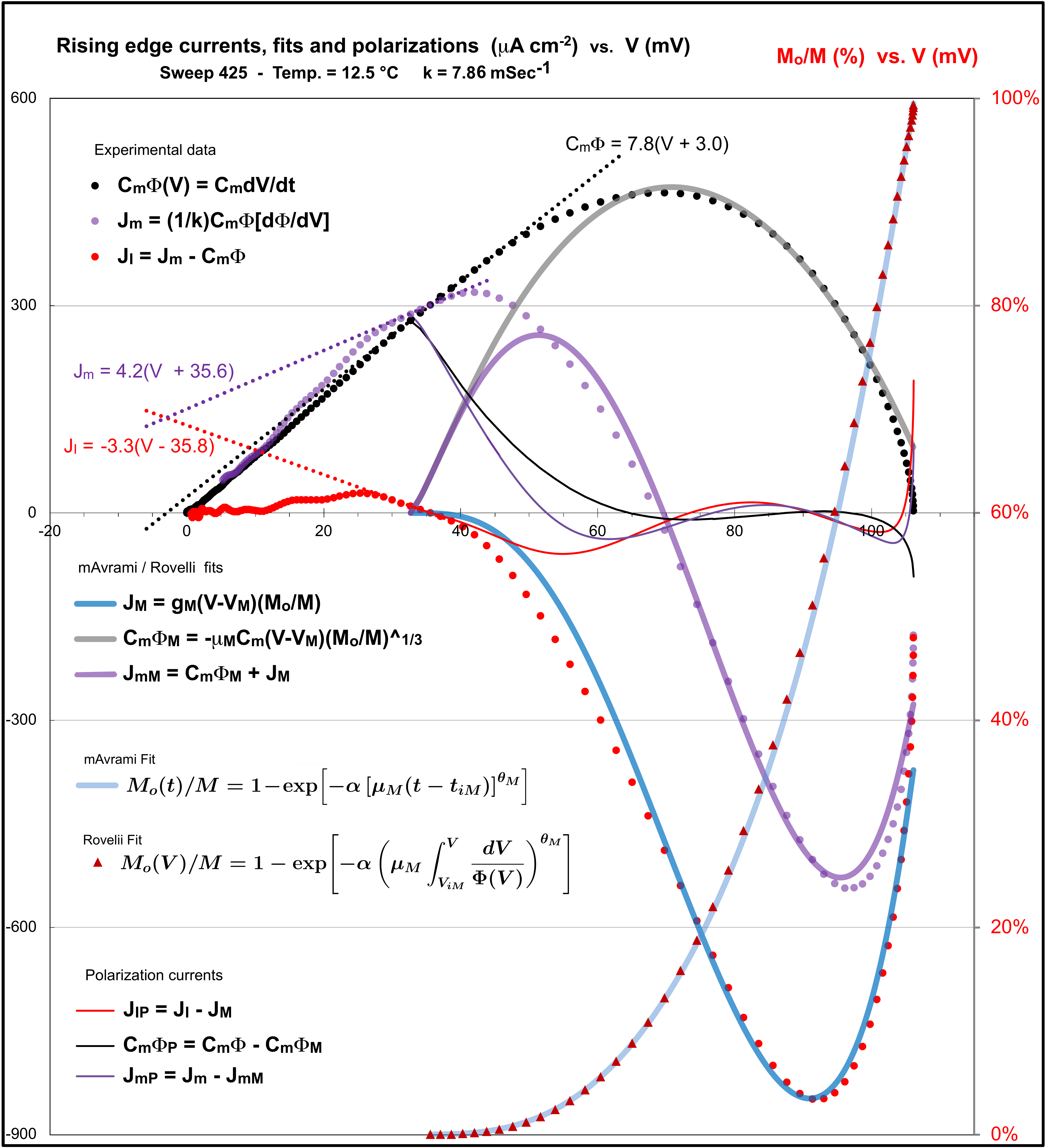}
\caption{\label{fig:RisingEdgeSweep425L}
\textbf{Rising-edge currents and channel dynamics in the phase-space
formulation of the propagated action potential.}
The capacitive current $C_m\Phi$, the geometry-converted
axial-current divergence $J_m$, and the total ionic current $J_I$ are
shown together with their charge-conserving mAvrami components
$(C_m\Phi_M,J_{mM},J_M)$ and complementary polarization components
$(C_m\Phi_P,J_{mP},J_{IP})$. At inception, the mAvrami components
vanish, while the polarization components are discontinuous but sum to
zero, preserving continuity of the reconstructed currents. As the
voltage increases, the ionic polarization current reverses sign and
the inward sodium activation current $J_M$ develops. The first ionic
zero-current crossing then occurs at $V\approx35.6$~mV, where
$d\Phi/dV\approx k$, confirming the phase-space cable equation. Near
the peak, the state-dependent decomposition components change branches,
while the experimentally reconstructed currents remain continuous. The
associated charge transfer is approximately
$10^{-8}\,\mathrm{C\,cm^{-2}}$, at which
$M_o/M\approx0.2$. The voltage-domain fit $M_o(V)/M$ is equivalent
branchwise to $M_o(t)/M$ through $dt=dV/\Phi(V)$
(Appendix~E). These features support the interpretation of a continuous
symmetry transition in the effective conductance at the peak. The curve
labeled ``Rovelli fit'' is obtained from the measured phase-space
velocity $\Phi(V)=dV/dt$, whereas the corresponding mAvrami fit is
obtained from the laboratory-time waveform. The two are distinct
representations of the same channel-completion law and are related
branchwise through $dt=dV/\Phi(V)$. The label acknowledges the
connection with Rovelli's relational formulation discussed in
Appendix~E; it does not attribute the fit itself to C. Rovelli.
}
\end{figure*}

\begin{figure*}
\includegraphics[width=2.1\columnwidth]{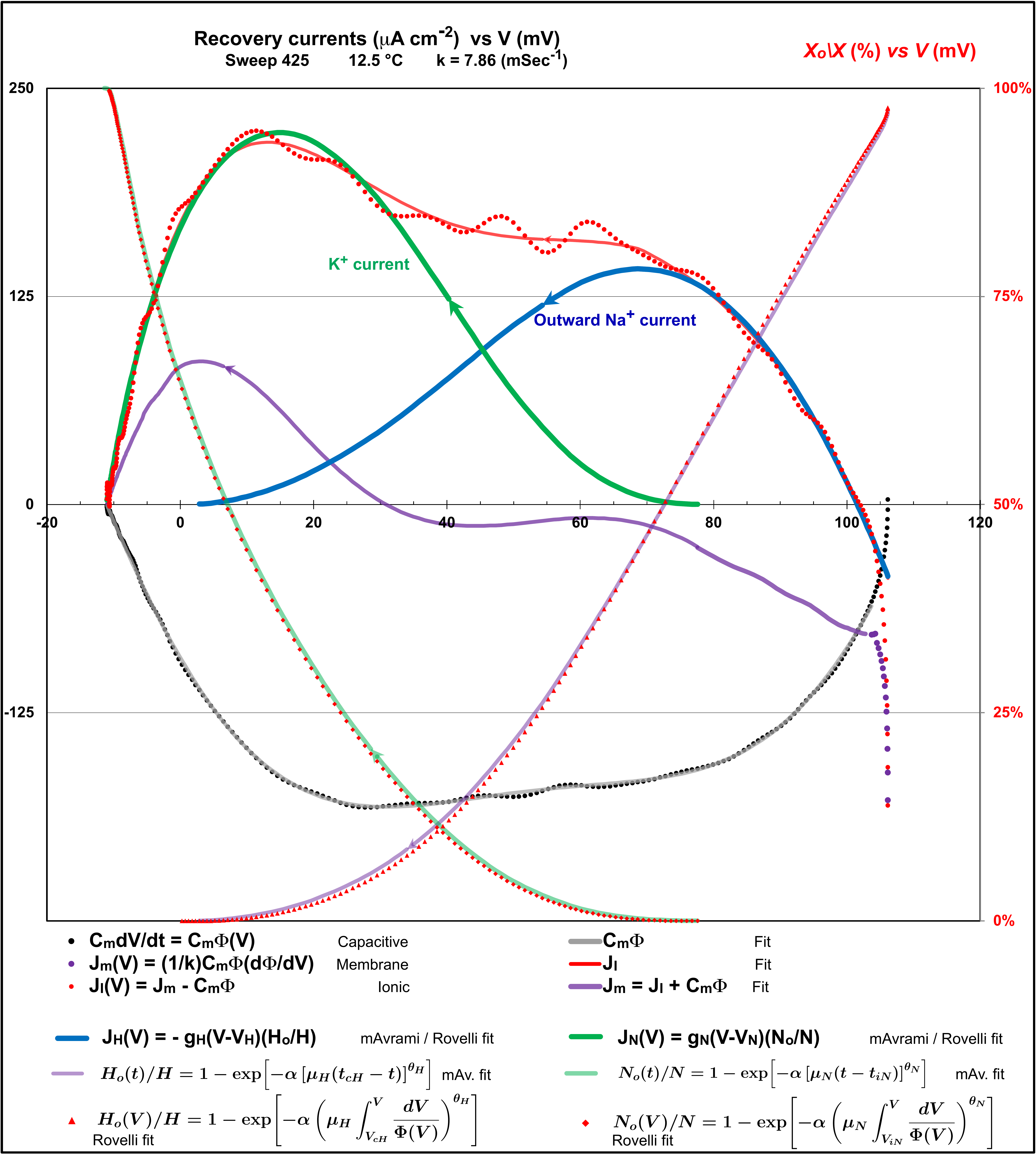}
\caption{\label{fig:RecoveryCurrentsSweep425L}
\textbf{Recovery currents and channel dynamics in the phase-space
formulation of the propagated action potential.}
The capacitive current $C_m\Phi(V)$, the geometry-converted
axial-current divergence $J_m(V)$, and the total ionic current $J_I(V)$
satisfy
$J_I=J_m-C_m\Phi$. A few millivolts below the peak, the second ionic
zero-current crossing occurs at $V\approx102$~mV. Decomposition of the
ionic current reveals effective sodium-dominated $J_H$ and
potassium-dominated $J_N$ contributions. The sodium component exhibits
a pronounced outward current during recovery. Solid curves represent
fits to the reconstructed data. The quasilinear slopes of the sodium
and potassium components determine their conductances, and their
intercepts with the zero-current axis determine their effective
reversal potentials. Channel fractions $X_o/X$ are shown on the right
axis and are represented by mAvrami and relational fits. The currents
and channel dynamics are displayed in a single phase-space
representation to preserve their intrinsic nonlinear coupling. The
curve labeled ``Rovelli fit'' is obtained from the measured phase-space
velocity $\Phi(V)=dV/dt$, whereas the corresponding mAvrami fit is
obtained from the laboratory-time waveform. The two are distinct
representations of the same channel-completion law and are related
branchwise through $dt=dV/\Phi(V)$. The label acknowledges the
connection with Rovelli's relational formulation discussed in
Appendix~E; it does not attribute the fit itself to C. Rovelli.
}
\end{figure*}

\begin{figure*}
\includegraphics[width=2.1\columnwidth]{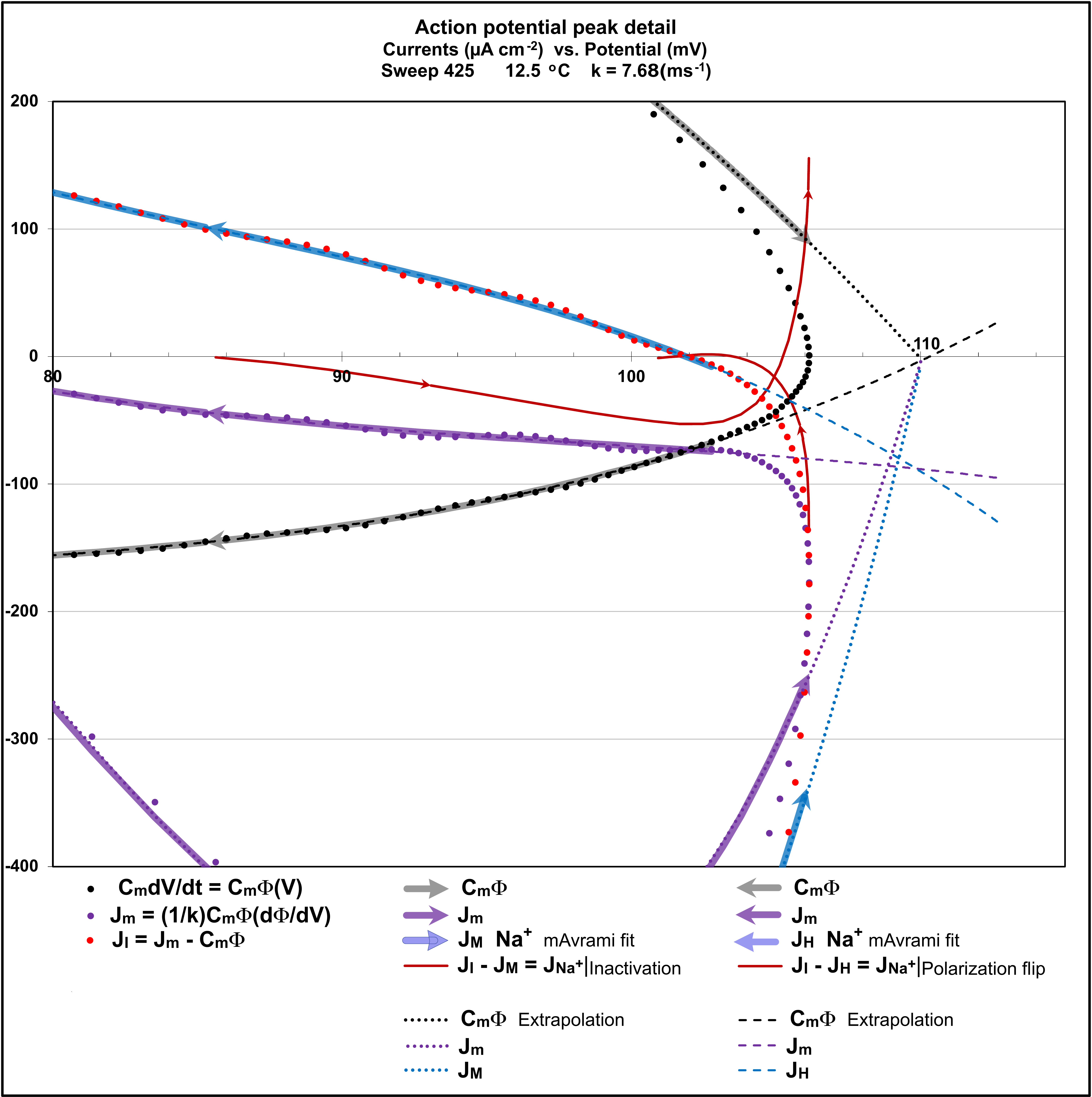}
\caption{\label{fig:APPeakSweep425L}
Currents in the vicinity of the action-potential peak $V_p$.
The capacitive current $C_m\Phi(V)$, the geometry-converted
axial-current divergence $J_m(V)$, and the total ionic current $J_I(V)$
satisfy
$J_I=J_m-C_m\Phi$. Arrows indicate the direction of time.
The mAvrami representations and associated ionic polarization
components change branches at $V_p$, marking a transition in the
decomposition while preserving continuity of the reconstructed
currents. Extrapolations of the three rising-edge mAvrami components and
the mAvrami capacitive recovery component intersect at the sodium
effective equilibrium potential $V_M$. By contrast, the recovery ionic
current crosses zero and extrapolates to an effective reversal
potential
$V_H\approx102\,\mathrm{mV}<V_M$, reflecting its relation to $J_m$
through the propagated charge balance.
}
\end{figure*}

\quad
The first crossing occurs during the transition from the initial
outward potassium-dominated current to the inward sodium-activation
current and primarily provides a consistency test of the
charge-conserving formulation
(Fig.~\ref{fig:RisingEdgeSweep425L}). The second occurs in the recovery
region and is more unexpected because it marks a reversal of the total
ionic current while the sodium completion fraction remains close to
unity
(Figs.~\ref{fig:RecoveryCurrentsSweep425L} and
\ref{fig:APPeakSweep425L}). Its interpretation as a sodium-dominated
outward component requires the continuity, reversal-potential, and
channel-completion analysis developed in Part~III.

\subsection{Quasilinear segments and parameter extraction}

\quad
The phase-space trajectories exhibit several quasilinear regions from
which physically meaningful parameters can be extracted. Four such
segments are particularly important: the foot and tail of the action
potential, where the three membrane-area quantities originate from and
return to the zero-current axis, respectively, and two shorter segments
associated with the two zero crossings of the total ionic current. The
first crossing occurs at the end of the action-potential foot and the
second shortly after the action-potential peak in the recovery region.
At these zero crossings,
\[
J_m=C_m\Phi.
\]

\quad
Throughout the component analysis, $J_I$ denotes the total ionic
current and is not restricted a priori to sodium and potassium alone.
The labels $K$, $M$, $H$, and $N$ identify effective dynamical
components whose dominant ionic associations are inferred from their
signs, quasilinear slopes, reversal potentials, and continuity along
the propagated trajectory. Chloride and other leakage currents are not
independently resolved by the present reconstruction; to the extent
that they contribute, they remain included in $J_I$. The ionic labels
should therefore be understood as dominant associations rather than as
an exhaustive separation of all ionic species.

\quad
At the foot of the action potential, where $\Phi(V)$ is approximately
linear in $V$, the three corresponding quantities become quasilinear
(Fig.~\ref{fig:RisingEdgeSweep425L}):
\begin{subequations}
\bq
\label{Foota}
C_m\Phi_K(V)
\approx
C_m\mu_KV,
\eq
\bq
\label{Footb}
J_{mK}
\approx
C_m\frac{\mu_K^2}{k}V,
\eq
\bq
\label{Footc}
J_K
\approx
C_m\mu_K
\left(
\frac{\mu_K}{k}-1
\right)V
\equiv
g_KV,
\eq
\bq
\label{Footd}
g_K
=
C_m\mu_K
\left(
\frac{\mu_K}{k}-1
\right).
\eq
\end{subequations}
Here $\mu_K$ and $g_K$ are the time rate and maximum conductance
associated with component $K$. The designation $K$ reflects the
historical association of the foot current with potassium-dominated
behaviour. In the present phase-space formulation, however, the
component is defined by its dynamical role in the propagated solution
rather than by an assumed exhaustive ionic identity.

\quad
Similarly, for the tail of the potassium-dominated recovery current
(Fig.~\ref{fig:RecoveryCurrentsSweep425L}),
\begin{subequations}
\bq
\label{Taila}
C_m\Phi_N(V)
\approx
-C_m\mu_N(V-V_N),
\eq
\bq
\label{Tailb}
J_{mN}
\approx
C_m\frac{\mu_N^2}{k}(V-V_N),
\eq
\bq
\label{Tailc}
J_N
\approx
C_m\mu_N
\left(
\frac{\mu_N}{k}+1
\right)
(V-V_N)
\equiv
g_N(V-V_N),
\eq
\bq
\label{Taild}
g_N
=
C_m\mu_N
\left(
\frac{\mu_N}{k}+1
\right).
\eq
\end{subequations}
Here
$\mu_N\approx10.8\,\mathrm{ms}^{-1}$ and
$g_N\approx25.6\,\mathrm{mS\,cm^{-2}}$
represent the time rate and maximum conductance of the
potassium-dominated recovery component $N$, while
$V_N\approx-10.9\,\mathrm{mV}$ is its effective reversal potential.
These relations connect the measured phase-space slopes, the
propagation parameter $k$, and the corresponding conductances and time
rates.

\quad
Related relations between ionic time rates and maximum conductances in
quasilinear regions of the nerve current were discussed by Scott
\cite{Scott}. In the present formulation, these relations follow
directly from the charge-conserving phase-space cable equation and
explicitly include the propagation parameter $k$.

\quad
The first ionic zero-current crossing defines a short quasilinear
segment visible in Fig.~\ref{fig:RisingEdgeSweep425L}, where the three
corresponding slopes are displayed.

\quad
Of particular importance is the recovery-region branch associated with
the second ionic zero-current crossing
(Figs.~\ref{fig:RecoveryCurrentsSweep425L} and
\ref{fig:APPeakSweep425L}). In Part~III, this branch is shown to contain
an additional current component,
\bq
\label{OutwardL}
J_H
\approx
-g_H(V-V_H),
\eq
whose phase-space trajectory remains continuous with the
sodium-dominated inward branch. This continuity, together with the
reversal-potential and channel-completion results, motivates its
interpretation as a sodium-dominated deactivation current during
recovery. From the figure,
\[
V_H\approx102\,\mathrm{mV},
\qquad
g_H\approx11.95\,\mathrm{mS\,cm^{-2}}.
\]

\quad
The inception of the inward sodium-dominated component $J_M$ is inferred
by extrapolating the negative-slope branch of the total ionic current.
Charge conservation requires the associated capacitive and $J_m$
components to originate at the same effective equilibrium potential.
In the extrapolated quasilinear region, the corresponding relations are
\begin{subequations}
\bq
\label{Sodiuma}
C_m\Phi_M(V)
\approx
-C_m\mu_M(V-V_M),
\eq
\bq
\label{Sodiumb}
J_{mM}
\approx
C_m\frac{\mu_M^2}{k}(V-V_M),
\eq
\bq
\label{Sodiumc}
J_M
\approx
C_m\mu_M
\left(
\frac{\mu_M}{k}+1
\right)
(V-V_M)
\equiv
g_M(V-V_M),
\eq
\bq
\label{Sodiumd}
g_M
=
C_m\mu_M
\left(
\frac{\mu_M}{k}+1
\right).
\eq
\end{subequations}

\quad
The values of $\mu_M$ and $g_M$ cannot be determined directly from the
unseparated phase-space trajectories because the additional
polarization current prevents the waveform from reaching the
extrapolated equilibrium potential $V_M$. In Part~III, introduction of
a functional form for the open-channel fraction permits separation of
the overlapping contributions. The resulting fits determine the time
rate $\mu_M$ and maximum conductance $g_M$.

\subsection{Summary of Part II}

\quad
Part~II establishes a self-consistent phase-space and correlational
formulation of the charge-conserving cable equation. From the measured
waveform $V(t)$ and the propagation parameter $k$, the axial Ohmic
current $J_z$, the geometry-converted axial-current divergence $J_m$,
the capacitive current, and the total ionic current are determined
through branchwise relations among $V$, $\Phi(V)$, and
$d\Phi(V)/dV$. Laboratory time is not an independent variable in these
relations, but temporal direction and elapsed time remain encoded
through $\Phi(V)$. Away from the waveform turning points at which
$\Phi(V)=0$, the theory predicts the total-ionic-current zero condition
$d\Phi(V)/dV=k$, and the experimental data confirm one such crossing
on each branch of the propagated action potential. The recovery-branch
crossing motivates the decomposition of $J_I(V)$ into activation,
polarization, and deactivation components developed in Part~III.
\section{\label{sec:level2}Phase-space reconstruction, channel kinetics, and relational parametrization}
\subsection{mAvrami representation of channel completion}

\quad
Part~II established a charge-conserving phase-space formulation of the
cable equation in which the axial Ohmic current $J_z$, the
geometry-converted axial-current divergence $J_m$, the capacitive
current, and the total ionic current are reconstructed directly from
the measured waveform $V(t)$. The propagation parameter $k$ follows
from the geometric and material properties of the propagated solution,
while quasilinear segments of the reconstructed current curves provide
effective slopes and reversal potentials. Complete separation of the
overlapping ionic and polarization components, and hence determination
of their individual time rates $\mu_X$, maximum conductances $g_X$,
and state fractions $X_o/X$, requires the additional kinetic
representation developed in Part~III.
\begin{subequations}
\label{eq:currents}

\begin{equation}
C_m\Phi_K(V)
\approx
C_m\mu_K(V-V_K),
\label{subeq:currentsa}
\end{equation}
\begin{equation}
J_K(V)
\approx
g_K(V-V_K),
\label{subeq:currentsb}
\end{equation}
\begin{equation}
C_m\Phi_M(V)
=
-C_m\mu_M(V-V_M)\frac{M_o}{M},
\label{subeq:currentsc}
\end{equation}
\begin{equation}
J_M(V)
=
g_M(V-V_M)\frac{M_o}{M},
\label{subeq:currentsd}
\end{equation}
\begin{equation}
J_{mM}(V)
=
J_M(V)+C_m\Phi_M(V),
\label{subeq:currentse}
\end{equation}
\newline
$J_{MP}(V)
=
J_I(V)-J_K(V)-J_M(V)$
\begin{equation}
\approx
\sum_{n=1}^{4}
g_{Pn}(V-V_{Pn})\frac{Pn_o}{Pn},
\label{subeq:currentsf}
\end{equation}
\begin{equation}
J_H(V)
=
-g_H(V-V_H)\frac{H_o}{H},
\label{subeq:currentsg}
\end{equation}
\begin{equation}
J_N(V)
=
g_N(V-V_N)\frac{N_o}{N},
\label{subeq:currentsh}
\end{equation}
\newline
$J_{HP}(V)
=
J_I(V)-J_H(V)-J_N(V)$
\begin{equation}
\approx
\sum_{n=5}^{6}
g_{Pn}(V-V_{Pn})\frac{Pn_o}{Pn}.
\label{subeq:currentsi}
\end{equation}
\end{subequations}

\noindent
The foot current $J_K$ is retained as the weak quasilinear component
reconstructed in Part~II and is not assigned an independent mAvrami
completion law. For the rising-edge polarization current,
$n=1,2$ denotes the post-inception polarization segment, with
$g_1=g_2=g_{12}$ and $V_{P1}=V_{P2}=V_{P12}$, whereas $n=3,4$
denotes the pre-peak polarization segment, with
$g_3=g_4=g_{34}$ and $V_{P3}=V_{P4}=V_{P34}$. For the recovery
polarization current, $n=5,6$.

\quad
In the Hodgkin--Huxley framework, the completion fractions are
represented phenomenologically by empirical gating variables. Here we
adopt a different representation: during steady propagation, each
channel-opening or channel-closing process follows a single S-shaped
completion law. We implement this choice using the Avrami equation,
\begin{equation}
\frac{X_o(t)}{X}
=1-\exp\left[-A_X (t-t_{iX})^{\theta_X}\right].
\end{equation}
Here $A_X$ is the Avrami parameter, $\theta_X$ is a dimensionless
exponent, and $t_{iX}$ denotes the inception time of the process. For
closing processes, the corresponding closure time is denoted by
$t_{cX}$.

\quad
The application of Avrami kinetics to axonal conductance has an early precedent in the work of Cope, who showed that the rise of the Hodgkin--Huxley potassium conductance following depolarization could be fitted by the Avrami equation \cite{Cope}. The present modified Avrami representation differs in that the temperature-dependent time rate is incorporated explicitly and the formulation is applied to sodium activation, sodium deactivation, potassium activation, and the associated polarization processes reconstructed from propagated action-potential data.

\quad
Fits across temperature reveal that the Avrami parameter $A_X$ may be
rewritten in terms of the time rate $\mu_X$ and exponent $\theta_X$ as
\begin{equation}
A_X
\approx
\alpha_X(\mu_X)^{\theta_X}.
\end{equation}
The empirical stability and interpretation of the dimensionless
coefficient $\alpha_X$ are discussed in the following subsection. This
scaling motivates the modified Avrami (mAvrami) completion law,
\begin{equation}
\frac{X_o(t)}{X}
=
1-\exp\left[
-\alpha_X
\left\{
\mu_X(t-t_{iX})
\right\}^{\theta_X}
\right].
\label{mAvrami}
\end{equation}
The parameter $\alpha_X$ is dimensionless. In the final constrained
fits developed below, the process-dependent coefficients are replaced
by the common value $\alpha_X\equiv\alpha$.

\quad
For opening processes, $X_o/X$ denotes an increasing completion
fraction. For closing processes, the corresponding quantity denotes
the fraction of the active state remaining before closure is complete.
Thus, $H_o/H$ and the closing polarization fractions decrease toward
zero as $t$ approaches their respective closure or completion times.
The completed fraction of the H-channel closing process is
$1-H_o/H$, with an analogous definition for the closing polarization
segments. This distinction is used when the monotone completion
functional is introduced in Appendix~E.

\quad
The completion fractions used for the ionic and polarization components
are therefore written as
\begin{subequations}
\label{mAvramiEQS}
\begin{equation}
\frac{M_o(t)}{M}
=
1-\exp\left[-\alpha\left(\mu_M (t-t_{iM})\right)^{\theta_M}\right]
\label{mAvramia}
\end{equation}
\begin{equation}
\frac{H_o(t)}{H}
=
1-\exp\left[-\alpha\left(\mu_H (t_{cH}-t)\right)^{\theta_H}\right]
\label{mAvramib}
\end{equation}
\begin{equation}
\frac{N_o(t)}{N}
=
1-\exp\left[-\alpha\left(\mu_N (t-t_{iN})\right)^{\theta_N}\right]
\label{mAvramic}
\end{equation}
\newline
$\frac{Pn_o(t)}{Pn}
=
1-\exp\left[-\alpha\left(\mu_{Pn}(t-t_{iPn})\right)^{\theta_{Pn}}\right],$
\begin{equation}
\qquad n=1,3,4
\label{mAvramid}
\end{equation}
$\frac{Pn_o(t)}{Pn}
=
1-\exp\left[-\alpha\left(\mu_{Pn}(t_{cPn}-t)\right)^{\theta_{Pn}}\right],$
\begin{equation}
\qquad n=2,5,6
\label{mAvramie}
\end{equation}

\end{subequations}
Here $t_{iM}$ and $t_{iN}$ are the inception times for sodium
M-channel and potassium N-channel opening, $t_{cH}$ is the closure
time for the sodium H-channel, $t_{iPn}$ are the inception times
for polarization segments $n=1,3,4$, and $t_{cPn}$ are the completion
times for polarization segments $n=2,5,6$.

\quad
Within the present framework, each reconstructed ionic or polarization
component is represented by a conductance factor, a driving-force
factor, and an mAvrami completion fraction. This parametrization provides a unified
description of channel opening, channel closing, and polarization
processes across the identified symmetry regions of the propagated
action potential.
\subsection{Universal scaling of channel kinetics}
\begin{figure*}
\centering
\includegraphics[width=2.1\columnwidth]{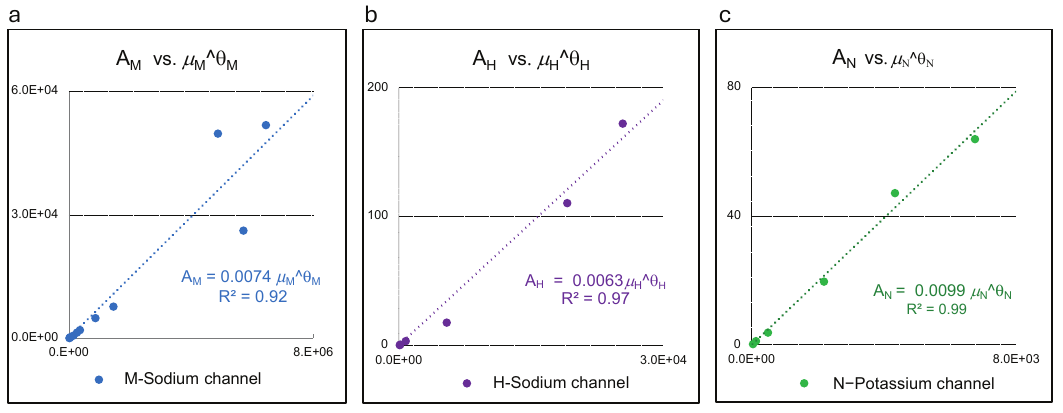}
\caption{\label{AvramiParL}
Avrami parameters $A_M$, $A_H$, and $A_N$ plotted against
$(\mu_M)^{\theta_M}$, $(\mu_H)^{\theta_H}$, and
$(\mu_N)^{\theta_N}$, respectively.
\newline
\textbf{(a) $A_M$ vs.\ $(\mu_M)^{\theta_M}$.}
Points correspond to measurements at 11 temperatures. While the rate
$\mu_M$ increases exponentially with temperature, the exponent
$\theta_M \approx 4$ and proportionality coefficient $\alpha_M$ remain
approximately constant. The fitted slope gives
$\alpha_M \approx 0.0074$, with values lying in the range
$\alpha_M \approx (0.0060$--$0.0075)$.
\newline
\textbf{(b) $A_H$ vs.\ $(\mu_H)^{\theta_H}$.}
Data from Sweeps 170, 220, 327, 525, 630, and 695 show that the scaling
$A_H \propto (\mu_H)^{\theta_H}$ also holds for the recovery current
$J_H$. The corresponding prefactor $\alpha_H$ is approximately
constant, with values in the range
$\alpha_H \approx (0.007$--$0.009)$, reflecting the distinct channel
state during recovery.
\newline
\textbf{(c) $A_N$ vs.\ $(\mu_N)^{\theta_N}$.}
For potassium N-channels, $\mu_N$ grows exponentially with temperature
while $\alpha_N$ and $\theta_N \approx 3$ remain approximately
constant. The prefactor lies in the range
$\alpha_N \approx (0.009$--$0.011)$.
Across all three channel types, the fitted proportionality coefficients
lie within the interval $\alpha_X \approx (0.006$--$0.011)$, with the
M- and H-channel values concentrated near
$(6.5$--$8.5)\times 10^{-3}$. This empirical collapse motivates fixing
$\alpha_X$ to a single constant in subsequent fits, reducing the number
of free parameters without degrading agreement with experiment.
}
\end{figure*}
\begin{figure*}
\centering
\includegraphics[width=2.1\columnwidth]{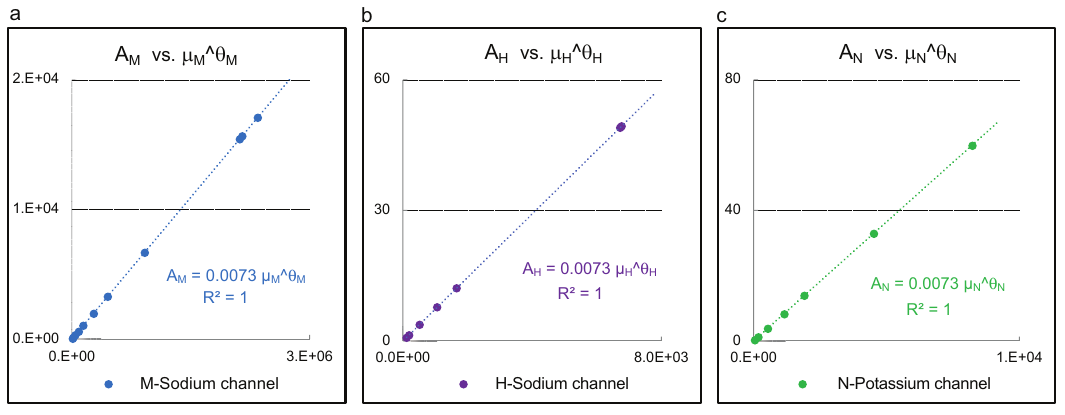}
\caption{\label{AlphaPar4L}
Fits of ionic-current parameters using fixed completion-law constants
$\alpha=0.007297$, $\theta_M=3.78$, $\theta_H=3$, and $\theta_N=3$
(see SM Fig.~\ref{fig:TableMHNL} for parameter values).
\newline
{\bf (a) $A_M$ vs.\ $(\mu_M)^{\theta_M}$.}
Points correspond to measurements at 11 temperatures. Although
$\mu_M$ increases exponentially with temperature, both $\theta_M$ and
$\alpha$ remain effectively temperature independent.
\newline
{\bf (b) $A_H$ vs.\ $(\mu_H)^{\theta_H}$.}
Data from Sweeps 170, 220, 327, 525, 630, and 695 show that the scaling
relation $A_H=\alpha(\mu_H)^{\theta_H}$ also describes the recovery
current $J_H$. Differences relative to M-channel parameters reflect the
change in channel state between activation and recovery.
\newline
{\bf (c) $A_N$ vs.\ $(\mu_N)^{\theta_N}$.}
For potassium N-channels, $\mu_N$ grows exponentially with temperature,
whereas $\alpha$ and $\theta_N=3$ remain approximately constant.
}
\end{figure*}

\quad
Fits across temperature reveal a robust empirical correlation between
the Avrami parameter $A_X$, the time rate $\mu_X$, and the exponent
$\theta_X$, as shown in Fig.~\ref{AvramiParL}:
\begin{equation}
A_X \approx \alpha_X(\mu_X)^{\theta_X}.
\label{eq:AXScaling}
\end{equation}
For all reconstructed channel processes, the Avrami coefficients $A_X$ exhibit a power-law
dependence on the corresponding rates, $A_X \propto
(\mu_X)^{\theta_X}$, with exponents $\theta_M \approx 4$ and
$\theta_N \approx 3$ that remain approximately constant across
temperatures.

\quad
A key feature of these fits is the approximate stability of the dimensionless prefactors \(\alpha_X\). When treated as independent fitting parameters, their values remain approximately temperature independent and lie within the interval \(\alpha_X\approx0.006\)--\(0.011\), with the M- and H-channel values concentrated near \(6.5\times10^{-3}\)--\(8.5\times10^{-3}\). The spread, particularly for the N component, does not establish exact universality, but it motivates testing a common dimensionless scaling constraint across the reconstructed processes.

\quad
Accordingly, subsequent fits impose the constrained ansatz \(\alpha_X\equiv\alpha\). In the final fits presented here, the common prefactor is fixed to
\begin{equation}
\alpha=0.007297\ldots .
\label{eq:commonAlpha}
\end{equation}
This constraint reduces the number of adjustable parameters without degrading agreement with experiment, as shown in Fig.~\ref{AlphaPar4L}. The numerical value equals the fine-structure constant to the quoted precision, but the fitting procedure should be regarded as a consistency test rather than as evidence that the prefactor has been derived from electromagnetic theory.

\quad
Preliminary unconstrained fits exhibited several closely spaced parameter sets with comparable residuals, consistent with partial compensation among the fitted parameters. Imposing the common prefactor selected a single stable solution and reduced this compensation. The constrained fit therefore indicates that a common dimensionless prefactor acts as a strong organizing condition on the combined kinetics. A quantitative comparison of the competing minima would be required to establish uniqueness.

\quad
The appearance of a separate dimensionless coefficient is natural once the characteristic rate scale is represented independently by \(\mu_X\). The noteworthy empirical observation is the clustering of this coefficient across distinct channel processes and temperatures. Its numerical proximity to the fine-structure constant may indicate a deeper structural relation, but its physical origin is not established by the present analysis.

\quad
The observed scaling motivates the use of the mAvrami completion law
introduced above throughout the remainder of this work. In the next
subsection, the resulting completion fractions are used to reconstruct
the propagated action potential and its constituent current components.

\subsection{Reconstruction of the propagated action potential}

\quad
We begin with the recovery region because its decomposition is most directly constrained by the observed quasilinear segments in Figs.~\ref{fig:RecoveryCurrentsSweep425L} and \ref{fig:APPeakSweep425L}. The reconstructed recovery trajectory contains two distinguishable ionic branches: the sodium-associated recovery component \(J_H\) and the outward potassium-dominated component \(J_N\). A smaller residual polarization current \(J_{HP}\) occupies the interval immediately following the action-potential peak.

\quad
In the recovery region, we write
\begin{equation}
J_I(V)=J_H(V)+J_N(V)+J_{HP}(V).
\label{RecoveryDecomposition}
\end{equation}
Here \(J_H\) is the mAvrami sodium-associated recovery current, \(J_N\) is the potassium-dominated recovery current, and \(J_{HP}\) is the recovery polarization current. In the narrow interval between the action-potential peak and the second zero-current crossing, \(J_N\) is small; within that interval, \(J_{HP}\approx J_I-J_H\). The general decomposition, however, retains all three contributions.

\quad
The rising-edge decomposition is less direct. The action-potential foot contains a quasilinear component, but the inward sodium-dominated current does not reach an evident fully developed quasilinear segment before the peak. This absence suggests that overlapping polarization currents remain important in the negative-resistance region. When points extending beyond the onset of this region are included in a single activation fit, the fitted Avrami exponent becomes unstable; restricting the fit to the activation branch yields an exponent close to, but below, four.

\quad
Using the completion fractions defined above, the measured ionic current is decomposed into a weak foot component and two rising-edge symmetry regions with distinct polarization kinetics:
\begin{equation}
J_I(V)=J_K(V)+J_M(V)+J_{MP}(V).
\label{RisingDecomposition}
\end{equation}
Here \(J_K\) is the weak foot current, \(J_M\) is the sodium M-channel activation current described by the mAvrami completion law, and \(J_{MP}\) is the associated ionic polarization current (Fig.~\ref{fig:RisingEdgeSweep425L}; SM Figs.~\ref{fig:RisingEdgeCurrentsSweep170L} and \ref{fig:Currents525_695L}). The polarization contribution consists of two post-inception segments, \(J_{P1}\) and \(J_{P2}\), extending from inception through the capacitive-current maximum, and two pre-peak segments, \(J_{P3}\) and \(J_{P4}\), spanning the negative-resistance region.

\quad
In the negative-resistance region, the pre-peak polarization current is resolved into two consecutive stages with a common fitted conductance and reversal potential, \(g_{P3}=g_{P4}\) and \(V_{P3}=V_{P4}<V_p\). Apart from the narrow polarization-flip interval, this conductance is comparable to \(g_M\). For most sweeps, the associated rates satisfy \(\mu_{P3}\approx\mu_M\) and \(\mu_{P4}\approx2\mu_M\). The latter relation characterizes the P4 polarization process that initiates sodium-channel inactivation; it should not be interpreted as a kinetic connector between activation and deactivation. Sweep~170 and Sweep~525 are exceptions: in Sweep~170 the ordering of the two rates is reversed (Fig.~\ref{fig:mAvramiSweep170L}), whereas Sweep~525 is described by a single pre-peak polarization segment with \(\mu_{P4}\approx1.36\mu_M\) (SM Fig.~\ref{fig:mAvramiLabFitSweep525L}).

\begin{figure*}
\centering
\includegraphics[width=2.1\columnwidth]{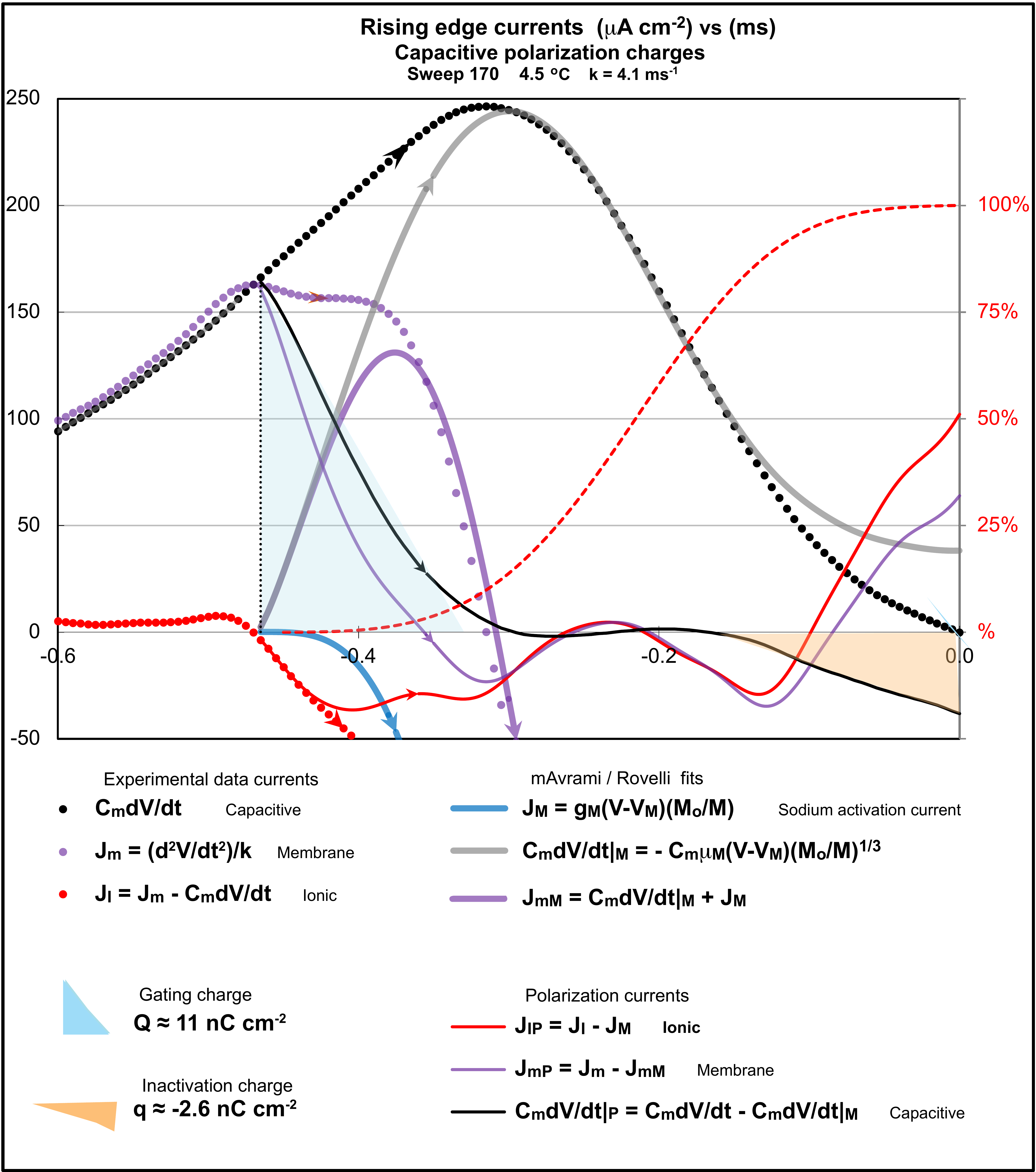}
\caption{\label{fig:GatingChargeSweep170L}
Rising-edge currents plotted versus time, showing the capacitive
current, $J_m$, and the total ionic current, together with their
corresponding polarization components and mAvrami fits. The fitted currents
$C_m\Phi_M$, $J_{mM}$, and $J_M$, as well as the fraction
$M_o(t)/M$, originate at the inception time $t=t_{iM}$, marking the
onset of M-channel opening. The polarization currents $C_m\Phi_{MP}$
and $J_{mMP}$ begin abruptly at this point and subsequently decay.
The integral of the first polarization segment of $C_m\Phi_{MP}$
corresponds to a transferred charge
$Q\approx11\,\mathrm{nC\,cm^{-2}}$, representing the gating charge displaced within the membrane electric
field during channel activation. This charge
displacement precedes significant channel opening and decreases near
the maximum rate of rise of the action potential, where the
open-channel fraction is $M_o/M \approx 0.2$. The magnitude of $Q$ is
approximately temperature independent across all sweeps. Near the
action-potential peak, prior to the transition to the recovery state, a
second polarization component of opposite sign transfers a smaller
charge $q \approx -2.6\,\mathrm{nC\,cm^{-2}}$.
}
\end{figure*}

\quad
The mAvrami representation of the sodium current yields quantitative
predictions for the fraction of open channels, the associated gating
charge, and the inception potential of the inward sodium current. These features are illustrated in Fig.~\ref{fig:GatingChargeSweep170L}
and SM Fig.~\ref{fig:GatingChargeSweep525-695L}.
\quad
The predicted gating-charge transfer is largely complete when only
$\sim20\%$ of the sodium channels are open, as inferred from the present
decomposition of the currents. The same representation reproduces these
features across temperatures, suggesting that the mAvrami
form provides a consistent description of the channel-opening dynamics.

\quad
The mAvrami form is an empirically constrained and parsimonious representation rather than an arbitrary fitting choice. It is consistent with several experimental observations: sodium inactivation follows and overlaps activation rather than appearing as a wholly independent process \cite{BezArm1977,ArmBez1977}; the reconstructed gating-current time course and integrated gating-charge transfer are of the magnitude reported experimentally \cite{ArmstrongBezanilla1,ArmstrongBezanilla,KeynesRojas}; and measurements associated with inactivation show little net gating-charge transfer \cite{BezArm1977,Bezanilla}. 

\quad
The displacement of gating charge before the development of substantial ionic conductance is well established experimentally. In squid potassium channels, the pronounced separation between the charge--voltage and conductance--voltage relations shows that a large fraction of the gating charge moves among nonconducting states \cite{Bezanilla}. Related sodium-channel experiments demonstrate that activation gating-current movement precedes and overlaps the development of sodium conductance. The present reconstruction extends these qualitative observations by predicting a quantitative correspondence for the propagated action potential: the activation gating charge is largely transferred when the reconstructed sodium open-channel fraction is approximately $M_o/M\approx0.2$. This numerical fraction is a result of the present decomposition and should not be attributed directly to the earlier voltage-clamp experiments.

\quad
Within the present framework, the small net inactivation-associated charge results from polarization contributions of opposite sign on the two sides of the pre-peak reversal. Their partial cancellation provides a possible explanation for the experimentally observed near-zero net transfer. Taken together, these constraints support the mAvrami law as a parsimonious representation consistent with the reconstructed current balance and the cited experimental observations. They do not establish that it is mathematically unique among all monotone saturating kinetic laws.

\quad
The resulting decomposition reconstructs the experimental currents
throughout the propagated action potential. The experimentally
reconstructed capacitive current, $J_m$, and total ionic current remain
continuous through the action-potential peak, whereas discontinuities
appear only in the branch-specific fitted components. The reconstruction
therefore remains constrained by the charge-conserving phase-space
cable equation rather than by independent voltage-clamp fitting
assumptions. It also reveals the symmetry transition and the separation
between activation and recovery reversal potentials examined in the
following subsection.
\subsection{Symmetry transition and outward sodium current}

\quad
The reconstruction described above exposes a branch-dependent structure of the propagated solution that is not represented by a single-valued voltage-clamp current relation. In phase space, the total ionic current vanishes when
\[
\frac{d\Phi}{dV}=k,
\]
and the experimental trajectory satisfies this condition on both the rising and recovery branches. The two zero-current crossings therefore provide direct evidence of a two-branch phase-space current structure.

\quad
This branch dependence is consistent with a hysteretic interpretation because the same membrane potential is traversed on the rising and recovery branches with different values of \(\Phi(V)\), different current decompositions, and different effective intercepts. The two crossings alone do not prove thermodynamic hysteresis; rather, they identify the branch-dependent structure whose hysteretic interpretation is examined in Appendix~B and
Fig.~\ref{fig:HysteresisL}.

\begin{figure}[t]
\includegraphics[width=\columnwidth]{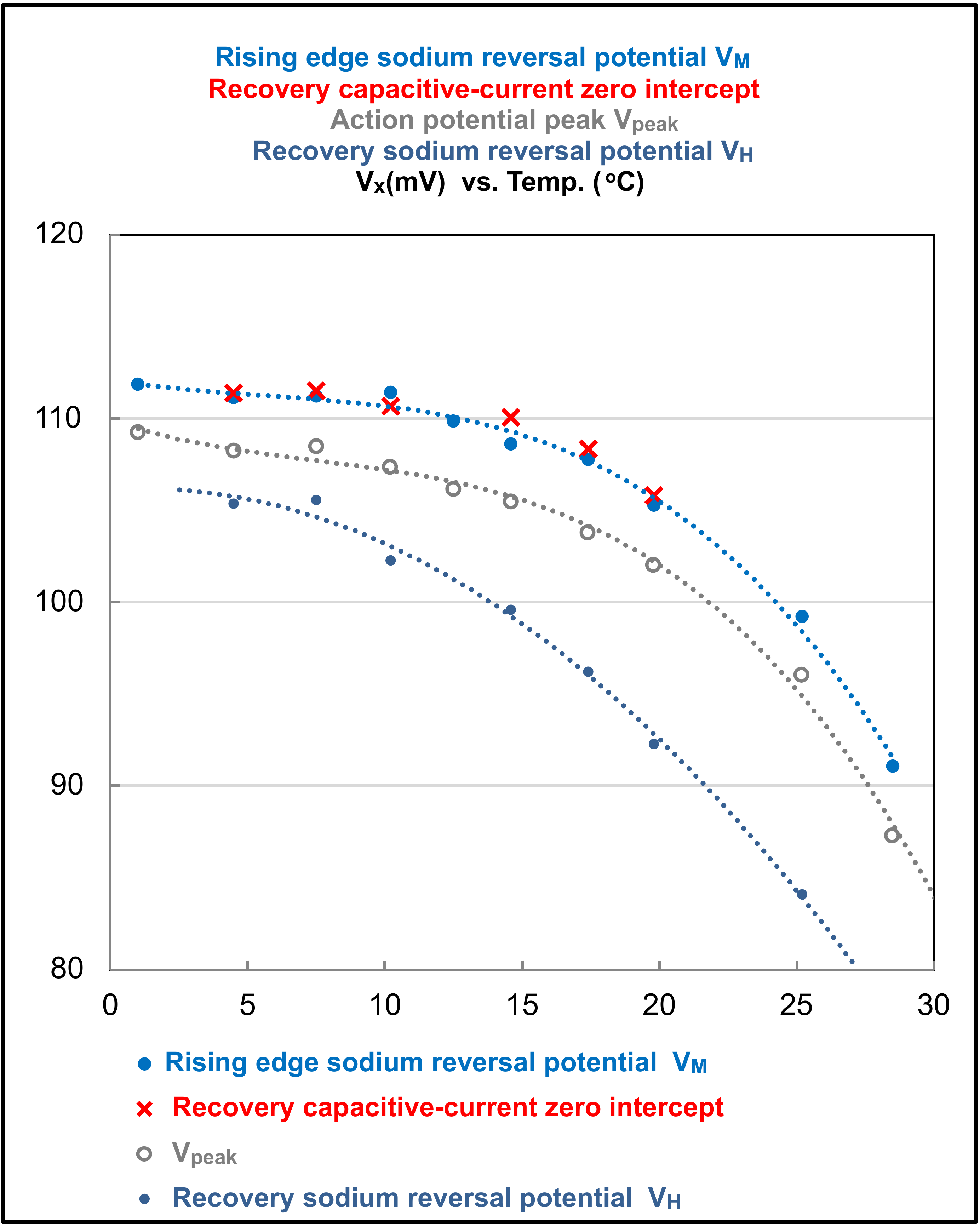}
\caption{\label{fig:RecoveryCCReversalL}
Extrapolated zero-current intercept of the recovery capacitive branch as a function of temperature. The mAvrami fits show that this intercept remains close to the rising-edge sodium equilibrium potential \(V_M\) over the measured temperature range. In contrast, the recovery ionic branch extrapolates to an effective intercept \(V_H\) below the action-potential peak \(V_p\). The separation of these intercepts is consistent with distinct activation and recovery channel states and with a shift in the effective recovery driving force.}
\end{figure}
\begin{figure*}
\centering
\includegraphics[width=2.1\columnwidth]{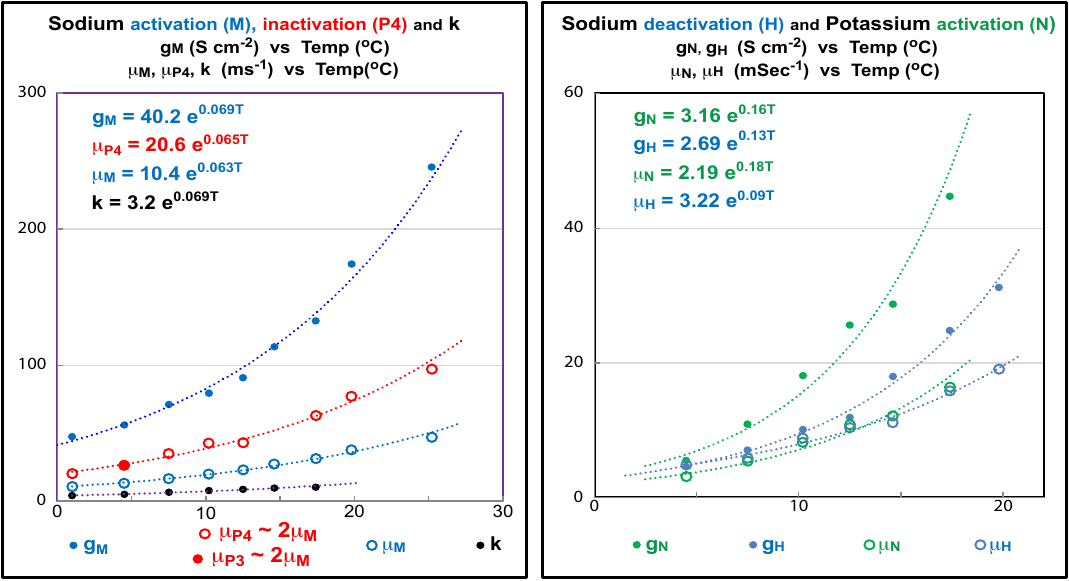}
\caption{\label{fig:MHSymmetriesL}
Parameters extracted from phase-space fits of the propagated action potential: maximum conductances \(g_X\) [mS\,cm\(^{-2}\)], time rates \(\mu_X\) [ms\(^{-1}\)], and propagation parameter \(k\) [ms\(^{-1}\)] as functions of temperature. See SM Fig.~\ref{fig:TableMHNL} for the corresponding parameter table.
\textbf{(Left panel) Sodium activation and inactivation-associated polarization.}
The maximum conductance \(g_M\), rates \(\mu_M\) and \(\mu_{P4}\), and propagation parameter \(k\) exhibit similar effective temperature dependences over most of the measured range. The approximate relation \(\mu_{P4}\approx2\mu_M\) characterizes the P4 polarization process associated with the onset of sodium-channel inactivation. Deviations occur in particular sweeps, including Sweeps~170 and 525.
\textbf{(Right panel) Sodium deactivation and potassium activation.}
The sodium H-channel and potassium N-channel parameters exhibit larger and distinct effective activation slopes. Parameters of \(J_{P5}\) and \(J_{P6}\) are not shown; representative values for Sweep~170 are given in Fig.~\ref{fig:mAvramiSweep170L}.}
\end{figure*}
\begin{figure*}
\includegraphics[width=2.1\columnwidth]{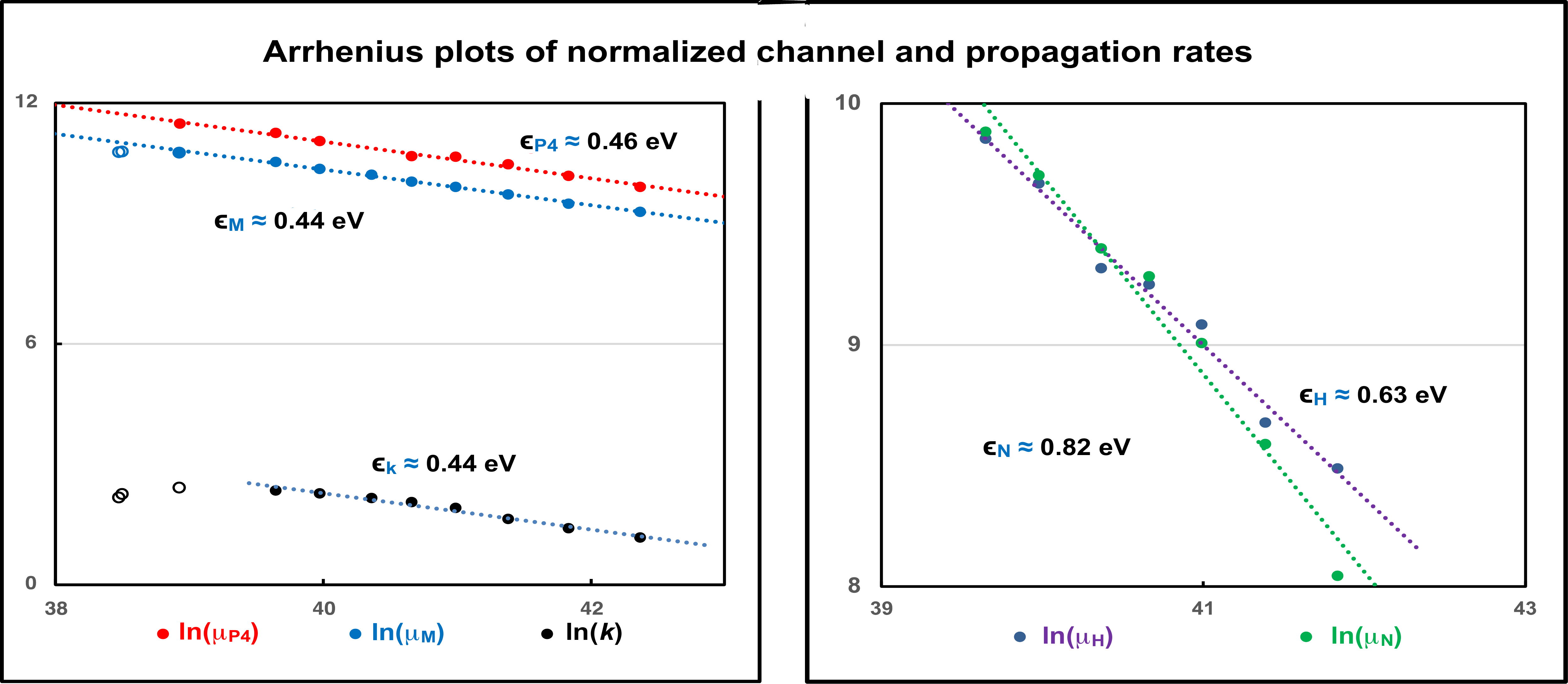}
\centering
\caption{\label{fig:BoltzmannRatesL}
Arrhenius representations of the reconstructed rates.
\textbf{(Left panel)}
Sodium activation and pre-peak polarization rates \(\mu_M\) and \(\mu_{P4}\), the propagation parameter \(k\), and the foot-current rate \(\mu_K\).
\textbf{(Right panel)}
Sodium deactivation and potassium activation rates \(\mu_H\) and \(\mu_N\).
The approximately linear relations yield distinct effective activation energies. The sodium activation rate \(\mu_M\), the P4 polarization rate, and the propagation parameter \(k\) have similar slopes, whereas sodium deactivation and potassium activation exhibit larger effective activation energies.}
\end{figure*}

\quad
Figure~\ref{fig:APPeakSweep425L} and
SM Fig.~\ref{fig:APPeakDetailSweep170L} show the currents near the
action-potential peak. The experimentally reconstructed capacitive
current, $J_m$, and total ionic current remain continuous through
$V_p$, whereas the mAvrami decomposition makes the transition explicit
through discontinuities in the fitted components. The recovery branch
$J_H$ extrapolates to an effective intercept $V_H$ below the
rising-edge sodium intercept $V_M$. The capacitive current and $J_m$
remain constrained by Eq.~\eqref{cableVa}. Together, the continuity of
the phase-space trajectory, the relation between the rising and
recovery branches, and the fitted intercepts support a sodium-dominated
interpretation of $J_H$, although these observations do not
independently establish ionic selectivity.

\quad
Direct ionic selectivity cannot be determined from the propagated
waveform alone. Nevertheless, the fitted intercepts and
channel-completion analysis further restrict the plausible
interpretations of the recovery current and support its identification
as a sodium-dominated deactivation component.

\quad
Outward sodium currents have been reported in cardiac cells
\cite{Wellis}. The mechanism proposed there involved a local increase
of sodium concentration near the membrane, although that interpretation
was later questioned by Carmeliet \cite{Carmeliet}. In the present
formulation, the outward recovery current follows from the phase-space
current balance and does not require a transient local accumulation of
sodium ions. During recovery, the capacitive current
$C_m\Phi$ is negative. At the second ionic zero-current crossing,
$J_m=C_m\Phi<0$. Beyond the crossing, $J_m$ is less negative than
$C_m\Phi$, so that
\[
J_I=J_m-C_m\Phi>0,
\]
even though $J_m$ remains negative until it changes sign near the tail
of the action potential. Capacitive discharge therefore contributes
directly to the outward sign of the reconstructed ionic current through
the local charge balance, although it does not itself drive sodium ions
outward.

\quad
The shift from the sodium-associated intercept \(V_M\) to the effective recovery intercept \(V_H\) may reflect the combined influence of the sodium concentration gradient and an evolving electrostatic contribution to the effective driving force. This interpretation is suggested by the propagated solution and its phase-space constraints rather than established by direct ionic-selectivity measurements. The experimental currents remain continuous across \(V_p\); the discontinuities occur only in the chosen mAvrami and polarization decomposition, which makes the transition explicit while preserving the charge-conserving total current. No separate polarization fit was attempted for the potassium activation current.

\subsection{Relational (clock-free) parametrization}

\quad
The correlational formulation used here follows directly from the steadily propagated solution and does not rely on a general theory of relational time. The comparison with Rovelli is interpretive rather than foundational: it places the independently derived action-potential formulation within a broader class of descriptions in which physical evolution is expressed through correlations among observable variables.

\quad
Because time and membrane potential are in one-to-one correspondence along each monotonic branch, with \(\Phi(V)=dV/dt\), the completion fractions may be written equivalently in phase space. Using \(dt=dV/\Phi(V)\), the mAvrami completion law becomes
\begin{equation}
\frac{X_o(V)}{X}
=
1-\exp\left[
-\alpha(\mu_X)^{\theta_X}
\left(
\int_{V_{iX}}^{V}
\frac{d\widetilde V}{\Phi(\widetilde V)}
\right)^{\theta_X}
\right],
\label{eq:f_VL}
\end{equation}
where $V_{iX}$ is the inception potential of an opening process $X$.
Closing processes are written analogously by integrating to the
potential corresponding to the closure time, with the orientation
chosen along the appropriate monotonic branch.

\quad
The designation ``clock-free'' is used only in the limited sense that
laboratory time is no longer an independently specified evolution
variable. Time is not eliminated; it remains recoverable branchwise from the measured trajectory. The completion fractions are therefore functionals of the measured phase-space trajectory rather than functions of an independently prescribed external clock. Because the waveform contains distinct rising and recovery branches, \(\Phi(V)\) is branch-valued globally but single-valued along each monotonic branch.

\quad
The resulting formulation is relational: channel completion is specified through correlations among \(V\), \(\Phi(V)\), and the completion fractions. This structure is independently derived from propagation and charge conservation and is conceptually consistent with Rovelli's view that physical evolution may be formulated through correlations among observables rather than through a privileged external time variable \cite{Rovelli1}. Further discussion is given in Appendix~E.

\subsection{Temperature dependence and activation energies}
 
\quad
The reconstructed time rates provide a direct means of examining the
temperature dependence of the propagated action potential. Because the
same phase-space and mAvrami representation is applied across all
experimental sweeps, the extracted rates $\mu_X$ and propagation
parameter $k$ can be compared on a common thermodynamic basis.

\quad
The quasilinear segments of $J_K$, $J_M$, $J_H$, and $J_N$
correspond to completion fractions close to unity, so their slopes
determine the maximum conductances $g_K$, $g_M$, $g_H$, and $g_N$.
Likewise, the slopes of the corresponding quasilinear branches of
$C_m\Phi$ determine the associated time rates
$\mu_K$, $\mu_M$, $\mu_H$, and $\mu_N$.
The temperature dependence of these parameters is summarized in
Fig.~\ref{fig:MHSymmetriesL}.

\quad
All extracted parameters exhibit systematic temperature dependence. Up to approximately \(20^\circ\mathrm{C}\), the sodium activation rate \(\mu_M\), the propagation parameter \(k\), and the pre-peak polarization rate \(\mu_{P4}\) display similar exponential temperature dependences. Through Eq.~\eqref{Sodiumd}, the maximum sodium conductance \(g_M\) follows a related trend. These observations are consistent with a common effective temperature-dependent contribution to sodium activation, pre-peak polarization, and propagation, although similar slopes alone do not establish a unique microscopic mechanism.

\quad
A further feature of Fig.~\ref{fig:MHSymmetriesL} is the approximate relation \(\mu_{P4}\approx2\mu_M\) at the onset of the transition from the M-channel state to the H-channel state. Although deviations occur in individual sweeps, the relation persists through much of the experimental temperature range. It characterizes a structured coupling between M-channel activation and the P4 polarization process that initiates sodium-channel inactivation.

\quad
To characterize the temperature dependence quantitatively, the extracted rates are plotted in Arrhenius form in Fig.~\ref{fig:BoltzmannRatesL}. We write
\begin{subequations}
\label{boltzmann3}
\begin{equation}
\mu_X
=
\kappa_X
\exp\left(-\frac{\epsilon_X}{k_B T}\right),
\end{equation}
\begin{equation}
\ln\left(\frac{\mu_X}{1\,\mathrm{ms}^{-1}}\right)
=
\ln\left(\frac{\kappa_X}{1\,\mathrm{ms}^{-1}}\right)
-
\frac{\epsilon_X}{k_B T}.
\end{equation}
\end{subequations}

\quad
The fitted numerical relations are
\begin{subequations}
\label{Boltzmann1}
\begin{equation}
\ln\left(\frac{\mu_K}{1\,\mathrm{ms}^{-1}}\right)
=
23.9-\frac{0.54\,\mathrm{eV}}{k_B T},
\end{equation}
\begin{equation}
\ln\left(\frac{\mu_M}{1\,\mathrm{ms}^{-1}}\right)
=
28.1-\frac{0.44\,\mathrm{eV}}{k_B T},
\end{equation}
\begin{equation}
\ln\left(\frac{\mu_{P4}}{1\,\mathrm{ms}^{-1}}\right)
=
29.3-\frac{0.46\,\mathrm{eV}}{k_B T},
\end{equation}
\begin{equation}
\ln\left(\frac{k}{1\,\mathrm{ms}^{-1}}\right)
=
19.9-\frac{0.44\,\mathrm{eV}}{k_B T},
\end{equation}
\begin{equation}
\ln\left(\frac{\mu_H}{1\,\mathrm{ms}^{-1}}\right)
=
34.9-\frac{0.63\,\mathrm{eV}}{k_B T},
\end{equation}
\begin{equation}
\ln\left(\frac{\mu_N}{1\,\mathrm{ms}^{-1}}\right)
=
42.5-\frac{0.82\,\mathrm{eV}}{k_B T}.
\end{equation}
\end{subequations}
\quad
The fits remain approximately linear up to \(20^\circ\mathrm{C}\), yielding effective activation energies of \(0.44\,\mathrm{eV}\) for sodium activation, \(0.46\,\mathrm{eV}\) for the P4 polarization process associated with sodium-channel inactivation, \(0.63\,\mathrm{eV}\) for sodium deactivation, and \(0.82\)--\(0.84\,\mathrm{eV}\) for potassium activation.

\quad
The propagation parameter \(k\) exhibits the same effective Arrhenius slope as the sodium activation rate \(\mu_M\), corresponding to approximately \(0.44\,\mathrm{eV}\). Because \(k\) is a derived property of the propagated solution rather than a microscopic channel rate, the shared slope indicates a common effective temperature dependence and is consistent with a shared rate-limiting contribution. It does not by itself establish that propagation and sodium activation arise from a single microscopic process.

\quad
A further compensation is observed for sodium activation. Between \(4.5^\circ\mathrm{C}\) and \(19.8^\circ\mathrm{C}\), the maximum sodium conductance \(g_M\) increases by approximately a factor of three, while the corresponding activation interval decreases by a similar factor. Because the voltage-dependent factor \([M_o(V)/M](V-V_M)\) changes more moderately over this range, this compensation suggests that the integrated sodium influx during activation may remain approximately constant. Direct numerical integration of the reconstructed current at each temperature would be required to establish this conclusion quantitatively.

\quad
Taken together, the temperature dependences of the conductances, time rates, polarization rate, and propagation parameter reveal a structured set of effective activation barriers linking activation, inactivation-associated polarization, deactivation, potassium recovery, and propagation.

\subsection*{G. Universality of channel kinetics}

\begin{figure*}
\centering
\includegraphics[width=2.08\columnwidth]{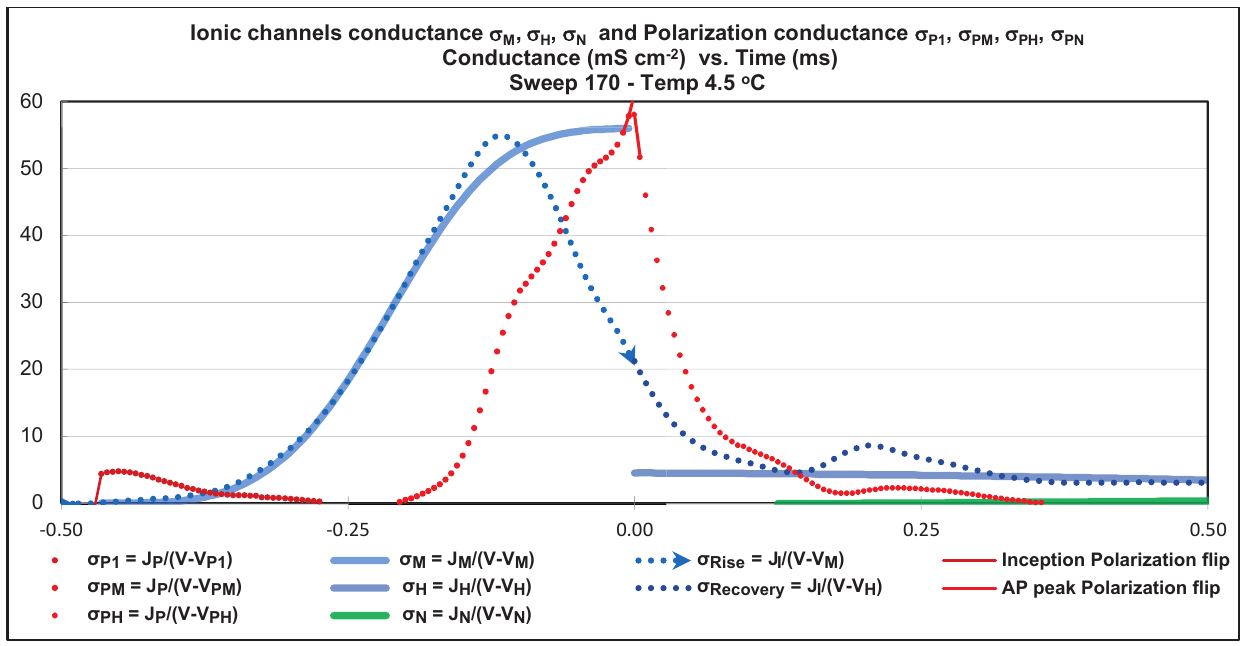}
\caption{\label{fig:ConductanceSweep170L}
Effective conductances and associated time rates for sodium and potassium current components in Sweep~170 (\(T=4.5^\circ\mathrm{C}\)). The plotted parameters organize the activation, polarization, deactivation, and recovery segments on the two branches of the propagated action potential. Discontinuities belong to the branch-specific decomposition; the reconstructed capacitive current, $J_m$, and total ionic current remain continuous through the action-potential peak.}
\end{figure*}

\begin{figure*}
\centering
\includegraphics[width=2.1\columnwidth]{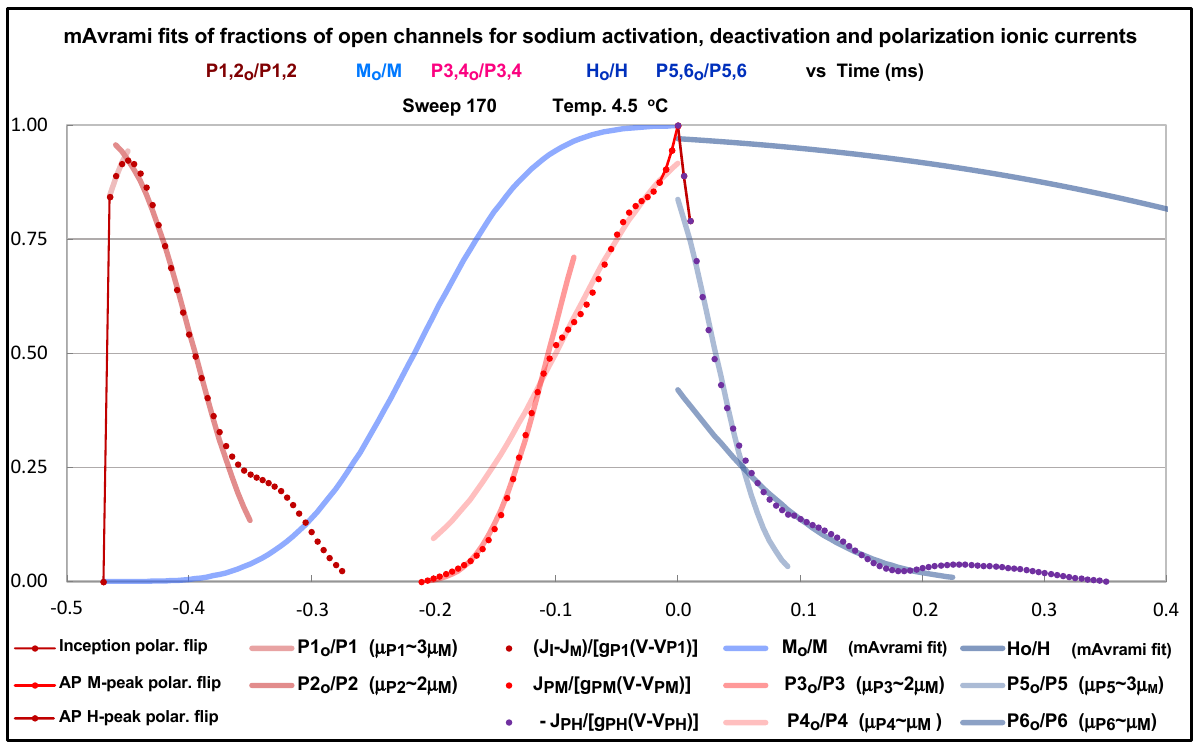}
\caption{\label{fig:mAvramiSweep170L}
mAvrami representations of sodium activation, rising-edge polarization, sodium deactivation, and recovery polarization fractions for Sweep~170 (\(T=4.5^\circ\mathrm{C}\)). The processes are represented by the same stretched-exponential form, with branch-specific rates and exponents. The figure displays the organization of the fitted completion variables through activation, polarization reversal, and sodium deactivation. Potassium activation is included in the associated parameter analysis but lies outside the plotted scope of this figure.}
\end{figure*}

\begin{figure*}
\includegraphics[width=2.09\columnwidth]{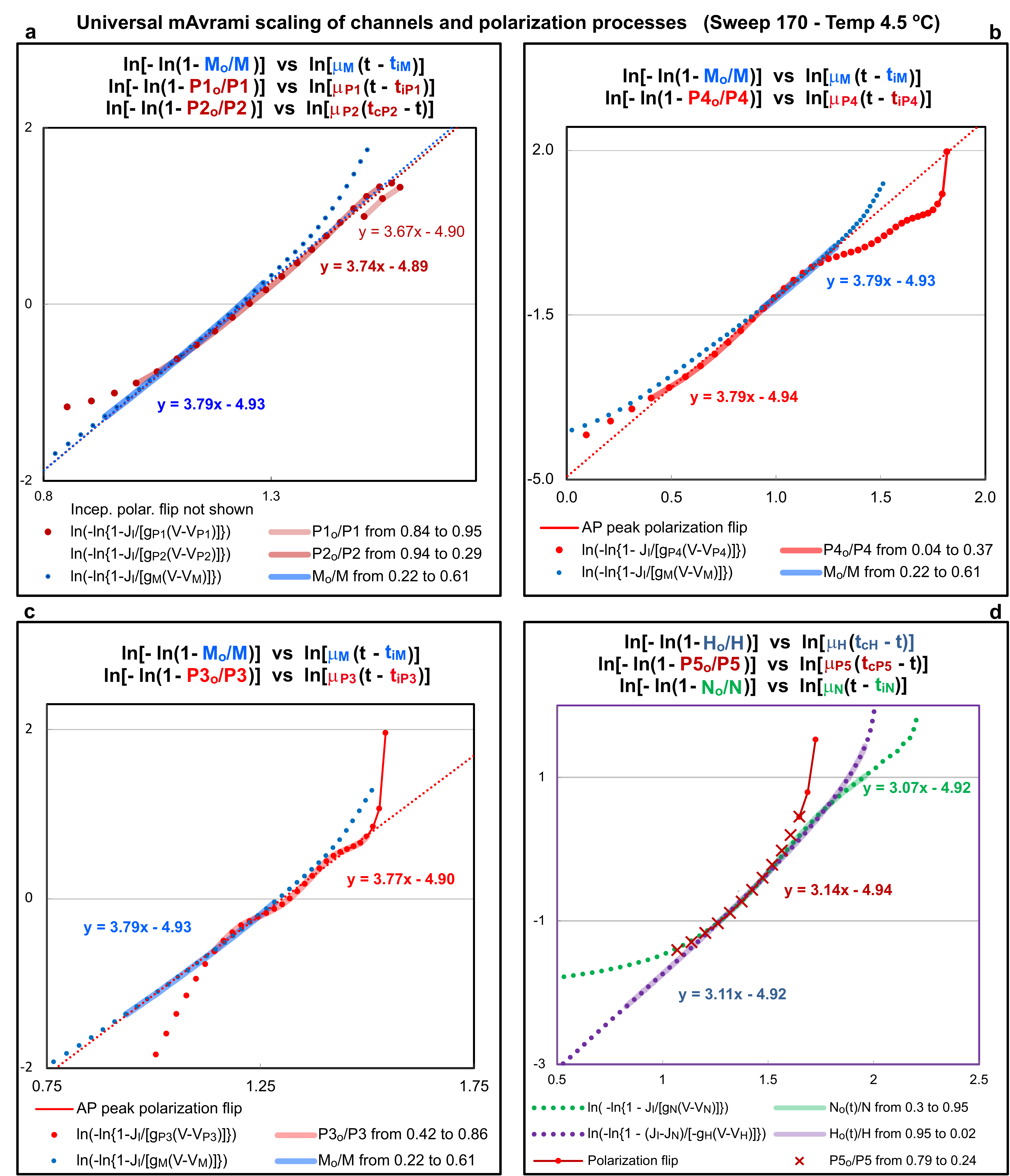}
\caption{\label{fig:mAvramiLabFitSweep170L}
Linearized mAvrami representations for Sweep 170 ($T=4.5^\circ$C).
Panels (a)--(c) plot
$\ln[-\ln(1-X_o/X)]$ versus $\ln[\mu_X(t-t_{iX})]$ for the sodium
activation fraction $M_o/M$ and the associated rising-edge polarization
fractions $Pn_{o}/Pn$ (n=1,3,4), and versus $\ln[\mu_{P2}(t_{cP2}-t)]$ for $P2_{o}/P2$ .
The activation and polarization fractions collapse onto nearly linear
relations with exponents $\theta\approx3.7$--$3.8$ and
$\ln\alpha\approx-4.94$ to $-4.89$. The inception polarization flip is
not shown explicitly. The polarization reorganization at the
action-potential peak begins in panel (c), where $P3$ is displayed.
\textbf{(d) Recovery region.}
The continuation of the peak polarization flip is represented by
$P5_{o}/P5$, together with the sodium deactivation and potassium
activation fractions $H_o/H$ and $N_o/N$. Allowing $\theta_{P5}$ to be
determined independently, as appropriate for the recovery branch,
gives $\theta_{P5}=3.11$ and $\ln\alpha_{P5}\approx-4.92$, improving
the fit relative to imposing the rising-edge value. The corresponding
recovery exponents are $\theta_H=3.14$ and $\theta_N=3.07$. Thus the
polarization transition through the action-potential peak is represented
across panels (c) and (d), while the recovery processes retain the same
linearized mAvrami form.
Taken together, the four panels and
Fig.~\ref{fig:ConductanceSweep170L} show that the reconstructed currents
remain continuous through the action-potential peak despite internal
reorganization, and that activation, polarization, sodium deactivation,
and potassium activation form a coherent kinetic sequence with closely
related dimensionless scaling.
}
\end{figure*}

\quad
The phase-space reconstruction reveals a degree of kinetic organization
that is not apparent when the principal current components are considered
independently. Figure~\ref{fig:ConductanceSweep170L} shows that the
extracted conductances and time rates remain continuously organized
through the action-potential peak, despite the internal reorganization
of the fitted components. Figure~\ref{fig:mAvramiSweep170L} shows that
the completion fractions of sodium activation, polarization, sodium
deactivation, and potassium activation are described within a common
mAvrami framework. Finally, Fig.~\ref{fig:mAvramiLabFitSweep170L} shows
that the corresponding linearized kinetics collapse onto nearly common
scaling relations.

\quad
Activation, polarization, deactivation, and potassium recovery are
therefore not isolated processes. They appear as components of a
coherent kinetic sequence within the propagated action potential.

\begin{figure*}
\centering
\includegraphics[width=2.1\columnwidth]{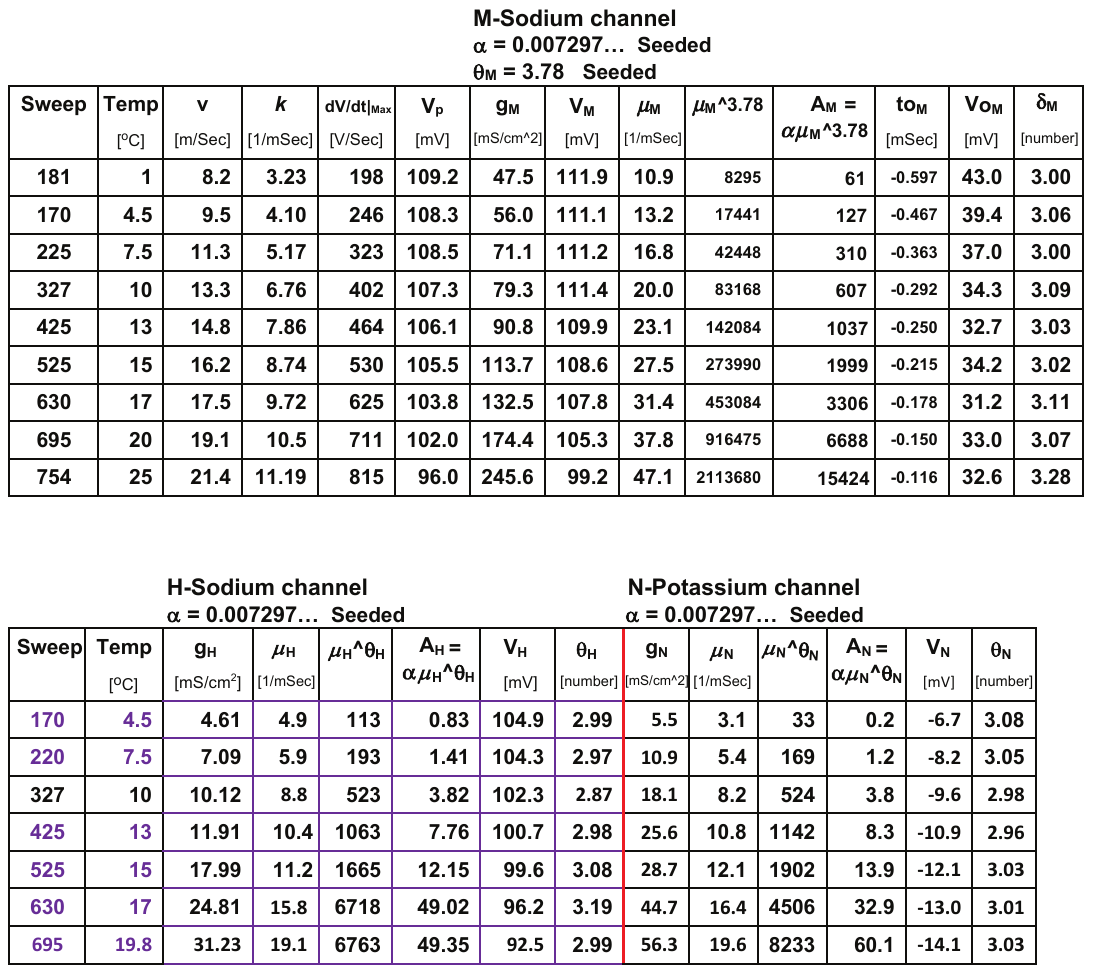}
\caption{\label{fig:TableMHNL}
Parameters obtained from mAvrami fits of the propagated action potential. A common dimensionless prefactor \(\alpha=0.007297\ldots\) is imposed in the final fits through \(A_X=\alpha(\mu_X)^{\theta_X}\), where \(\mu_X\) is the corresponding time rate and \(\theta_X\) is the mAvrami exponent.
\textbf{(a) M-sodium activation parameters.}
The activation process is characterized by \(\theta_M\approx3.78\).
\textbf{(b) H-sodium deactivation parameters.}
The recovery current satisfies \(A_H=\alpha(\mu_H)^{\theta_H}\), with \(\theta_H\approx3\). The fitted values of \(\mu_H\) and \(g_H\) differ from their M-channel counterparts, consistent with a transition between activation and recovery channel states.
\textbf{(c) N-potassium activation parameters.}
The potassium activation current satisfies \(A_N=\alpha(\mu_N)^{\theta_N}\), with \(\theta_N\approx3\). Together, the M-, H-, and N-channel fits support a common dimensionless scaling ansatz across the reconstructed processes.}
\end{figure*}

\subsection*{H. Summary of Part~III}

\quad
Part~III reconstructs the propagated action potential as a sequence
of coupled channel-state and polarization processes constrained by the
charge-conserving phase-space cable equation. The experimentally
reconstructed capacitive current, $J_m$, and total ionic current remain
continuous along the action-potential trajectory, although their
branch-specific decomposition into activation, polarization,
deactivation, and recovery components changes at inception and at the
action-potential peak.

\quad
The sodium activation, sodium deactivation, potassium activation, and polarization components are represented by modified Avrami relations. Their fitted parameters satisfy the empirical scaling \(A_X\approx\alpha_X(\mu_X)^{\theta_X}\). The approximate clustering of the dimensionless prefactors motivates final constrained fits with a common value \(\alpha=0.007297\ldots\). This numerical identification remains empirical; neither exact universality nor a fundamental physical origin is established by the present analysis.

\quad
The reconstruction reveals a branch-dependent reorganization near the action-potential peak. The rising sodium current \(J_M\) extrapolates to \(V_M\), whereas the recovery branch \(J_H\) extrapolates to the lower effective intercept \(V_H\). Together with the continuity of the phase-space trajectory and the channel-completion analysis, this separation supports the interpretation of \(J_H\) as a sodium-dominated outward recovery current associated with the $M\rightarrow H$ transition. Direct ionic selectivity is not determined by the propagated-waveform data alone.

\quad
The completion laws may be written either in laboratory time or branchwise along the phase-space trajectory through \(dt=dV/\Phi(V)\). The membrane potential therefore provides an intrinsic ordering variable on each monotonic branch, and the kinetics may be expressed through correlations among \(V\), \(\Phi(V)\), and the completion fractions. Taken together, the results reveal a coherent phase-space and kinetic organization extending across activation, inactivation-associated polarization, deactivation, potassium recovery, and propagation.

\section{Discussion}
\label{sec:discussion}

\quad
The present work establishes a charge-conserving phase-space formulation of the steadily propagated action potential in which the capacitive, ionic, and axoplasmic (Ohmic) currents are reconstructed directly from the experimentally measured waveform. The formulation originated from the observation that the waveform of the steadily propagated action-potential is preserved under Galilean transformation. The same invariance is inherited by the reconstructed current components, indicating that the essential physics of steady propagation resides in the propagated waveform itself rather than in a particular temporal parametrization. This observation naturally leads to a phase-space description in which the membrane potential serves as the intrinsic ordering variable and laboratory time is reconstructed from the propagated trajectory.

\quad
Several experimental observations follow directly from the charge-conserving formulation. The reconstructed total ionic current exhibits two zero crossings determined by the condition
\[
\frac{d\Phi(V)}{dV}=k,
\]
where $k$ is the propagation determined by the axonal geometry and axoplasmic resistivity. Both crossings are present in the Rosenthal--Bezanilla steady-propagation data. The second occurs during the recovery phase, where the reconstructed ionic current reverses sign while the sodium-channel completion fraction remains close to unity.

\quad
Within the present interpretation, the recovery component is most
consistently interpreted as a sodium-dominated outward current with an
effective reversal potential $V_H$ below the rising-edge sodium reversal
potential $V_M$. This behaviour is not attributed simply to channel
closure. Rather, it reflects a branch-dependent reorganization of the
ionic driving-force structure during propagation, associated with the
spatial divergence of $J_z$ and the resulting local charge
redistribution. The recovery-current structure therefore emerges from
the charge-conserving propagated solution rather than from an additional
phenomenological current introduced to reproduce the experimental
waveform.

\quad
An equally important consequence concerns the continuity of the propagated solution. Although the activation, deactivation, and recovery components reorganize at the action-potential peak, the total ionic current remains continuous, as required by charge conservation. The propagated action potential therefore exhibits the characteristics of a continuous phase transition in which the peak corresponds to a change in the symmetry of the sodium-channel population rather than to a discontinuity of the underlying dynamics. Independent experimental observations of coupled thermal and volumetric changes accompanying excitation in nerve cells, fibers, and synapses \cite{TasakiPhase} provide a broader physiological context in which phase-transition-like behaviour in excitable membranes is not unprecedented, although they do not by themselves establish the microscopic origin of the transition proposed here.

\quad
A second principal result concerns the kinetics of channel completion. The modified Avrami (mAvrami) representation provides a unified empirical description of sodium activation ($M$), sodium deactivation ($H$), potassium activation ($N$), and the associated polarization processes. Across these distinct channel processes, the fitted exponents remain nearly temperature independent, while the extracted microscopic time rates exhibit consistent Arrhenius behaviour. These observations indicate that the kinetics are governed by common structural constraints extending across ion species, channel states, and temperatures.

\quad
Particularly noteworthy is the appearance of an approximately invariant dimensionless prefactor that remains nearly constant across the different channel processes. Because the characteristic time scale is determined independently by the measured ionic time rates, this dimensionless coefficient emerges explicitly rather than being absorbed into a conventional kinetic rate constant. Its numerical proximity to the fine-structure constant,
$\alpha = 0.007297\ldots$, is an empirical observation. Whether this correspondence reflects a deeper microscopic principle or an effective phenomenological parametrization cannot be determined from the present work and awaits a more fundamental theoretical derivation. More generally, however, the emergence of a stable dimensionless parameter suggests that propagated channel kinetics possess an underlying scaling structure extending beyond individual gating processes.

\quad
The phase-space formulation also reveals a broader relational structure. Along each monotonic branch of the propagated trajectory, the exact one-to-one correspondence
\[
dt=\frac{dV}{\Phi(V)}\]

The resulting relational ("clock-free") parametrization expresses the completion kinetics in terms of correlations among experimentally observable variables. Time is not eliminated physically; rather, it is reconstructed from the propagated trajectory itself. The resulting description depends only upon the correlations among the membrane potential, the phase-space velocity, and the channel completion fractions, without requiring an independently prescribed external clock.

\quad
From this perspective, the propagated action potential possesses quantities that remain invariant under the transformation between temporal and phase-space descriptions. These include the propagated trajectory, the completion fractions, the continuity of the total ionic current, and the empirical dimensionless scaling parameters. Together, these relational invariants provide a unified framework connecting the charge-conserving cable equation, the current decomposition, the mAvrami formulation, and the experimentally observed scaling relations within a single dynamical description.

\quad
The present formulation also suggests a broader methodological perspective. In many biological systems, the experimentally accessible information consists primarily of correlations among measurable variables, whereas time serves chiefly as an external laboratory parameter. The propagated action potential demonstrates that, whenever a one-to-one correspondence exists among observable quantities, an equivalent relational formulation may arise naturally. Whether similar relational structures underlie more complex biological phenomena, including collective neural dynamics, sleep--wake transitions, or other large-scale physiological processes, remains an open question for future investigation.

\subsection{Possible collective-state extension}

\quad
The reconstructed $M\rightarrow H$ transition shows, within the present
formulation, that the sodium-channel population is not specified by
membrane potential alone. Activation and recovery states attained at
the same membrane potential may differ in phase-space velocity,
effective reversal potential, conductance, time rate, and channel
completion fraction. The observed transition therefore identifies
internal state structure that is not reducible to $V$. It also raises a
question beyond the present reconstruction: might the observed $H$
state be one member of a larger family of nearby configurations $H_n$,
where proximity refers to the extended state space
$\{V,\Phi,\boldsymbol{f}\}$ rather than necessarily to membrane
potential alone?

\quad
In such an extension, activation of a local excitable unit could be
followed by one of several state transitions,
\begin{equation}
M_i\longrightarrow H_{n_i},
\qquad
n_i=1,\ldots,q,
\label{eq:PossibleHnStates}
\end{equation}
with the selected state depending on the preceding trajectory and on
the states of coupled units. 

\quad
The action-potential foot may provide a particularly sensitive region
for such state selection. There, the net ionic current is the small
difference among several overlapping contributions, with a predominant
potassium component. Small history-dependent changes in channel
availability could therefore alter the local current balance and, in a
more general neuron, create or remove stable near-rest configurations
without requiring a large change in resting membrane potential.
Repeated firing could move slow channel variables between basins of
attraction, so that the selected configuration would be expressed in
the threshold, latency, or foot trajectory of a later response. This
possibility should not be confused with the occurrence of several
nearby minima in a numerical fit: physical multistability would require
distinct stable states under the same external conditions and
reproducible switching between them.For a collection of $N$ units, the
resulting configuration could be represented by
\begin{equation}
\boldsymbol{\mathcal H}
=
\left(
H_{n_1},H_{n_2},\ldots,H_{n_N}
\right).
\label{eq:CollectiveHState}
\end{equation}
If coupling makes the stability or transition probability of each
local state depend on correlations with the others, particular
configurations $\boldsymbol{\mathcal H}^{\,\mu}$ could become stable
collective attractors. A transient propagated input could then write a
state by moving the system into the basin of one such attractor, while
a later partial or related input could retrieve the state by returning
the system toward the same correlated configuration. This would be
Hopfield-like in the restricted sense of collective attractors, basins
of attraction, and state-dependent retrieval \cite{Hopfield, Hopfield2}; it
does not imply that the proposed channel dynamics obey a particular
Hopfield energy function or symmetric coupling rule.

\quad
The present experiments were performed on the giant squid axon, a
system specialized for reliable action-potential propagation and not
itself regarded as a substrate of memory storage. The relevance of the
reconstructed $M\rightarrow H$ transition is therefore not that the
giant axon stores memory, but that it may reveal a more general
channel-state organization that could acquire a different functional
role in other neurons or in coupled neuronal populations.

\quad
The present results do not demonstrate a multiplicity of $H_n$ states,
their persistence after completion of the action-potential cycle, or
the coupling required to stabilize collective configurations. They
identify only initial elements of such a possibility: structurally
distinct channel states, an organized transition between them, and
completion variables capable of distinguishing states attained at the
same membrane potential. Any contribution of such states to memory
would not replace established synaptic and molecular mechanisms of
storage and consolidation \cite{Kandel}, but could provide an
additional microscopic state variable whose relation to those
mechanisms remains to be established. Evidence for such a mechanism
would require reproducible history-dependent responses under the same
initial conditions, persistence of the internal state after recovery,
reversible writing and resetting, collective lifetimes exceeding those
of isolated units, and retrieval from incomplete input. The resulting
research question is therefore specific: can the structurally distinct
channel states reconstructed from propagated action potentials be
coarse-grained into interacting variables whose correlations support
stable collective attractors and state-dependent retrieval?

\section{Conclusions}
\label{sec:conclusions}

\quad
The present work establishes a charge-conserving phase-space
formulation of the steadily propagated action potential in which the
axial Ohmic current $J_z$, its geometry-converted divergence $J_m$, the
capacitive current, and the total ionic current are reconstructed from
the experimentally measured propagation waveform and the propagation
parameter. The propagated action potential is thereby represented as a
self-consistent dynamical system whose electrical structure can be
determined without introducing voltage-clamp currents as independent
inputs.

\quad
The formulation provides a closed correlational description in which
the current balance is determined by relations among $V$, $\Phi(V)$,
and $d\Phi(V)/dV$. Laboratory time is not an independent variable in
these relations, although temporal direction and elapsed time remain
recoverable along each monotonic branch from the sign of $\Phi(V)$ and
the identity
\[
dt=\frac{dV}{\Phi(V)}.
\]

\quad
Several structural features of steady propagation follow directly.
At points where $\Phi(V)\neq0$, the total ionic current has one zero
crossing on each branch at
\[
\frac{d\Phi(V)}{dV}=k.
\]
The recovery-branch crossing is followed by an outward total ionic
current whose continuity, effective reversal potential, channel
completion, and temperature dependence support a sodium-dominated
interpretation. The experimentally reconstructed currents remain
continuous through the action-potential peak, whereas their
state-dependent decomposition changes. These observations support the
interpretation of the peak as a continuous $M\rightarrow H$ transition
between distinct sodium-channel regimes rather than as a discontinuity
of the propagated dynamics.

\quad
The modified Avrami representation provides a unified empirical
description of sodium activation, sodium deactivation, potassium
activation, and the associated polarization processes. The nearly
temperature-independent completion exponents, together with the
Arrhenius behaviour of the characteristic time rates, reveal a common
kinetic organization across distinct channel processes and
temperatures. The appearance of a stable dimensionless prefactor whose
numerical value is close to the fine-structure constant remains an
empirical observation. Whether this correspondence reflects a deeper
microscopic principle or an effective phenomenological scaling awaits a
fundamental theoretical derivation.

\quad
The structurally distinct channel states reconstructed here also raise
a specific question for future investigation: can such states be
coarse-grained into interacting variables whose correlations support
stable collective attractors and state-dependent retrieval? The
present giant-squid-axon results identify channel-state structure and an
organized transition, but do not demonstrate persistent multistability,
collective stabilization, or a mechanism of memory storage.

\quad
Finally, the phase-space formulation admits an exact relational
representation on each monotonic branch. The temporal and phase-space
descriptions preserve the same propagated trajectory, current balance,
completion fractions, and dimensionless completion parameters. The
description is ``clock-free'' only in the restricted sense that
laboratory time is reconstructed from experimentally observable
correlations rather than prescribed as an independent evolution
variable.

\quad
More broadly, the propagated action potential provides an
experimentally grounded example of how charge conservation, symmetry,
and relational parametrization can reveal dynamical organization that
is not evident in a conventional time-domain description alone.
Whether comparable correlational structures are useful in more complex
physiological systems remains an open question.
\begin{acknowledgments}

\quad	NKJ wants to thank Prof. J. J. C. Rosenthal making the present work possible by giving him in 1999 the electronic copy of Wa05097a.dat file containing the experimental data collected in Reference [3]. He is greatly indebted to him and to Prof. Francisco Bezanilla for sharing their data.
NKJ wants to thank Dr. Bogdan Mihaila for recommending further study of the role and the meaning of temperature independent dimensionless constants $\alpha_X$.
 NKJ wants to thank Dr. Andrew Perlman for asking a question about Hodgkin-Huxley equations close to 55 years ago. 
 NKJ and FC want to thank the Santa Fe Institute for its hospitality during the work on the paper. NKJ and FC want to thank Prof. Geoffrey West, Prof. Benjamin Drukarch and Prof. Avadh Behari Saxena for helpful comments. In particular we are grateful for the encouraging acknowledgement by Prof. Bezanilla of the second outward current upon observing the recovery region graph at our poster at the Biophysics Symposium in San Francisco.
\end{acknowledgments}
\bibliographystyle{apsrev4-2}
\bibliography{Jurisic-Cooper-Physics-of-the-Propagating-Action-Potential-JULY-18-2026-copy}

\begin{thebibliography}{33}%
\makeatletter
\providecommand \@ifxundefined [1]{%
 \@ifx{#1\undefined}
}%
\providecommand \@ifnum [1]{%
 \ifnum #1\expandafter \@firstoftwo
 \else \expandafter \@secondoftwo
 \fi
}%
\providecommand \@ifx [1]{%
 \ifx #1\expandafter \@firstoftwo
 \else \expandafter \@secondoftwo
 \fi
}%
\providecommand \natexlab [1]{#1}%
\providecommand \enquote  [1]{``#1''}%
\providecommand \bibnamefont  [1]{#1}%
\providecommand \bibfnamefont [1]{#1}%
\providecommand \citenamefont [1]{#1}%
\providecommand \href@noop [0]{\@secondoftwo}%
\providecommand \href [0]{\begingroup \@sanitize@url \@href}%
\providecommand \@href[1]{\@@startlink{#1}\@@href}%
\providecommand \@@href[1]{\endgroup#1\@@endlink}%
\providecommand \@sanitize@url [0]{\catcode `\\12\catcode `\$12\catcode
  `\&12\catcode `\#12\catcode `\^12\catcode `\_12\catcode `\%12\relax}%
\providecommand \@@startlink[1]{}%
\providecommand \@@endlink[0]{}%
\providecommand \url  [0]{\begingroup\@sanitize@url \@url }%
\providecommand \@url [1]{\endgroup\@href {#1}{\urlprefix }}%
\providecommand \urlprefix  [0]{URL }%
\providecommand \Eprint [0]{\href }%
\providecommand \doibase [0]{https://doi.org/}%
\providecommand \selectlanguage [0]{\@gobble}%
\providecommand \bibinfo  [0]{\@secondoftwo}%
\providecommand \bibfield  [0]{\@secondoftwo}%
\providecommand \translation [1]{[#1]}%
\providecommand \BibitemOpen [0]{}%
\providecommand \bibitemStop [0]{}%
\providecommand \bibitemNoStop [0]{.\EOS\space}%
\providecommand \EOS [0]{\spacefactor3000\relax}%
\providecommand \BibitemShut  [1]{\csname bibitem#1\endcsname}%
\let\auto@bib@innerbib\@empty
\bibitem [{\citenamefont {Hodgkin}\ \emph {et~al.}(1952)\citenamefont
  {Hodgkin}, \citenamefont {Huxley},\ and\ \citenamefont
  {Katz}}]{Hodgkin1952a}%
  \BibitemOpen
  \bibfield  {author} {\bibinfo {author} {\bibfnamefont {A.~L.}\ \bibnamefont
  {Hodgkin}}, \bibinfo {author} {\bibfnamefont {A.~F.}\ \bibnamefont
  {Huxley}},\ and\ \bibinfo {author} {\bibfnamefont {B.}~\bibnamefont {Katz}},\
  }\href@noop {} {\bibfield  {journal} {\bibinfo  {journal} {J. Physiol.
  (London)}\ }\textbf {\bibinfo {volume} {116}},\ \bibinfo {pages} {424}
  (\bibinfo {year} {1952})}\BibitemShut {NoStop}%
\bibitem [{\citenamefont {Hodgkin}\ and\ \citenamefont
  {Huxley}(1952)}]{Hodgkin1952b}%
  \BibitemOpen
  \bibfield  {author} {\bibinfo {author} {\bibfnamefont {A.~L.}\ \bibnamefont
  {Hodgkin}}\ and\ \bibinfo {author} {\bibfnamefont {A.~F.}\ \bibnamefont
  {Huxley}},\ }\href@noop {} {\bibfield  {journal} {\bibinfo  {journal} {J.
  Physiol. (London)}\ }\textbf {\bibinfo {volume} {117}},\ \bibinfo {pages}
  {500} (\bibinfo {year} {1952})}\BibitemShut {NoStop}%
\bibitem [{\citenamefont {Rosenthal}\ and\ \citenamefont
  {Bezanilla}(2000)}]{RB}%
  \BibitemOpen
  \bibfield  {author} {\bibinfo {author} {\bibfnamefont {J.~J.~C.}\
  \bibnamefont {Rosenthal}}\ and\ \bibinfo {author} {\bibfnamefont
  {F.}~\bibnamefont {Bezanilla}},\ }\href@noop {} {\bibfield  {journal}
  {\bibinfo  {journal} {Biol. Bull.}\ }\textbf {\bibinfo {volume} {199}},\
  \bibinfo {pages} {135} (\bibinfo {year} {2000})}\BibitemShut {NoStop}%
\bibitem [{\citenamefont {Jurisic}(1987)}]{Jurisic1987}%
  \BibitemOpen
  \bibfield  {author} {\bibinfo {author} {\bibfnamefont {N.}~\bibnamefont
  {Jurisic}},\ }\href@noop {} {\bibfield  {journal} {\bibinfo  {journal}
  {Biophysical Journal}\ }\textbf {\bibinfo {volume} {51}},\ \bibinfo {pages}
  {817, 823} (\bibinfo {year} {1987})}\BibitemShut {NoStop}%
\bibitem [{\citenamefont {Jurisic}\ and\ \citenamefont
  {Perez}(1988)}]{Jurisic1988}%
  \BibitemOpen
  \bibfield  {author} {\bibinfo {author} {\bibfnamefont {N.}~\bibnamefont
  {Jurisic}}\ and\ \bibinfo {author} {\bibfnamefont {E.}~\bibnamefont
  {Perez}},\ }\href@noop {} {\bibfield  {journal} {\bibinfo  {journal}
  {Mathematical Biosciences}\ }\textbf {\bibinfo {volume} {90}},\ \bibinfo
  {pages} {71} (\bibinfo {year} {1988})}\BibitemShut {NoStop}%
\bibitem [{\citenamefont {Avrami}(1939)}]{Avrami1}%
  \BibitemOpen
  \bibfield  {author} {\bibinfo {author} {\bibfnamefont {M.}~\bibnamefont
  {Avrami}},\ }\href@noop {} {\bibfield  {journal} {\bibinfo  {journal}
  {Journal of Chemical Physics}\ }\textbf {\bibinfo {volume} {7}},\ \bibinfo
  {pages} {1103} (\bibinfo {year} {1939})}\BibitemShut {NoStop}%
\bibitem [{\citenamefont {Avrami}(1940)}]{Avrami2}%
  \BibitemOpen
  \bibfield  {author} {\bibinfo {author} {\bibfnamefont {M.}~\bibnamefont
  {Avrami}},\ }\href@noop {} {\bibfield  {journal} {\bibinfo  {journal}
  {Journal of Chemical Physics}\ }\textbf {\bibinfo {volume} {8}},\ \bibinfo
  {pages} {212} (\bibinfo {year} {1940})}\BibitemShut {NoStop}%
\bibitem [{\citenamefont {Avrami}(1941)}]{Avrami3}%
  \BibitemOpen
  \bibfield  {author} {\bibinfo {author} {\bibfnamefont {M.}~\bibnamefont
  {Avrami}},\ }\href@noop {} {\bibfield  {journal} {\bibinfo  {journal}
  {Journal of Chemical Physics}\ }\textbf {\bibinfo {volume} {9}},\ \bibinfo
  {pages} {177} (\bibinfo {year} {1941})}\BibitemShut {NoStop}%
\bibitem [{\citenamefont {Cope}(1977)}]{Cope}%
  \BibitemOpen
  \bibfield  {author} {\bibinfo {author} {\bibfnamefont {F.~W.}\ \bibnamefont
  {Cope}},\ }\href@noop {} {\bibfield  {journal} {\bibinfo  {journal} {Physiol.
  Chem. \& Physics}\ }\textbf {\bibinfo {volume} {9}},\ \bibinfo {pages} {155}
  (\bibinfo {year} {1977})}\BibitemShut {NoStop}%
\bibitem [{\citenamefont {Rovelli}(2018)}]{Rovelli1}%
  \BibitemOpen
  \bibfield  {author} {\bibinfo {author} {\bibfnamefont {C.}~\bibnamefont
  {Rovelli}},\ }\href@noop {} {\emph {\bibinfo {title} {The Order of Time}}}\
  (\bibinfo  {publisher} {Riverhead Books},\ \bibinfo {address} {New York},\
  \bibinfo {year} {2018})\BibitemShut {NoStop}%
\bibitem [{\citenamefont {Scott}(1975)}]{Scott}%
  \BibitemOpen
  \bibfield  {author} {\bibinfo {author} {\bibfnamefont {A.~C.}\ \bibnamefont
  {Scott}},\ }\href@noop {} {\bibfield  {journal} {\bibinfo  {journal} {Rev.
  Mod. Phys.}\ }\textbf {\bibinfo {volume} {47}},\ \bibinfo {pages} {487}
  (\bibinfo {year} {1975})}\BibitemShut {NoStop}%
\bibitem [{\citenamefont {Bezanilla}\ and\ \citenamefont
  {Armstrong}(1977)}]{BezArm1977}%
  \BibitemOpen
  \bibfield  {author} {\bibinfo {author} {\bibfnamefont {F.}~\bibnamefont
  {Bezanilla}}\ and\ \bibinfo {author} {\bibfnamefont {C.~M.}\ \bibnamefont
  {Armstrong}},\ }\href@noop {} {\bibfield  {journal} {\bibinfo  {journal} {J.
  Gen. Physiol.}\ }\textbf {\bibinfo {volume} {70}},\ \bibinfo {pages} {549}
  (\bibinfo {year} {1977})}\BibitemShut {NoStop}%
\bibitem [{\citenamefont {Armstrong}\ and\ \citenamefont
  {Bezanilla}(1977)}]{ArmBez1977}%
  \BibitemOpen
  \bibfield  {author} {\bibinfo {author} {\bibfnamefont {C.~M.}\ \bibnamefont
  {Armstrong}}\ and\ \bibinfo {author} {\bibfnamefont {F.}~\bibnamefont
  {Bezanilla}},\ }\href@noop {} {\bibfield  {journal} {\bibinfo  {journal} {J.
  Gen. Physiol.}\ }\textbf {\bibinfo {volume} {70}},\ \bibinfo {pages} {567}
  (\bibinfo {year} {1977})}\BibitemShut {NoStop}%
\bibitem [{\citenamefont {Armstrong}\ and\ \citenamefont
  {Bezanilla}(1973)}]{ArmstrongBezanilla1}%
  \BibitemOpen
  \bibfield  {author} {\bibinfo {author} {\bibfnamefont {C.~M.}\ \bibnamefont
  {Armstrong}}\ and\ \bibinfo {author} {\bibfnamefont {F.}~\bibnamefont
  {Bezanilla}},\ }\href@noop {} {\bibfield  {journal} {\bibinfo  {journal}
  {Nature, Lond.}\ }\textbf {\bibinfo {volume} {242}},\ \bibinfo {pages} {459}
  (\bibinfo {year} {1973})}\BibitemShut {NoStop}%
\bibitem [{\citenamefont {Armstrong}\ and\ \citenamefont
  {Bezanilla}(1974)}]{ArmstrongBezanilla}%
  \BibitemOpen
  \bibfield  {author} {\bibinfo {author} {\bibfnamefont {C.~M.}\ \bibnamefont
  {Armstrong}}\ and\ \bibinfo {author} {\bibfnamefont {F.}~\bibnamefont
  {Bezanilla}},\ }\href@noop {} {\bibfield  {journal} {\bibinfo  {journal} {J.
  Gen. Physiol.}\ }\textbf {\bibinfo {volume} {63}},\ \bibinfo {pages} {533}
  (\bibinfo {year} {1974})}\BibitemShut {NoStop}%
\bibitem [{\citenamefont {Keynes}\ and\ \citenamefont
  {Rojas}(1974)}]{KeynesRojas}%
  \BibitemOpen
  \bibfield  {author} {\bibinfo {author} {\bibfnamefont {R.~D.}\ \bibnamefont
  {Keynes}}\ and\ \bibinfo {author} {\bibfnamefont {D.}~\bibnamefont {Rojas}},\
  }\href@noop {} {\bibfield  {journal} {\bibinfo  {journal} {J. Physiol.,
  Lond.}\ }\textbf {\bibinfo {volume} {239}},\ \bibinfo {pages} {393} (\bibinfo
  {year} {1974})}\BibitemShut {NoStop}%
\bibitem [{\citenamefont {Bezanilla}(2018)}]{Bezanilla}%
  \BibitemOpen
  \bibfield  {author} {\bibinfo {author} {\bibfnamefont {F.}~\bibnamefont
  {Bezanilla}},\ }\href {https://doi.org/10.1085/jgp.201812090} {\bibfield
  {journal} {\bibinfo  {journal} {J. Gen. Physiol.}\ }\textbf {\bibinfo
  {volume} {150}},\ \bibinfo {pages} {911} (\bibinfo {year}
  {2018})}\BibitemShut {NoStop}%
\bibitem [{\citenamefont {Wellis}\ \emph {et~al.}(1990)\citenamefont {Wellis},
  \citenamefont {DeFelice},\ and\ \citenamefont {Mazzanti}}]{Wellis}%
  \BibitemOpen
  \bibfield  {author} {\bibinfo {author} {\bibfnamefont {D.~P.}\ \bibnamefont
  {Wellis}}, \bibinfo {author} {\bibfnamefont {L.~J.}\ \bibnamefont
  {DeFelice}},\ and\ \bibinfo {author} {\bibfnamefont {M.}~\bibnamefont
  {Mazzanti}},\ }\href@noop {} {\bibfield  {journal} {\bibinfo  {journal}
  {Biophys. J.}\ }\textbf {\bibinfo {volume} {57}},\ \bibinfo {pages} {41}
  (\bibinfo {year} {1990})}\BibitemShut {NoStop}%
\bibitem [{\citenamefont {Carmeliet}(1992)}]{Carmeliet}%
  \BibitemOpen
  \bibfield  {author} {\bibinfo {author} {\bibfnamefont {E.}~\bibnamefont
  {Carmeliet}},\ }\href@noop {} {\bibfield  {journal} {\bibinfo  {journal}
  {Cardiovascular Research}\ }\textbf {\bibinfo {volume} {26}},\ \bibinfo
  {pages} {433} (\bibinfo {year} {1992})}\BibitemShut {NoStop}%
\bibitem [{\citenamefont {Tasaki}(1999{\natexlab{a}})}]{TasakiPhase}%
  \BibitemOpen
  \bibfield  {author} {\bibinfo {author} {\bibfnamefont {I.}~\bibnamefont
  {Tasaki}},\ }\href@noop {} {\bibfield  {journal} {\bibinfo  {journal}
  {Ferroelectrics}\ }\textbf {\bibinfo {volume} {220}},\ \bibinfo {pages} {305}
  (\bibinfo {year} {1999}{\natexlab{a}})}\BibitemShut {NoStop}%
\bibitem [{\citenamefont {Hopfield}(1982)}]{Hopfield}%
  \BibitemOpen
  \bibfield  {author} {\bibinfo {author} {\bibfnamefont {J.~J.}\ \bibnamefont
  {Hopfield}},\ }\href@noop {} {\bibfield  {journal} {\bibinfo  {journal}
  {Proc. Natl. Acad. Sci. USA}\ }\textbf {\bibinfo {volume} {79}},\ \bibinfo
  {pages} {2534} (\bibinfo {year} {1982})}\BibitemShut {NoStop}%
\bibitem [{\citenamefont {Hopfield}(1984)}]{Hopfield2}%
  \BibitemOpen
  \bibfield  {author} {\bibinfo {author} {\bibfnamefont {J.~J.}\ \bibnamefont
  {Hopfield}},\ }\href@noop {} {\bibfield  {journal} {\bibinfo  {journal}
  {Proceedings of the National Academy of Sciences}\ }\textbf {\bibinfo
  {volume} {81}},\ \bibinfo {pages} {3088} (\bibinfo {year}
  {1984})}\BibitemShut {NoStop}%
\bibitem [{\citenamefont {Kandel}(2001)}]{Kandel}%
  \BibitemOpen
  \bibfield  {author} {\bibinfo {author} {\bibfnamefont {E.~R.}\ \bibnamefont
  {Kandel}},\ }\href@noop {} {\bibfield  {journal} {\bibinfo  {journal}
  {Bioscience Reports}\ }\textbf {\bibinfo {volume} {21}} (\bibinfo {year}
  {2001})}\BibitemShut {NoStop}%
\bibitem [{\citenamefont {Landau}\ and\ \citenamefont
  {Lifshitz}(1980)}]{Landau}%
  \BibitemOpen
  \bibfield  {author} {\bibinfo {author} {\bibfnamefont {L.~D.}\ \bibnamefont
  {Landau}}\ and\ \bibinfo {author} {\bibfnamefont {E.~M.}\ \bibnamefont
  {Lifshitz}},\ }\href@noop {} {\emph {\bibinfo {title} {Statistical
  Physics}}},\ \bibinfo {edition} {3rd}\ ed.\ (\bibinfo  {publisher} {Pergamon
  Press},\ \bibinfo {year} {1980})\ p.~\bibinfo {pages} {2}\BibitemShut
  {NoStop}%
\bibitem [{\citenamefont {Tasaki}(1999{\natexlab{b}})}]{TasakiHeat}%
  \BibitemOpen
  \bibfield  {author} {\bibinfo {author} {\bibfnamefont {I.}~\bibnamefont
  {Tasaki}},\ }\href@noop {} {\bibfield  {journal} {\bibinfo  {journal} {Jpn.
  J. Physiol.}\ }\textbf {\bibinfo {volume} {49}},\ \bibinfo {pages} {125}
  (\bibinfo {year} {1999}{\natexlab{b}})}\BibitemShut {NoStop}%
\bibitem [{\citenamefont {Rabe}\ \emph {et~al.}(2007)\citenamefont {Rabe},
  \citenamefont {Ahn},\ and\ \citenamefont {Triscone}}]{Rabe}%
  \BibitemOpen
  \bibinfo {editor} {\bibfnamefont {K.}~\bibnamefont {Rabe}}, \bibinfo {editor}
  {\bibfnamefont {C.~H.}\ \bibnamefont {Ahn}},\ and\ \bibinfo {editor}
  {\bibfnamefont {J.-M.}\ \bibnamefont {Triscone}},\ eds.,\ \href@noop {}
  {\emph {\bibinfo {title} {Physics of Ferroelectrics: A Modern
  Perspective}}},\ \bibinfo {series} {Topics in Applied Physics}, Vol.\
  \bibinfo {volume} {105}\ (\bibinfo  {publisher} {Springer-Verlag},\ \bibinfo
  {year} {2007})\BibitemShut {NoStop}%
\bibitem [{\citenamefont {von Hippel}(1970)}]{Hippel}%
  \BibitemOpen
  \bibfield  {author} {\bibinfo {author} {\bibfnamefont {A.~R.}\ \bibnamefont
  {von Hippel}},\ }in\ \href@noop {} {\emph {\bibinfo {booktitle} {Proceedings,
  Second Int. Meeting on Ferroelectricity 1969}}},\ Vol.~\bibinfo {volume}
  {28}\ (\bibinfo {year} {1970})\ \bibinfo {note} {supplement 1}\BibitemShut
  {NoStop}%
\bibitem [{\citenamefont {Leuchtag}(1987)}]{Leuchtag1}%
  \BibitemOpen
  \bibfield  {author} {\bibinfo {author} {\bibfnamefont {H.~R.}\ \bibnamefont
  {Leuchtag}},\ }\href@noop {} {\bibfield  {journal} {\bibinfo  {journal} {JJ.
  theor. Biol.}\ }\textbf {\bibinfo {volume} {127}},\ \bibinfo {pages} {321}
  (\bibinfo {year} {1987})}\BibitemShut {NoStop}%
\bibitem [{\citenamefont {Leuchtag}(1995)}]{Leuchtag2}%
  \BibitemOpen
  \bibfield  {author} {\bibinfo {author} {\bibfnamefont {H.~R.}\ \bibnamefont
  {Leuchtag}},\ }\href@noop {} {\bibfield  {journal} {\bibinfo  {journal}
  {Biophysical Chemistry}\ }\textbf {\bibinfo {volume} {53}},\ \bibinfo {pages}
  {197} (\bibinfo {year} {1995})}\BibitemShut {NoStop}%
\bibitem [{\citenamefont {Palti}\ and\ \citenamefont {Jr.}(1969)}]{Palti}%
  \BibitemOpen
  \bibfield  {author} {\bibinfo {author} {\bibfnamefont {Y.}~\bibnamefont
  {Palti}}\ and\ \bibinfo {author} {\bibfnamefont {W.~J.~A.}\ \bibnamefont
  {Jr.}},\ }\href@noop {} {\bibfield  {journal} {\bibinfo  {journal} {J. Membr.
  Biol.}\ }\textbf {\bibinfo {volume} {1}},\ \bibinfo {pages} {431} (\bibinfo
  {year} {1969})}\BibitemShut {NoStop}%
\bibitem [{\citenamefont {Hodgkin}(1975)}]{Hodgkin1975}%
  \BibitemOpen
  \bibfield  {author} {\bibinfo {author} {\bibfnamefont {A.~L.}\ \bibnamefont
  {Hodgkin}},\ }\href@noop {} {\bibfield  {journal} {\bibinfo  {journal}
  {Philos. Trans. R. Soc. Lond. B}\ }\textbf {\bibinfo {volume} {270}},\
  \bibinfo {pages} {297} (\bibinfo {year} {1975})}\BibitemShut {NoStop}%
\bibitem [{\citenamefont {Adrian}(1975)}]{Adrian}%
  \BibitemOpen
  \bibfield  {author} {\bibinfo {author} {\bibfnamefont {R.~H.}\ \bibnamefont
  {Adrian}},\ }\href@noop {} {\bibfield  {journal} {\bibinfo  {journal} {Proc.
  R. Soc. Lond. B}\ }\textbf {\bibinfo {volume} {189}},\ \bibinfo {pages} {81}
  (\bibinfo {year} {1975})}\BibitemShut {NoStop}%
\bibitem [{\citenamefont {Sangrey}\ \emph {et~al.}(2004)\citenamefont
  {Sangrey}, \citenamefont {Friesen},\ and\ \citenamefont {Levy}}]{Sangrey1}%
  \BibitemOpen
  \bibfield  {author} {\bibinfo {author} {\bibfnamefont {T.~D.}\ \bibnamefont
  {Sangrey}}, \bibinfo {author} {\bibfnamefont {W.~O.}\ \bibnamefont
  {Friesen}},\ and\ \bibinfo {author} {\bibfnamefont {W.~B.}\ \bibnamefont
  {Levy}},\ }\href@noop {} {\bibfield  {journal} {\bibinfo  {journal} {J.
  Neurophysiol.}\ }\textbf {\bibinfo {volume} {91}},\ \bibinfo {pages} {2541}
  (\bibinfo {year} {2004})}\BibitemShut {NoStop}%
\end{thebibliography}%


\begin{appendices}
\section{Current definitions and the conventional ``membrane current'' terminology}
\label{app:currentDefinitions}

The terminology used for currents in cable theory can obscure the
physical distinction between longitudinal conduction through the
axoplasm and the ionic and capacitive processes that enter the local
membrane charge balance. This appendix defines the current quantities
used in the present work and explains the restricted sense in which the
conventional designation ``membrane current'' is retained.

\subsection{Axial Ohmic current and its divergence}

\quad
Let $J_z(z,t)$ denote the axial current density flowing longitudinally
through the axoplasm. It is defined per unit cross-sectional area of the
axon and obeys Ohm's law,
\begin{equation}
J_z(z,t)
=
-\frac{1}{R_i}\frac{\partial V(z,t)}{\partial z},
\label{app:OhmCurrent}
\end{equation}
where $R_i$ is the axoplasmic resistivity. Thus, $J_z$ is the axial
Ohmic current in the present formulation. It flows longitudinally
through the axoplasm, not across the membrane.

\quad
For a cylindrical axon of radius $R$, spatial variation of the axial
current produces a local imbalance between the current entering and
leaving an infinitesimal axonal segment. We denote the corresponding
quantity by
\begin{equation}
J_m(z,t)
\equiv
-\frac{R}{2}\frac{\partial J_z(z,t)}{\partial z}.
\label{app:AxoplasmicDivergence}
\end{equation}
The geometric factor $R/2$ is the ratio of axonal cross-sectional area
to membrane area per unit length,
$\pi R^2/(2\pi R)$. It therefore converts the axial-current divergence
to the membrane-area normalization used for the capacitive and ionic
currents. The quantity $J_m$ has the dimensions of current density, but
does not represent an independently transported radial current. We
refer to it as the \emph{geometry-converted axial-current divergence}.
It is determined by the spatial variation of $J_z$ and is not the axial
current $J_z$ itself. For the propagated solution, the value of $J_m$ is therefore fixed by
Ohm's law, axonal geometry, and the voltage waveform
\cite{Jurisic1987}.
\subsection{Charge conservation at the membrane}

\quad
The two contributions to the local membrane charge balance are the
capacitive current
\begin{equation}
J_C(z,t)
=
C_m\frac{\partial V(z,t)}{\partial t},
\label{app:CapacitiveCurrent}
\end{equation}
and the total ionic current $J_I(z,t)$. Both quantities are expressed
per unit membrane area. The capacitive current describes the change in
membrane charge, whereas the total ionic current represents net charge
transfer through ionic pathways across the membrane. The sign
convention used in this work takes positive current as outward. The
total ionic current includes all ionic species contributing to the
propagated waveform; sodium and potassium labels introduced later
denote dominant effective components rather than an exhaustive
separation of ionic species.

\quad
The spatial variation of $J_z$ produces a local imbalance between axial
inflow and outflow within an axonal segment. When normalized to the
membrane area of that segment, this imbalance is the
geometry-converted axial-current divergence $J_m$. Local charge
conservation requires $J_m$ to equal the sum of the capacitive and total
ionic currents,
\begin{equation}
J_m(z,t)
=
C_m\frac{\partial V(z,t)}{\partial t}
+
J_I(z,t).
\label{app:CurrentBalance}
\end{equation}
Equivalently,
\begin{equation}
J_I(z,t)
=
J_m(z,t)
-
C_m\frac{\partial V(z,t)}{\partial t}.
\label{app:IonicDifference}
\end{equation}
Equation~\eqref{app:IonicDifference} is the current identity used
throughout the manuscript. It states that, for the propagated solution,
the total ionic current is reconstructed by subtracting the capacitive
current from $J_m$.

\quad
The quantities in Eq.~\eqref{app:CurrentBalance} have the same units but
different physical meanings. The axial current $J_z$ is defined per unit
axonal cross-sectional area and flows longitudinally through the
axoplasm. By contrast, $J_m$, $C_m\partial V/\partial t$, and $J_I$ are
expressed per unit membrane area. The capacitive current describes the
change in membrane charge, whereas the total ionic current represents
net ionic charge transfer across the membrane. The quantity $J_m$
represents the local supply or removal of charge associated with the
spatial variation of the axial current; it does not represent an
independently transported radial current.
\subsection{Meaning of the conventional designation ``membrane current''}

\quad
In the electrophysiological and cable-theory literature, the sum
\begin{equation}
C_m\frac{\partial V}{\partial t}+J_I
\label{app:ConventionalMembraneCurrent}
\end{equation}
is commonly called the ``membrane current.'' For a propagated cable
solution, local charge conservation makes this sum identically equal to
$J_m$, the geometry-converted axial-current divergence. The conventional
designation is therefore consistent with the current balance, but it can
obscure the origin of $J_m$ in the spatial variation of $J_z$.

\quad
The distinction is particularly important when propagated and
voltage-clamp descriptions are compared. In voltage-clamp experiments,
the conventional designation ``membrane current'' refers to the sum of
the capacitive current associated with membrane charging and the ionic
current across the membrane, inferred from the external electrode
current required to impose the commanded membrane potential. In a
propagated action potential, the same capacitive-plus-ionic current is
supplied locally by the axial-current divergence,
\[
J_m=-\frac{R}{2}\frac{\partial J_z}{\partial z}
   =C_m\frac{\partial V}{\partial t}+J_I.
\]
The same local charge-balance relation therefore underlies both
descriptions, but the current is supplied externally in the
voltage-clamp experiment and by axial-current divergence during
propagation. For an unforced space-clamped action potential, no
axial-current divergence or continuing external current contributes.
Consequently, at the action-potential peak,
$\partial V/\partial t=0$ and $J_I=0$, so the ionic components must sum
to zero. At the peak of a propagated action potential,
$\partial V/\partial t=0$ but $J_m$ need not vanish; therefore,
$J_I=J_m$, and no cancellation of the ionic components is required.

\quad
For this reason, the symbol $J_m$ is retained in the equations and
figures to preserve continuity with conventional cable notation. In
descriptive prose, it is referred to as the
\emph{geometry-converted axial-current divergence}, or simply as $J_m$
after its meaning has been established. The phrase ``membrane current''
is used only when referring to conventional nomenclature, historical
figure labels, or the identity
$J_m=C_m\partial V/\partial t+J_I$. Neither $J_z$ nor $J_m$ should be
described simply as ``the axoplasmic current'': $J_z$ is the axial
Ohmic current, whereas $J_m$ is the geometry-converted divergence of
$J_z$, expressed on the membrane-area basis used in the local charge
balance.

\quad
Thus, the phase-space reconstruction preserves the same structural
distinction: $J_z$ is the axial Ohmic current; $J_m$ is the
geometry-converted divergence of $J_z$, expressed on the membrane-area
basis; $C_m\Phi$ is the capacitive current; and $J_I$ is the total ionic
current. The definition
\[
J_m=-\frac{R}{2}\frac{\partial J_z}{\partial z}
\]
and the balance
\[
J_m=C_m\Phi+J_I
\]
together express local charge conservation. They do not imply identical
physical origins: $J_z$ describes longitudinal conduction through the
axoplasm, $J_m$ represents the local imbalance of axial current,
$C_m\Phi$ describes membrane charging, and $J_I$ represents ionic
charge transfer across the membrane.

\quad
For the steadily propagating solution developed in Part~II, these
definitions lead to the phase-space expressions for $J_z(V)$, $J_m(V)$,
and $J_I(V)$. Those relations are derived there directly from Ohm's law,
steady propagation, and charge conservation and are not repeated here.
The purpose of this appendix is to distinguish the axial Ohmic current
$J_z$ from its geometry-converted divergence $J_m$ and to clarify the
restricted use of the conventional designation ``membrane current.''

\section{\label{sec:levelA}Action-potential hysteresis and effective polarization states}

\subsection{Electrodynamic origin of the hysteretic trajectory}

\quad
The steadily propagated action potential follows distinct depolarizing
and recovery branches. At a membrane potential attained on both
branches,
\[
V_{\uparrow}=V_{\downarrow},
\]
the phase-space velocities generally differ:
\[
\Phi_{\uparrow}(V)>0,
\qquad
\Phi_{\downarrow}(V)<0.
\]
The two points therefore have the same membrane potential but do not
represent the same dynamical state.

\quad
The membrane capacitance $C_m$ is held constant throughout the action
potential. Consequently, the capacitive charge density,
\begin{equation}
q_C(V)=C_mV,
\label{eq:hystCapCharge}
\end{equation}
is a single-valued function of membrane potential. The capacitive
trajectory retraces the same line during charging and recovery and
cannot by itself generate hysteresis:
\begin{equation}
\oint V\,dq_C
=
C_m\oint V\,dV
=
0.
\label{eq:hystCapArea}
\end{equation}
The hysteretic behaviour must therefore originate in the propagated
current balance and in the internal channel states rather than in a
history-dependent membrane capacitance. 
\quad
Although the membrane capacitance is constant and
$q_C=C_mV$ is single-valued, the capacitive current is branch
dependent because it is proportional to $\Phi(V)$. On the depolarizing
branch, $C_m\Phi>0$, whereas on the recovery branch,
$C_m\Phi<0$. Consequently,
\[
J_I(V)=J_m(V)-C_m\Phi(V)
\]
shows that capacitive discharge shifts the reconstructed ionic current
in the outward direction during recovery. The capacitive element does
not itself possess hysteresis, but its branch-dependent current,
together with the branch dependence of $J_m$, contributes to the
separation of the ionic-current trajectories in the $(V,J_I)$ plane.

\quad
Let $J_m$ denote the geometry-converted axial-current divergence defined
in Appendix~A. Charge conservation gives
\begin{equation}
J_m(V)
=
-\frac{R}{2}\frac{\partial J_z}{\partial z}
=
J_I(V)+C_m\Phi(V)
=
\frac{C_m}{k}
\Phi(V)\frac{d\Phi(V)}{dV},
\label{eq:hystAxialDivergence}
\end{equation}
where $J_I$ is the total ionic current and $k$ is the propagation
parameter. At the same membrane potential, the two branches generally
satisfy
\begin{equation}
J_m^{\uparrow}(V)
\neq
J_m^{\downarrow}(V),
\label{eq:hystBranchJm}
\end{equation}
because their phase-space velocities and local trajectory geometries
differ. It follows that
\begin{equation}
J_I^{\uparrow}(V)
=
J_m^{\uparrow}(V)-C_m\Phi_{\uparrow}(V)
\neq
J_m^{\downarrow}(V)-C_m\Phi_{\downarrow}(V)
=
J_I^{\downarrow}(V).
\label{eq:hystBranchJI}
\end{equation}
Thus, the branch dependence of $J_m$ and $C_m\Phi$, constrained by local
charge conservation, provides the electrodynamic basis of the
propagated current--voltage hysteresis.

\quad
The axial-current divergence alone does not specify the internal state
of the membrane. That information is supplied by the channel-completion
variables. The complete propagated trajectory may therefore be
represented in the extended state space
\begin{equation}
\Gamma_{\mathrm{mP}}
=
\{V,\Phi,\boldsymbol{f}\},
\label{eq:hystStateSpace}
\end{equation}
where $\boldsymbol{f}$ denotes the set of mAvrami completion fractions.
At the same membrane potential,
\begin{equation}
\boldsymbol{f}_{\uparrow}(V)
\neq
\boldsymbol{f}_{\downarrow}(V),
\label{eq:hystBranchStates}
\end{equation}
so the depolarizing and recovery branches correspond to different
channel configurations. The hysteresis loop appears when this
continuous higher-dimensional trajectory is projected onto a reduced
plane such as $(V,J_I)$ or onto the effective-equilibrium-potential
representation introduced below.

\subsection{Symmetry change at the action-potential peak}

\quad
At action-potential inception, the channel decomposition changes from
the recovery $H$ symmetry to the activation $M$ symmetry, and the inward
sodium activation current $J_M$ is initiated with effective reversal
potential $V_M$. At the action-potential peak $V_p$, the sodium-channel
description changes from the $M$ branch to the $H$ branch while the
sodium completion fraction remains close to unity. The
sodium-dominated recovery current is characterized by the lower
effective reversal potential $V_H$.

\quad
The experimentally reconstructed capacitive current, total ionic
current, and geometry-converted axial-current divergence $J_m$ remain
continuous at $V_p$. The individual components of the mAvrami
decomposition, however, change branches there. In particular, the
pre-peak polarization-current segment $J_{P4}$ and the first post-peak
segment $J_{P5}$ satisfy
\begin{equation}
\lim_{V\rightarrow V_p^-}J_{P4}(V)
\neq
\lim_{V\rightarrow V_p^+}J_{P5}(V).
\label{eq:hystP45Discontinuity}
\end{equation}
This discontinuity belongs to the decomposition, not to the
experimentally reconstructed total ionic current.

\quad
The $J_{P4}\rightarrow J_{P5}$ change is consistent with a change of
constitutive branch at the $M\rightarrow H$ transition. A finite
discontinuity of polarization current would not, by itself, require a
discontinuity of an underlying polarization charge; it would instead
correspond to a change in its rate of evolution. The present work does
not integrate the polarization-current segments to reconstruct such a
charge and therefore does not assume this stronger interpretation.

\quad
The simultaneous changes in sodium conductance, ionic time rate,
effective reversal potential, and polarization-current branch support
the interpretation of the action-potential peak as a continuous
symmetry transition between distinct sodium-channel states. The
experimentally reconstructed currents remain continuous while the
state-dependent decomposition changes. This behaviour is consistent
with the general description of a continuous phase transition in which
the state symmetry changes without a discontinuity in the observable
trajectory \cite[p.~445]{Landau}.
\subsection{Effective-equilibrium-potential hysteresis}

\quad
The branch-dependent ionic driving-force structure is summarized by the
effective sodium and potassium equilibrium-potential shifts
\begin{equation}
\Delta V_{\mathrm{Na}}
=
V_M-V_H,
\qquad
\Delta V_{\mathrm{K}}
=
V_K-V_N.
\label{eq:hystPotentialShifts}
\end{equation}
Here $V_M$ and $V_H$ characterize the sodium activation and recovery
branches, respectively, while $V_N$ and $V_K$ characterize the
corresponding potassium states during recovery and return toward rest.

\quad
Across the analyzed temperatures, $\Delta V_{\mathrm{Na}}$ and
$\Delta V_{\mathrm{K}}$ are approximately linearly correlated. Their
temperature dependences have similar slopes, and the average ratio for
the six sweeps is
\begin{equation}
\frac{V_M-V_H}{V_K-V_N}
\approx 0.9.
\label{eq:hystShiftRatio}
\end{equation}
These correlations suggest that the sodium and potassium
effective-equilibrium-potential shifts are not merely independent local
fitting parameters, but reflect a coordinated reorganization of the
propagated ionic driving-force structure.

\begin{figure*}
\centering
\includegraphics[width=2.1\columnwidth]{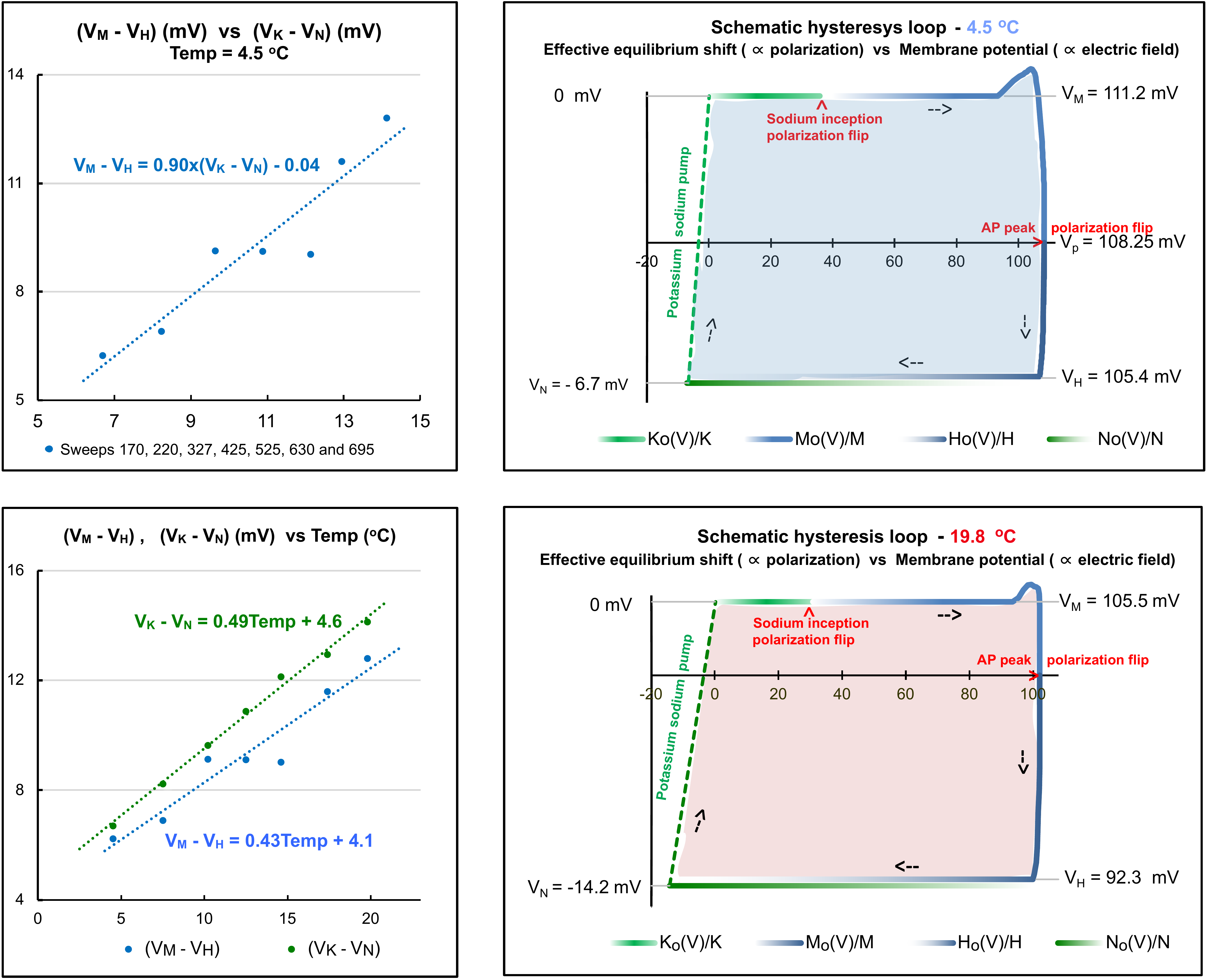}
\caption{\label{fig:HysteresisL}
Effective-equilibrium-potential representation of action-potential
hysteresis.
{\bf (a)} Correlation between the sodium shift $(V_M-V_H)$ and the
potassium shift $(V_K-V_N)$ over the analyzed temperatures.
{\bf (b)} Schematic reduced hysteresis trajectory at
$4.5\,^\circ\mathrm{C}$.
{\bf (c)} Temperature dependence of $(V_K-V_N)$ and $(V_M-V_H)$.
{\bf (d)} Schematic reduced hysteresis trajectory at
$19.8\,^\circ\mathrm{C}$.
In panels (b) and (d), the horizontal coordinate is the membrane
potential and is proportional to the transmembrane electric field for
fixed effective membrane thickness. The vertical coordinate is the
branch-dependent effective-equilibrium-potential shift, used as a
collective state coordinate associated with polarization-related and
channel-symmetry changes. Arrows indicate the direction of traversal
during the action potential. Because the vertical coordinate has not
been calibrated as polarization charge, the enclosed geometrical area
is an index of the schematic hysteretic excursion and is not
interpreted here as a direct measurement of dissipated energy.
}
\end{figure*}

\quad
Figure~\ref{fig:HysteresisL} displays these relations. Panels (b) and
(d) are schematic reduced representations of the action-potential cycle
at $4.5\,^\circ\mathrm{C}$ and $19.8\,^\circ\mathrm{C}$. The horizontal
coordinate is the membrane potential and, for an approximately fixed
membrane thickness, is proportional to the transmembrane electric
field,
\begin{equation}
E_m\simeq\frac{V}{d_m}.
\label{eq:hystField}
\end{equation}
The vertical coordinate is the branch-dependent
effective-equilibrium-potential shift. It is used as a
polarization-related collective state coordinate, not as a directly
measured thermodynamic polarization.

\quad
On the common coordinate scales used in Fig.~\ref{fig:HysteresisL}, the
schematic loop at $19.8\,^\circ\mathrm{C}$ encloses a larger geometrical
area than that at $4.5\,^\circ\mathrm{C}$. This illustrates a larger
excursion of the effective driving-force state at the higher
temperature. It does not, however, establish that the heat released is
larger by the same factor. A quantitative energetic interpretation
would require an independently established constitutive relation
between the effective-equilibrium-potential shift and a polarization
charge. Heat production accompanying electrical excitation has been
measured experimentally \cite{TasakiHeat}, but the present schematic
loop is not a calorimetric measurement.
\subsection{Relation to polarization and ferroelectric-like behaviour}

\quad
A thermodynamic ferroelectric hysteresis loop is conventionally
represented in terms of polarization and electric field. For a thin
membrane, an analogous area-based representation would use the
polarization charge per unit membrane area $q_P$ and the membrane
potential $V$. The corresponding loop would then be $(q_P,V)$, or
equivalently $(P,E_m)$ in a volume description. The phrase
``polarization hysteresis'' is used here in this conventional structural
sense \cite{Rabe}; the present analysis does not identify the axonal
membrane with a crystalline ferroelectric material.

\quad
The present analysis identifies polarization-current components within
the decomposition of the total ionic current and shows that these
components change branch at channel-symmetry transitions. It does not,
however, establish the constitutive relation
\begin{equation}
J_P=\frac{dq_P}{dt}.
\label{eq:hystConstitutiveQuestion}
\end{equation}
Accordingly, the effective-equilibrium-potential shifts in
Fig.~\ref{fig:HysteresisL} are treated as polarization-related state
coordinates rather than as measurements of $q_P$ or $P$.

\quad
If Eq.~\eqref{eq:hystConstitutiveQuestion} is established in a
subsequent analysis, the polarization charge could be reconstructed on
each monotonic branch of the measured phase-space trajectory:
\begin{equation}
q_P(V)-q_P(V_i)
=
\int_{V_i}^{V}
\frac{J_P(\widetilde V)}
{\Phi(\widetilde V)}
\,d\widetilde V.
\label{eq:hystFutureCharge}
\end{equation}
The branchwise integrals would then be joined through the trajectory
turning points. Integration of all polarization-current segments would
test whether the discontinuous decomposition components assemble into
a continuous and approximately closed charge trajectory. A resulting
$q_P$--$V$ loop would permit the associated electrical work per
membrane area to be evaluated from
\begin{equation}
W_P^{(A)}
=
\oint V\,dq_P.
\label{eq:hystFutureEnergy}
\end{equation}
This reconstruction is not required for the present results and is left
for future work.

\quad
The constant value of $C_m$ is important in this interpretation. The
observed hysteresis is not attributed to a changing membrane
capacitance. Rather, it arises from the branch-dependent propagated
current balance together with the evolving internal channel states.
The ordinary capacitive contribution remains reversible, whereas the
ionic and polarization-related components follow different activation
and recovery paths.

\quad
Restoration of the membrane potential does not by itself establish
restoration of the species-specific ionic gradients. Their slower,
ATP-dependent recovery may provide a metabolic closure of the complete
cycle whose quantitative relation to a reconstructed
polarization-charge hysteresis loop remains to be determined.

\quad
Ferroelectric behaviour in biological excitation has previously been
proposed by von Hippel \cite{Hippel} and Leuchtag \cite{Leuchtag1},
including a possible ferroelectric interpretation of sodium-channel
behaviour \cite{Leuchtag2} based in part on the capacitance measurements
of Palti \cite{Palti}. The present results do not establish microscopic
ferroelectricity of the axonal membrane or sodium channel. They
demonstrate branch-dependent electrical hysteresis and effective
equilibrium potentials, identify polarization-current components, and
support the interpretation of an $M\rightarrow H$ symmetry transition.
These observations motivate the more limited description
\emph{ferroelectric-like}, while the quantitative identification of a
polarization order parameter remains an open constitutive problem.

\section{\label{sec:levelH}Optimal sodium-channel density}

\quad
Hodgkin \cite{Hodgkin1975} proposed that an optimal sodium-channel
density maximizes the velocity of action-potential propagation. He
assumed that the maximum sodium conductance $g_{\mathrm{Na}}$, identified
here with $g_M$, is proportional to the surface density $M$ of sodium
channels. His calculation yielded an optimal density of approximately
$1000\,\mu\mathrm{m}^{-2}$, about twice the value subsequently estimated
by Keynes and Rojas \cite{KeynesRojas}. Adrian \cite{Adrian} calculated
the maximum propagation velocity using modified Hodgkin--Huxley
equations that included a sodium gating current and obtained velocities
substantially below the observed value. Sangrey \emph{et al.}
\cite{Sangrey1} later used a reparametrized Hodgkin--Huxley model and
found that the predicted optimum remained robust after refinement of
the original formulation.

\quad
Following Hodgkin, we assume that the maximum sodium conductance is
proportional to the sodium-channel surface density:
\bq
\label{Density1}
g_M=g^*_{\mathrm{Na}}M,
\eq
where $g^*_{\mathrm{Na}}$ is the single-channel conductance at a given
temperature. Increasing $M$ increases the propagation velocity, but
also increases the membrane capacitance associated with the channels.
In the quasilinear region in which the sodium channels are fully open,
the maximum sodium conductance is given by Eq.~\eqref{Sodiumd}:
\bq
\label{Density2}
g_M
=
\mu_M
\left(
\frac{\mu_M}{k}+1
\right)
C_m.
\eq
As shown in Fig.~\ref{fig:MHSymmetriesL}(a), the temperature
dependences of $g_M$, $k$, and $\mu_M$ are approximately parallel up to
about $20\,^\circ\mathrm{C}$. Over this temperature range, $\mu_M$ and
$k$ maintain the approximately constant ratio
\bq
\label{Density3}
\frac{\mu_M}{k}\approx 3.2.
\eq
It is therefore reasonable to assume, over this range, that the
time-rate parameter $\mu_M$ has the same dependence on $R_i$, $R$, $v$,
and $C_m$ as the propagation parameter $k$. Using
Eq.~\eqref{Density3} together with Eq.~\eqref{propagationkL},
Eq.~\eqref{Density2} becomes
\bq
\label{Density4}
g_M
\approx
13.44\,kC_m
=
26.88\,v^2C_m^2\frac{R_i}{R},
\eq
and therefore
\bq
\label{Density5}
v^2
\approx
\frac{g^*_{\mathrm{Na}}R}{26.88R_i}
\frac{M}
{\left(C_0+MC^*_{\mathrm{Na}}\right)^2},
\eq
where $C^*_{\mathrm{Na}}$ is the effective capacitance associated with
a single sodium channel and $C_0$ is the membrane capacitance in the
absence of sodium channels.

\quad
Assuming that $C_0$ is independent of $M$, the propagation velocity is
maximal when
\bq
\label{Density6}
M
\approx
\frac{C_m}{2C^*_{\mathrm{Na}}}.
\eq
For $C^*_{\mathrm{Na}}=8\times10^{-18}\,\mathrm{F}$ and
$C_m=1\,\mu\mathrm{F\,cm^{-2}}$, this gives
\[
M\approx625\,\mu\mathrm{m}^{-2}.
\]
Keynes and Rojas \cite{KeynesRojas} estimated
$M\approx500\,\mu\mathrm{m}^{-2}$ using
$g^*_{\mathrm{Na}}=2.5\,\mathrm{pS}$ and
$C^*_{\mathrm{Na}}=8\times10^{-18}\,\mathrm{F}$.

\section{\label{sec:levelF}Methods and Fitting Procedure}

\quad
The experimental data analyzed in this work were originally recorded
by J.~J.~C. Rosenthal and F. Bezanilla
[Biol. Bull. \textbf{199}, 135--143 (2000)] and were provided directly
to Nikola K. Jurisic by Prof. Rosenthal in digital form. The raw
potential-versus-time recordings are not publicly archived and remain
the intellectual property of the original investigators. The processed
current-versus-potential datasets derived from these recordings,
together with the numerical analyses supporting the findings of this
article, are available from the corresponding author upon reasonable
request.

\quad
The Rosenthal--Bezanilla data contain action-potential recordings
$V(t)$ obtained at two positions along the axon over a range of
temperatures. The present analysis uses the recording obtained at the
position farther from the stimulus. Rosenthal measured the axoplasmic
resistivity at $18.5\,^\circ\mathrm{C}$ and found the result to be
consistent with the commonly used temperature dependence
\bq
R_i
=
51.05
\times
1.35^{-(T-6.3\,^\circ\mathrm{C})/(10\,^\circ\mathrm{C})}
\,
\Omega\,\mathrm{cm}.
\eq

\quad
The experimental action potentials consist of discrete voltage values
sampled at fixed time intervals. The noise level depends on temperature
and on the local rate of change of the waveform. Gaussian smoothing was
used to reduce noise while limiting the loss of waveform detail. The
low-temperature recordings contain more data points and exhibit greater
noise than the high-temperature recordings and therefore required
larger smoothing widths. At a given temperature, the recovery region
was noisier and required greater smoothing than the rising edge.

\begin{figure*}
\includegraphics[width=2.1\columnwidth]{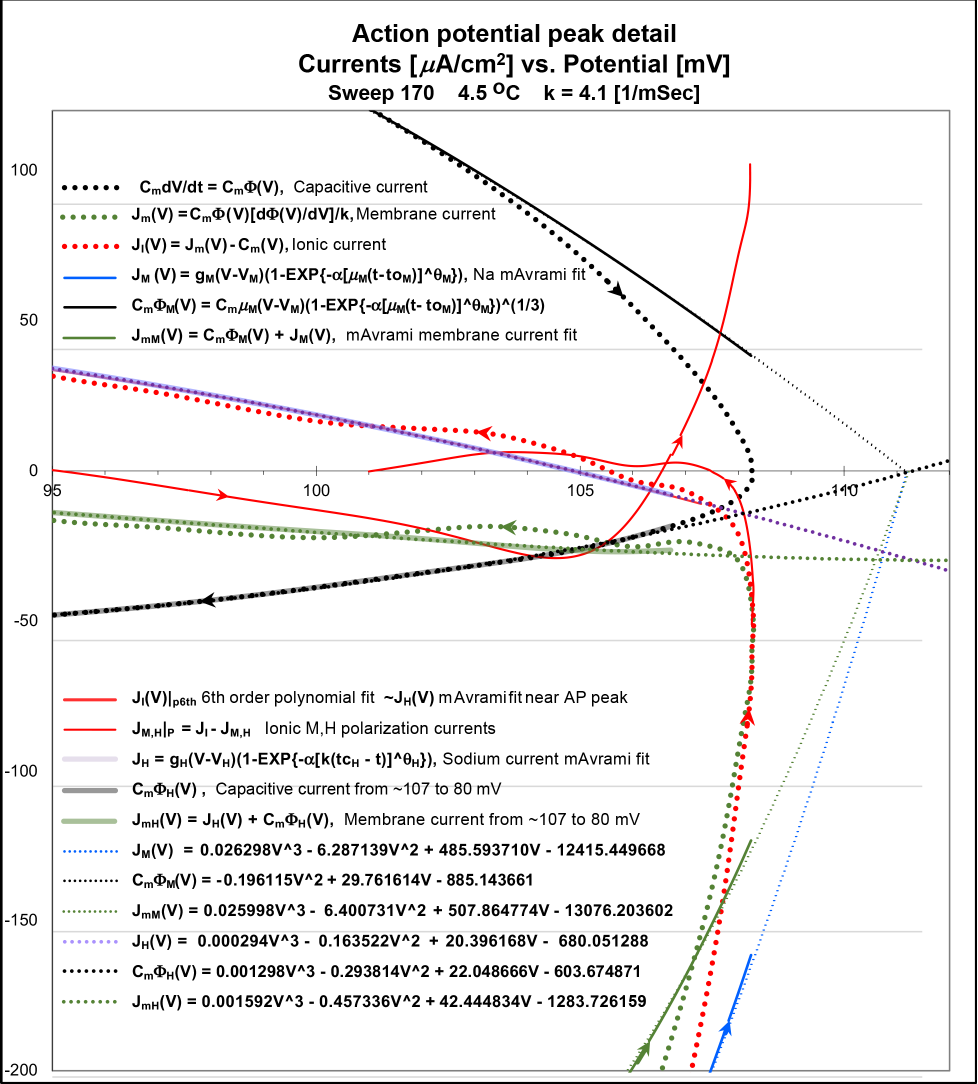}
\caption{\label{fig:APPeakDetailSweep170L}
Currents in the vicinity of the action-potential peak $V_p$. 
The experimental capacitive, membrane, and ionic currents remain continuous through $V_p$ (arrows indicate time direction). 
The mAvrami representations and the associated ionic polarization component exhibit discontinuities at $V_p$, marking the transition point in the decomposition while preserving continuity of the total current.
} 
\end{figure*}

\quad
Approximate time-rate parameters $\mu_K$, $\mu_M$, and $\mu_N$ and
maximum conductances $g_K$, $g_M$, and $g_N$ can be read from
Figs.~\ref{fig:RisingEdgeSweep425L},
\ref{fig:RecoveryCurrentsSweep425L}, and
\ref{fig:APPeakSweep425L}, respectively. Alternatively, if the
corresponding time rate is known, the maximum conductance can be
determined from Eqs.~\eqref{Footd}, \eqref{Sodiumd}, and
\eqref{Taild}, and vice versa. No corresponding relation was used for
the $H$-channel current. The time rate $\mu_H$ and maximum conductance
$g_H$ were therefore obtained separately from
Fig.~\ref{fig:APPeakDetailSweep170L}.

\quad
Quasilinear phase-space segments were interpreted as regions in which
the corresponding channel population is fully open. These regions can
therefore be fitted by the ionic-current expressions
Eqs.~\eqref{subeq:currentsd}, \eqref{subeq:currentsg}, and
\eqref{subeq:currentsh}. The corresponding open-channel fractions were
fitted with the mAvrami relations
Eqs.~\eqref{mAvramia}, \eqref{mAvramib}, and \eqref{mAvramic}, which
describe completion from zero to unity through a sigmoidal time
dependence.

\quad
The mAvrami fits are particularly sensitive to the time parameters
$t_{iM}$, $t_{iN}$, and $t_{cH}$ and to the corresponding time rates
$\mu_M$, $\mu_N$, and $\mu_H$. Here $t_{iM}$ and $t_{iN}$ are inception
times, whereas $t_{cH}$ is the $H$-channel closure time. The fits are
less sensitive to $\alpha_X$ and $\theta_X$. The values of all
$\alpha_X$ were therefore initialized with the fine-structure constant,
\[
\alpha=0.007297352,
\]
and the three mAvrami exponents were initialized with
$\theta_X=3.78$.
\quad
The quasilinear segment of the recovery sodium current is clearly
resolved and crosses the zero-current axis below, but close to, the
action-potential peak. By contrast, the inward current on the rising
edge does not display a comparably clear quasilinear segment. Inclusion
of points beyond the beginning of the negative-resistance region in the
fit of the open-channel fraction caused the fitted exponent $\theta_M$
to increase rapidly. This behaviour is consistent with an additional
polarization process superimposed on the inward sodium current
throughout the negative-resistance region and extending to shortly
before the action-potential peak.

\quad
The fitting interval for the $M$ component was therefore chosen to
extend from the action-potential inflection point, at which the
capacitive current reaches its maximum, to the beginning of the
negative-resistance region. This procedure determines the sodium
effective reversal potential $V_M$, the time rate $\mu_M$, the maximum
conductance $g_M$ through Eq.~\eqref{Sodiumd}, and the inception time
$t_{iM}$ used in Eqs.~\eqref{subeq:currentsd} and
\eqref{mAvramia}. The difference between the reconstructed total ionic
current $J_I$ and the fitted sodium activation current $J_M$ defines
the polarization current,
\[
J_P
=
\sum_n J_{Pn}
=
J_I-J_M.
\]

\quad
The completion fraction of each polarization process was similarly
fitted with the mAvrami relation Eq.~\eqref{mAvramid}. The values of
$\alpha_{Pn}$ were initialized with the fine-structure constant, and
the exponents $\theta_{Pn}$ were initialized with the value $3.78$. In
the absence of independent constraints, the polarization time rates
$\mu_{Pn}$ were treated as independent fitting parameters. The fitted
values show that, with some exceptions, the polarization time rates are
close to multiples of the sodium activation time rate $\mu_M$.

\quad
Most fitted parameters vary smoothly with temperature. The values of
$g_{P4}$, however, vary substantially from one sweep to another because
the fitted $P4$ component includes polarization reversals. Before a
polarization reversal, the polarization conductance satisfies
$g_{P4}\approx g_M$ at all analyzed temperatures.

\quad
The fitting forms for the capacitive current in
Eq.~\eqref{subeq:currentsc} and the sodium activation current in
Eq.~\eqref{subeq:currentsd} differ through the exponent $1/3$ in the
capacitive factor $(M_o/M)^{1/3}$. Both fitting forms use the same
effective reversal potential $V_M$, while the time rate $\mu_M$ and
maximum conductance $g_M$ are related by Eq.~\eqref{Sodiumd}. The
factor $(M_o/M)^{1/3}$ provides a good fit to the capacitive current
$C_m\Phi_M$. We did not test whether an analogous factor
$(X_o/X)^{1/3}$ also applies to the recovery currents.

\quad
The recovery-region fits were obtained by first approximating the
smoothed total ionic current $J_I$ with a sixth-order polynomial,
denoted by $J_I|_{\mathrm{p6th}}$. The fitted interval extended from the
potassium reversal potential $V_N$ to a point close to the effective
reversal potential $V_H$, excluding the narrow region containing the
polarization reversal. The recovery region displays an evident
superposition of two current components.

\quad
The potassium current $J_N$ was extracted first because it exhibits a
clear quasilinear segment terminating at its reversal potential. This
segment determines the maximum conductance $g_N$ and time rate $\mu_N$.
The completion fraction $N_o/N$ was then fitted, and the full recovery
potassium current $J_N$ was reconstructed. The recovery current $J_H$
was obtained by subtracting $J_N$ from the sixth-order polynomial fit:
\[
J_H
=
J_I|_{\mathrm{p6th}}-J_N.
\]

\quad
The sixth-order polynomial fit crosses the zero-current axis at the
effective reversal potential $V_H$, where a fraction of the sodium
channels has already deactivated as the membrane potential has
decreased from its peak. The maximum conductance $g_H$ and time rate
$\mu_H$ are determined at the action-potential peak from
\[
g_H
=
\left.
\frac{dJ_H(V)}{dV}
\right|_{V=V_p},
\qquad
\mu_H
=
\left.
\frac{d\Phi_H(V)}{dV}
\right|_{V=V_p}.
\]
The fit was completed by optimizing $\mu_H$ and $t_{cH}$.

\begin{figure*}
\includegraphics[width=2.1\columnwidth]{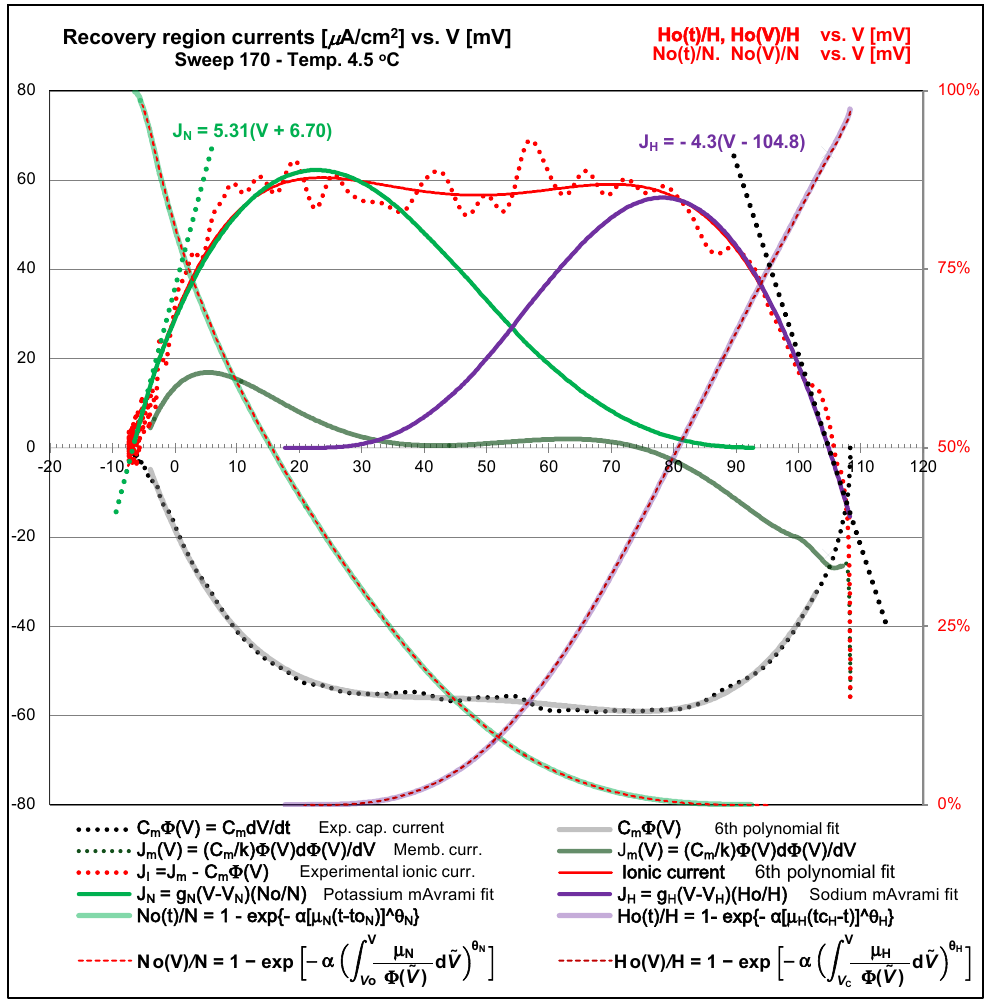}
\caption{\label{fig:RecoveryCurrentsSweep170L}
\textbf{Recovery currents and channel dynamics in the phase-space formulation of the propagating action potential, reconstructed from the charge-conserving phase-space cable equation.}
 The capacitive current $C_m\Phi(V)$, membrane current $J_m(V)$, and ionic current $J_I(V)$ satisfy $J_I = J_m - C_m\Phi$. Decomposition of the ionic current reveals distinct sodium ($J_H$) and potassium ($J_N$) contributions. The sodium current exhibits a pronounced outward component in the recovery region. Solid curves represent fits to the reconstructed data (symbols). Channel fractions $X_o/X$ are shown on the right axis and are described by mAvrami/Rovelli kinetics. The currents and channel dynamics are shown in a single phase-space representation to preserve their intrinsic nonlinear coupling.}\end{figure*}

\quad
All fits were performed using Gaussian smoothing and Microsoft Excel
Solver. The results are sensitive to the number of included data
points, with the sensitivity increasing at higher temperatures because
fewer points were recorded. The reconstructed recovery current $J_H$
also displays greater granularity in the lower-temperature sweeps. In
particular, a prominent feature occurs just above $55\,\mathrm{mV}$
(see SM Fig.~\ref{fig:RecoveryCurrentsSweep170L} and
SM Fig.~\ref{fig:Currents525_695L}(b)).

\quad
Although individual fits contain some uncertainty and depend on the
selected fitting interval, the aggregate behaviour across temperatures
is coherent and internally consistent. Sweep~525 is the principal
exception: its fitted values of $V_M$ and $\mu_{P4}$ are outliers,
whereas $g_M$, $\mu_M$, $t_{iM}$, $\delta_M$, and $Q_g$ remain
consistent with the temperature trends observed in the other sweeps.

\section{\label{sec:levelR}Relational parametrization of channel completion and its relation to Rovelli}

This appendix develops a branchwise correlational formulation of the modified Avrami description of ion-channel kinetics. The formulation follows directly from steady propagation, charge conservation, and Ohm's law: along each monotonic branch, the measured relation $\Phi(V)=dV/dt$ permits the channel dynamics to be expressed through correlations among experimentally accessible quantities. Laboratory time is not eliminated but remains recoverable from the propagated trajectory. The independently derived formulation provides the basis for the subsequent comparison with Rovelli's relational interpretation of physical dynamics.

\subsection*{E1. Relational phase-space formulation}

\quad
In the charge-conserving cable formulation developed in the main text, the membrane potential $V$ and its phase-space velocity
\begin{equation}
\Phi(V)=\frac{dV}{dt}
\end{equation}
determine the branchwise phase-space trajectory. The total ionic current is
\begin{equation}
J_I(V)
=
C_m\Phi(V)
\left(
\frac{1}{k}\frac{d\Phi(V)}{dV}-1
\right),
\label{eq:JI}
\end{equation}
where $C_m$ is the membrane capacitance and $k$ is the propagation constant.

\quad
Along each monotonic branch of the propagated action potential, the membrane potential is a one-to-one function of time. Consequently,
\begin{equation}
dt=\frac{dV}{\Phi(V)},
\label{eq:dt}
\end{equation}
which permits the independent variable to be changed from time to membrane potential without loss of branchwise information.

\quad
Equation~(\ref{eq:dt}) establishes an exact correspondence between the temporal and phase-space descriptions along each monotonic branch. Any trajectory-dependent quantity that is single-valued along a given branch can therefore be written equivalently as a function of membrane potential by replacing the temporal integral with its phase-space counterpart. The resulting formulation depends on experimentally observable correlations among $V$, $\Phi(V)$, and the state variables describing channel completion.

\quad
This branchwise transformation introduces no additional physical assumption. It provides the mathematical basis for expressing the modified Avrami completion laws through correlations among $V$, $\Phi(V)$, and the channel-state variables.

\subsection*{E2. Relational mAvrami kinetics}

\quad
Let $f\in[0,1]$ denote the completion fraction of a channel process, for example the fraction of open sodium or potassium channels along a specified branch of the propagated action potential. In the modified Avrami representation, the completion fraction is written in time as
\begin{equation}
f(t)
=
1-\exp\left[
-\left(
\int_{t_i}^{t}\mu(\tau)\,d\tau
\right)^n
\right],
\qquad n>0,
\label{eq:mAvrami_time}
\end{equation}
where $t_i$ is the inception time, $\mu$ is an effective microscopic time rate, and $n$ is the Avrami exponent.

\quad
The logarithmic exponent of the mAvrami law is therefore
\begin{equation}
-\ln[1-f(t)]
=
\left(
\int_{t_i}^{t}\mu(\tau)\,d\tau
\right)^n.
\label{eq:mAvrami_exponent_t}
\end{equation}
This form separates the monotone completion variable $f$ from the accumulated kinetic measure appearing in the exponent.

\quad
Using Eq.~(\ref{eq:dt}), the same kinetic law can be written without choosing time as the independent variable. Along any monotonic branch of the propagated trajectory,
\begin{equation}
dt=\frac{dV}{\Phi(V)},
\end{equation}
and therefore
\begin{equation}
f(V)
=
1-\exp\left[
-\left(
\int_{V_i}^{V}
\frac{\mu(\widetilde V)}{\Phi(\widetilde V)}
\,d\widetilde V
\right)^n
\right],
\label{eq:mAvrami_V}
\end{equation}
where $V_i$ is the inception potential of the same channel process. Equivalently,
\begin{equation}
-\ln[1-f(V)]
=
\left(
\int_{V_i}^{V}
\frac{\mu(\widetilde V)}{\Phi(\widetilde V)}
\,d\widetilde V
\right)^n.
\label{eq:mAvrami_exponent_V}
\end{equation}

\quad
For the channel components analyzed in the main text, the normalized completion law above is extended by the empirical dimensionless prefactor $\alpha$:
\begin{equation}
f_X(t)
=
1-\exp\left[
-\alpha
\left(
\mu_X\int_{t_{iX}}^{t}d\tau
\right)^{\theta_X}
\right],
\label{eq:f_X_t}
\end{equation}
or, equivalently,
\begin{equation}
f_X(V)
=
1-\exp\left[
-\alpha
\left(
\mu_X\int_{V_{iX}}^{V}
\frac{d\widetilde V}{\Phi(\widetilde V)}
\right)^{\theta_X}
\right],
\label{eq:f_X_V}
\end{equation}
where $X$ labels the channel process, $\mu_X$ is its characteristic time rate, $\theta_X$ is the corresponding Avrami exponent, and $\alpha$ is the dimensionless empirical prefactor. The same value of $\alpha$ is used across the fitted channel components, within the accuracy of the experimental parametrization.

\quad
At the level of the experimental data, the phase-space integral in Eq.~(\ref{eq:f_X_V}) is evaluated numerically as
\begin{widetext}
\begin{equation}
f_X(V_N)
=
1-\exp\left[
-\alpha
\left(
\mu_X
\sum_{j=1}^{N}
\frac{1}{2}
\left[
\frac{1}{\Phi(V_{j-1})}
+
\frac{1}{\Phi(V_j)}
\right]
(V_j-V_{j-1})
\right)^{\theta_X}
\right].
\label{eq:f_X_V_numeric}
\end{equation}
\end{widetext}

\quad
The potential-domain integral is evaluated separately on each monotonic branch. If an integration limit approaches a turning point at which $\Phi=0$, the integral is obtained from the corresponding time increments or from its branchwise limiting value; the singular endpoint is not inserted directly into the discrete $1/\Phi$ expression.

\quad
Equations~(\ref{eq:f_X_t}) and (\ref{eq:f_X_V}) are not distinct kinetic assumptions. They are two fitting representations of the same completion process. The temporal form uses the laboratory-time waveform, whereas the correlational form uses the measured phase-space velocity $\Phi(V)$ and membrane potential as the branchwise ordering variable. The two representations are compared directly in Figs.~\ref{fig:RisingEdgeSweep425L} and \ref{fig:RecoveryCurrentsSweep425L}. In these figures, the phase-space representation is labeled ``Rovelli fit'' as a succinct acknowledgment of its connection with the relational formulation developed in this appendix; the label does not attribute the fitting procedure itself to Rovelli. The modified Avrami kinetics may therefore be expressed through the observable correlations among $f_X$, $V$, and $\Phi(V)$.

\subsection*{E3. Intrinsic ordering and a monotone completion functional}

\quad
The correlational formulation developed above permits the channel kinetics to be described through observable relations among the completion fraction, membrane potential, and phase-space velocity. An immediate consequence is the existence of a natural monotone quantity associated with the completion process.

\quad
Define
\begin{equation}
\Lambda(V)
=
\int_{V_i}^{V}
\frac{\mu(\widetilde V)}{\Phi(\widetilde V)}
\,d\widetilde V
\end{equation}

\begin{equation}
\mathcal{S}(V)
\equiv
-\ln[1-f(V)]
=
\Lambda(V)^n.
\label{eq:S_f}
\end{equation}

\quad
Along the physical trajectory,
\begin{equation}
\frac{d\Lambda}{dt}
=
\frac{\mu(V)}{\Phi(V)}\frac{dV}{dt}
=
\mu(V)\geq 0.
\end{equation}
Consequently,
\begin{equation}
\frac{d\mathcal{S}}{dt}
=
\Phi(V)\frac{d\mathcal{S}}{dV}
=
n\Lambda^{n-1}\mu(V)
\geq 0.
\label{eq:S_monotone}
\end{equation}
Thus, $\mathcal{S}$ increases in the temporal direction along either branch, even though $d\mathcal{S}/dV$ changes sign when the orientation of the trajectory changes.

\quad
The quantity $\mathcal{S}=-\ln(1-f)$ therefore provides an intrinsic ordering of the channel-completion process. The ordering is determined by correlations among the completion fraction, membrane potential, and phase-space velocity rather than by treating laboratory time as an independently specified state variable.

\quad
Because $\mathcal{S}$ is a logarithmic monotone functional of completion, it is structurally analogous to an entropy-like ordering variable. The analogy is limited: $\mathcal{S}$ is not identified with thermodynamic entropy, and its monotonicity expresses directed channel completion rather than a general thermodynamic law.

\quad
The same functional is obtained whether the completion process is represented in laboratory time or branchwise in membrane potential. The correlational reformulation therefore preserves the directed ordering of the kinetics while retaining laboratory time as a reconstructable parameter.

\subsection*{E4. Relation to Rovelli's relational formulation}

\quad
The correlational representation developed above follows directly from steady propagation, charge conservation, and Ohm's law and does not rely on a general theory of relational time. Its comparison with Rovelli is therefore interpretive rather than foundational. The independently derived formulation is nevertheless conceptually consistent with Rovelli's relational viewpoint, in which the physical content of a theory may be expressed through correlations among observable quantities rather than through evolution with respect to an externally privileged clock \cite{Rovelli1}.

\quad
Along each monotonic branch, membrane potential serves as an intrinsic ordering variable, while temporal direction is retained in the sign of $\Phi(V)$. Laboratory time is not eliminated but can be reconstructed through
\begin{equation}
dt=\frac{dV}{\Phi(V)}.
\end{equation}
The formulation is therefore relational rather than timeless.

\quad
The same propagated solution may be represented in laboratory time or, branchwise, in membrane potential. Preserved under this change of parametrization are the geometric phase-space trajectory and the physical relations among the measured waveform, the reconstructed currents, and the channel completion fractions. In the reduced state representation
\begin{equation}
\Gamma=\{V,\Phi,\{f_X\}\},
\end{equation}
the observable dynamics are encoded in correlations among the state variables, while $t$ or $V$ specifies how the same physical evolution is traversed.

\quad
The common dimensionless prefactor found in the completion laws is an empirical property of the reconstructed kinetics and is not a consequence of relational parametrization itself. Rewriting an equivalent completion law in $t$ or in $V$ leaves the prefactor unchanged. The nontrivial empirical observation is its recurrence across channel components and temperatures. Whether this numerical stability reflects a deeper structural property remains an open question.

\quad
The propagated action potential thus provides a concrete example of a physical process whose current structure and channel evolution can be formulated through observable correlations without assigning a fundamental role to an external clock. Temporal ordering remains encoded in the orientation of the trajectory and is recoverable from the measured phase-space relation.

\subsection*{E5. Continuity at the action-potential peak and branchwise parametrization}

Let $t_\ast$ denote the time at which the propagated action potential reaches its maximum,
\begin{equation}
V(t_\ast)=V_{\mathrm{peak}},
\qquad
\Phi(t_\ast)=0.
\end{equation}

\quad
For a smooth measured waveform, the charge-conserving reconstruction yields a total ionic current that remains continuous across the peak:
\begin{equation}
\lim_{t\uparrow t_\ast}J_I(t)
=
J_I(t_\ast)
=
\lim_{t\downarrow t_\ast}J_I(t).
\label{eq:continuity}
\end{equation}

\quad
The peak is a turning point of the membrane-potential waveform, not a reversal of the spatial propagation of the action potential. As the waveform passes through the peak, the phase-space velocity changes from
\begin{equation}
\Phi>0
\quad\longrightarrow\quad
\Phi=0
\quad\longrightarrow\quad
\Phi<0.
\end{equation}
The rising and recovery branches therefore meet at the same point of the continuous $(V,\Phi)$ trajectory.

\quad
Membrane potential cannot serve as a regular local trajectory parameter exactly at the peak because $\Phi(V_{\mathrm{peak}})=0$. The relation
\begin{equation}
dt=\frac{dV}{\Phi(V)}
\end{equation}
must consequently be applied separately on the two monotonic branches and interpreted through its limiting behaviour as $V$ approaches $V_{\mathrm{peak}}$. Laboratory time remains regular through the turning point, while the $V$-based representation changes from the rising branch to the recovery branch.

\quad
The factored phase-space expressions must likewise be interpreted in this limiting sense. Away from the turning point,
\begin{equation}
\Phi(V)\frac{d\Phi(V)}{dV}
=
\frac{d\Phi}{dt}
=
\frac{d^2V}{dt^2},
\end{equation}
and the equality extends to the peak through the corresponding continuous limit. Consequently, neither $J_m$ nor $J_I$ is required to vanish at the peak merely because its phase-space expression contains the factor $\Phi(V)$. Their limiting values are determined by the curvature of the measured waveform.

\quad
Continuity of the reconstructed total ionic current density does not
require every component of its mathematical decomposition to remain
continuous. The activation, polarization, and recovery components may
change discontinuously when the representation passes from one
branch-specific decomposition to another, while their sum preserves the
continuity of the total ionic current density. Such component
discontinuities characterize the decomposition and do not imply a
discontinuity of the underlying local charge balance.
\quad
The action-potential peak is therefore a regular turning point of the physical evolution but a boundary between the two branchwise $V$-parametrizations. Whether the process is represented in laboratory time or separately on the rising and recovery branches in membrane potential, the same continuous phase-space trajectory and the same charge-conserving current balance are recovered.

\subsection*{E6. Broader methodological implication}

\quad
More generally, the present formulation raises the question of whether other biological processes admit useful intrinsic ordering variables when experimentally accessible observables form reproducible correlated trajectories. Suitably chosen relational variables might then provide natural descriptions of collective neural dynamics, sleep--wake transitions, or other large-scale biological processes.

\quad
The present work does not establish such descriptions for these more complex systems. It demonstrates only that, for the steadily propagated action potential, a relational parametrization arises naturally from the measurable correspondence among membrane potential, phase-space velocity, and channel-state variables.

\end{appendices}

\subsection*{SUPPLEMENTAL MATERIALS - Note on notation in the Supplemental Material}

Several Supplemental Material figures were prepared during earlier stages of the analysis and retain minor differences in graphical notation from the main text. These differences do not indicate different physical quantities or fitting procedures. In particular, graphical expressions such as $Xo/X$ and $Pio/Pi$ correspond to the main-text notation $X_o/X$ and $Pi_o/Pi$, respectively. The symbol $J_m$, occasionally identified in a graph by the conventional label `membrane current,'' denotes the axoplasmic-current-divergence current density defined in the main text. Labels such as `mAvrami/Rovelli fit'' or ``Rovelli fit'' denote the relational phase-space representation of the mAvrami completion law obtained through $dt=dV/\Phi(V)$; they do not denote a separate kinetic law or attribute the fitting procedure to Rovelli. Where graphical notation differs, the definitions and notation of the main text govern.

\begin{figure*}
\includegraphics[width=2.1\columnwidth]{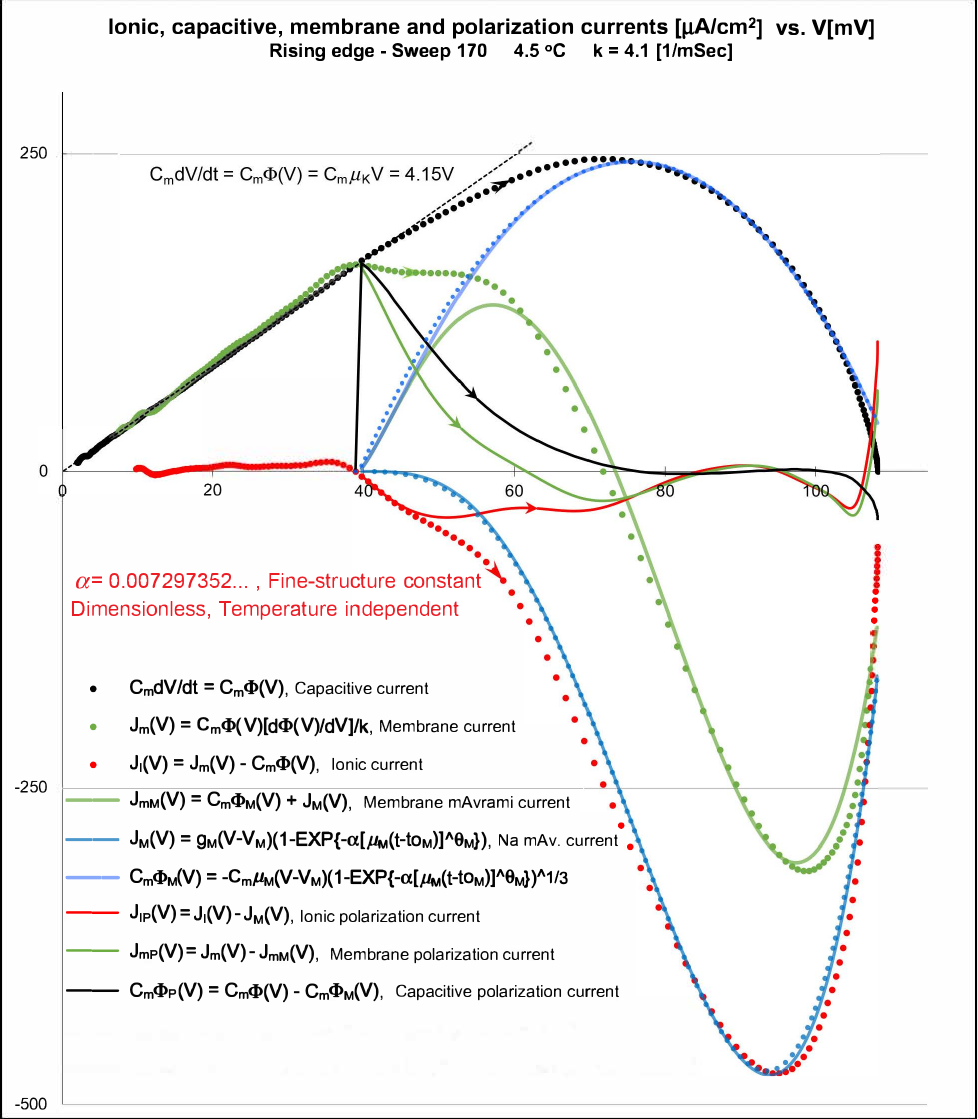}
\caption{\label{fig:RisingEdgeCurrentsSweep170L}The rising edge of the action potential covers the region from the resting potential to the peak of the action potential. Capacitive, membrane and ionic currents, and their parsing into mAvrami fits of the activation currents and the polarization portions are displayed. Note that $\mu_M$ and $g_M$ are related by equation Eq. \eqref {Sodiumd}. Fig.~ \protect \ref{fig:mAvramiSweep170L} displays detailed mAvrami fits of fractions of open channels for activation and polarization currents. At the inception point the three activation fits of $C_m\Phi_M$, $J_{mM}$ and $J_M$ are zero and the corresponding polarization currents $C_m\Phi_{P}$, $J_{mP}$ and $J_{IP}$ are discontinuous. At 4 $^oC$ the inception potential appears to coincide with the potential at which the ionic current crosses the zero current axis. At higher temperatures, the inception precedes by an increasing number of millivolts. At the peak of the action potential all currents are discontinuous. See Figures \protect\ref{fig:RisingEdgeSweep425L} and SM \protect\ref{fig:Currents525_695L} for sweeps at higher temperatures. Arrows $\rightarrow$ indicate the direction of time. Note that at the zero ionic current $d\Phi/dV=4.15$ $mSec^{-1}$ $\approx{k}$, the propagation constant.
} 
\end{figure*}

\begin{figure*}
\centering
\includegraphics[width=2.06\columnwidth]{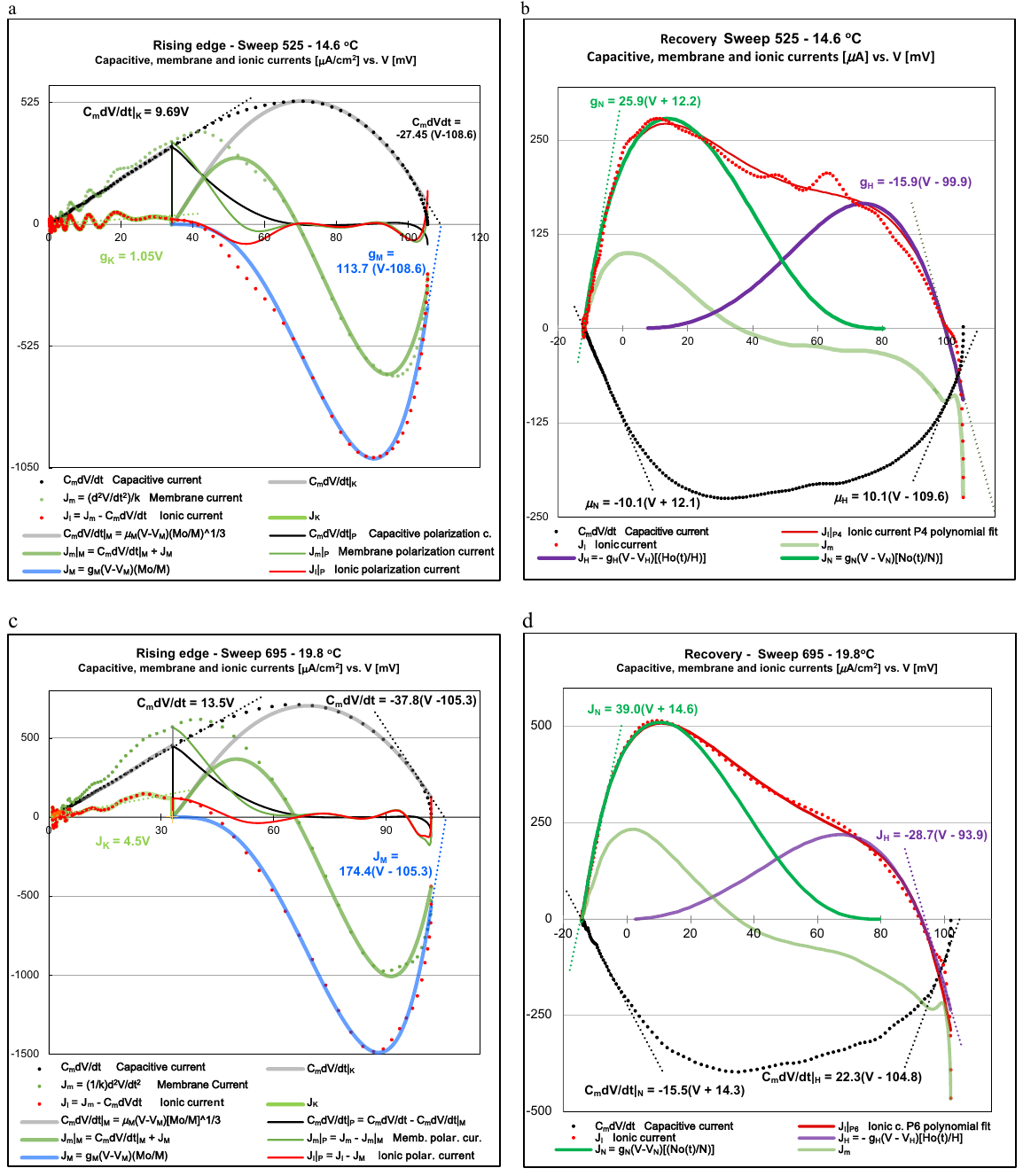}
\caption{Currents and quasi linear segments. 
Note that all currents look almost the same (scale) at the three temperatures (including Fig.~ \protect\ref{fig:RisingEdgeCurrentsSweep170L} and Fig.~\protect \ref{fig:RecoveryCurrentsSweep170L}) except the currents $J_K$ and $J_H$ which increase and decrease relatively as compared with other currents.
{\bf (a), (c),} Capacitive, membrane and ionic currents are displayed with their parsing into mAvrami fits and corresponding polarization currents. The linear slope of the current $J_K$ and the corresponding slope of the capacitive current linear segments are the time rate constant $\mu_K$ and maximum conductance $g_K$.
The slope of the linear segment of the ionic current $J_M$ is the maximum conductance $g_M$ for the sodium M-channel. The corresponding capacitive current linear slope is -$\mu_M$ where $\mu_M$ is the time rate constant of the M-channel. The polarization currents are discontinuous at the inception of the mAvrami fits of activation currents. {\bf (b), (d),} The linear slopes of sodium's and potassium's capacitive currents are $\mu_H$ and -$\mu_N$ corresponding to rate constants $\mu_H$ and $\mu_N$. The linear slopes of $J_H$ and $J_N$ are $-g_H$ and $g_N$ where $g_H$ and $g_N$ are the maximum conductance of the sodium H-channel and potassium N-channel respectively.
While slopes of $J_H $ and capacitive current $C_m\Phi_H$ intercept the potential axis at different points, $C_m\Phi_H$ intercepts the axis at $V \approx V_M$. }
 \label{fig:Currents525_695L} 
\end{figure*}
\begin{figure*}
\centering
\includegraphics[width=2.08\columnwidth]{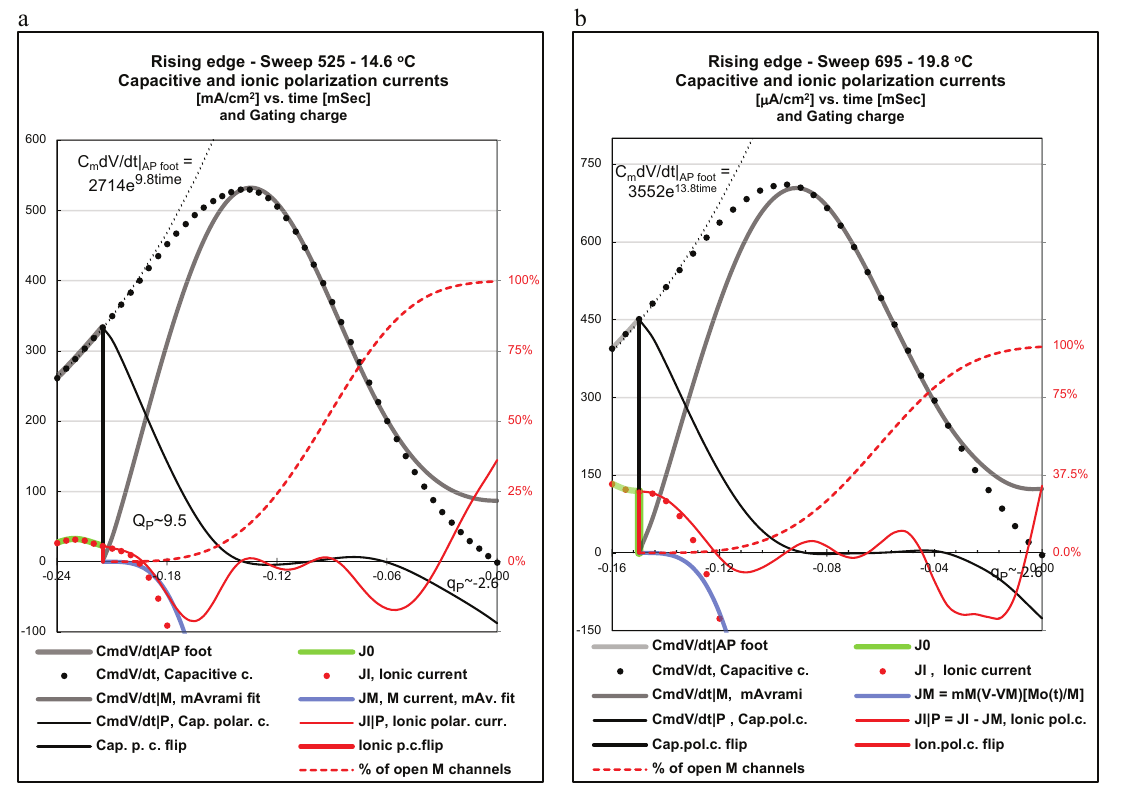}
\caption{The rising edge mAvrami fits of capacitive and ionic currents and the corresponding polarizations currents are plotted against time. The three mAvrami fitted currents $C_m\Phi_M$, $J_{mM}$, $J_M$, and the $M_o(t)/M $ curve begin and intersect the zero-current axis at the inception point $t=t_{iM}$, the time at which sodium M-channels start to open ($J_{mM}$ is not displayed, $J_{mM}$=$C_m\Phi_M$ + $J_M$). Capacitive and membrane polarization currents $C_m\Phi_{MP}$ and $J_{mMP}$, start with the polarization flip at $t=t_{iM}$ and then decay. The surface under the first segment of capacitive polarization curve $C_m\Phi_{MP}$ is equal to the polarization charge $Q_P$ $\approx$ $10x10^{-9}$ Coulomb/cm$^2$ moved across the membrane (a.k.a. gating charge). The flip from H-channel symmetry to M-channel symmetry and the transfer of charges across but within the membrane in conjunction with the ionic polarization current precedes the opening of sodium channels. In the present theory the motion of gating charges tapers off at approximately the maximum rate of rise of the action potential at about 60 mV when the fraction of open sodium channels is about $20\%$. The value of the charge $Q_P$ remains approximately the same at all temperatures. Towards the peak of the action potential, preceding the flip from M-channel symmetry to H-channel symmetry at the peak, there is an opposite charge transfer across and also within the membrane amounting to $q_P$ $\approx$ -2.6$x10^{-9}$ Coulomb/cm$^2$. 
}
 \label{fig:GatingChargeSweep525-695L} 
\end{figure*}
\begin{figure*}
\includegraphics[width=2.1\columnwidth]{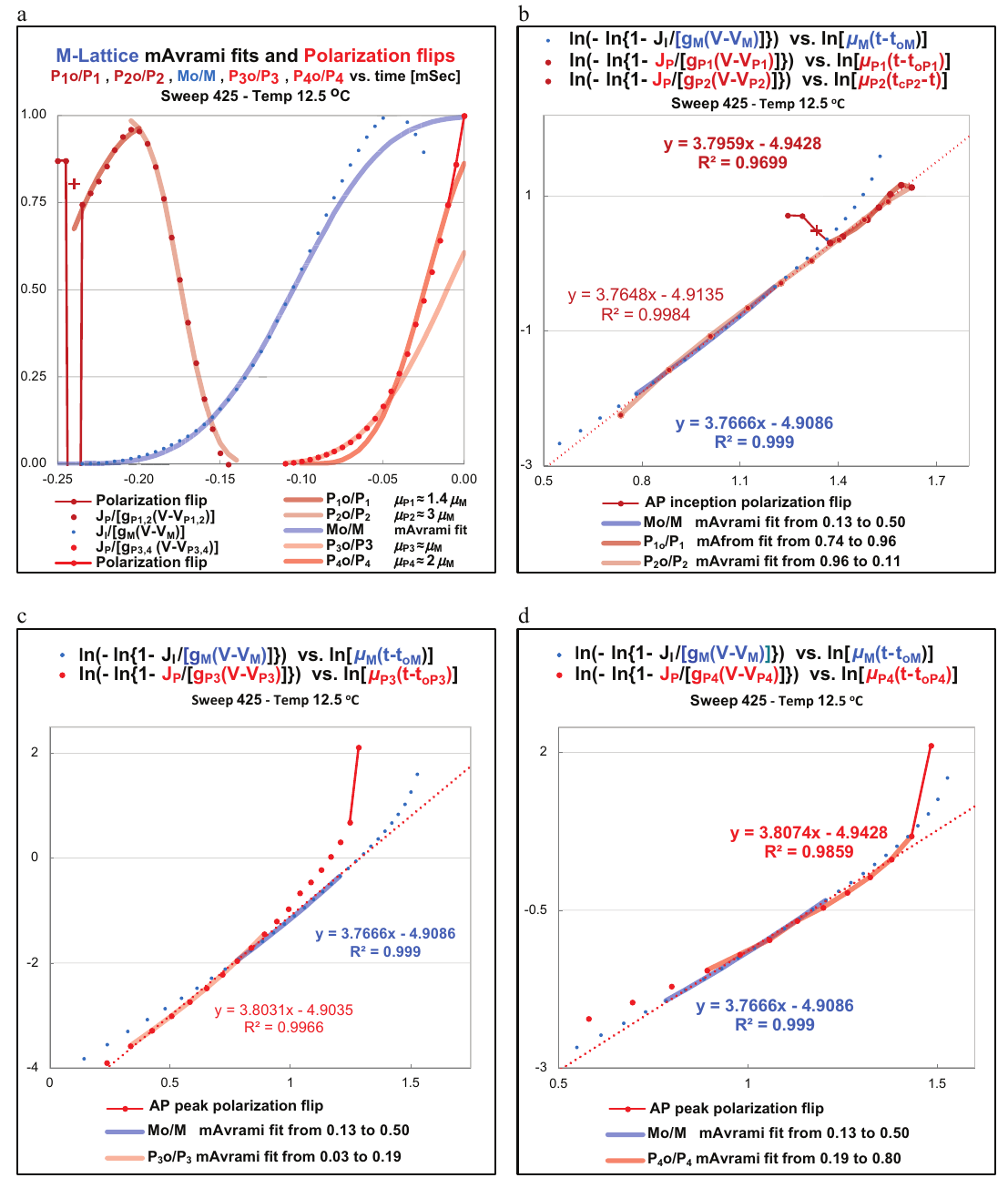}
\caption{Rising-edge mAvrami fits of the fraction of open sodium M channels, $M_o/M$, and of the polarization completion fractions $Pn_{lo}/Pn$. Inception and AP-peak polarization flips are displayed. Parameters $\alpha_M$ and $\alpha_{Pn}$ from Eq.~\eqref{mAvrami} are seeded with the fine-structure constant, $\alpha = 0.007297352$, and the parameters $\theta_M$ and $\theta_{Pn}$ are seeded with the value $3.78$. \textbf{Note:} $g_{P1}=g_{P2}$, $V_{P1}=V_{P2}$, $g_{P3}=g_{P4}$, and $V_{P3}=V_{P4}$. \textbf{(a)} Both the inception and AP-peak polarization segments consist of two concatenated portions with different time rates. Inception polarization-flip interpolation is indicated by `+'. The polarization time rates $\mu_{P2,3,4}$ are close to multiples of the sodium M-channel time rate $\mu_M$; see also Fig.~\ref{fig:mAvramiLabFitSweep170L}. \textbf{(b)--(d)} $\ln \alpha = \ln(0.007297352\ldots) = -4.920243\ldots$.}
\label{fig:mAvramiLabFitSweep425L}
\end{figure*}
\begin{figure*}
\includegraphics[width=2.09\columnwidth]{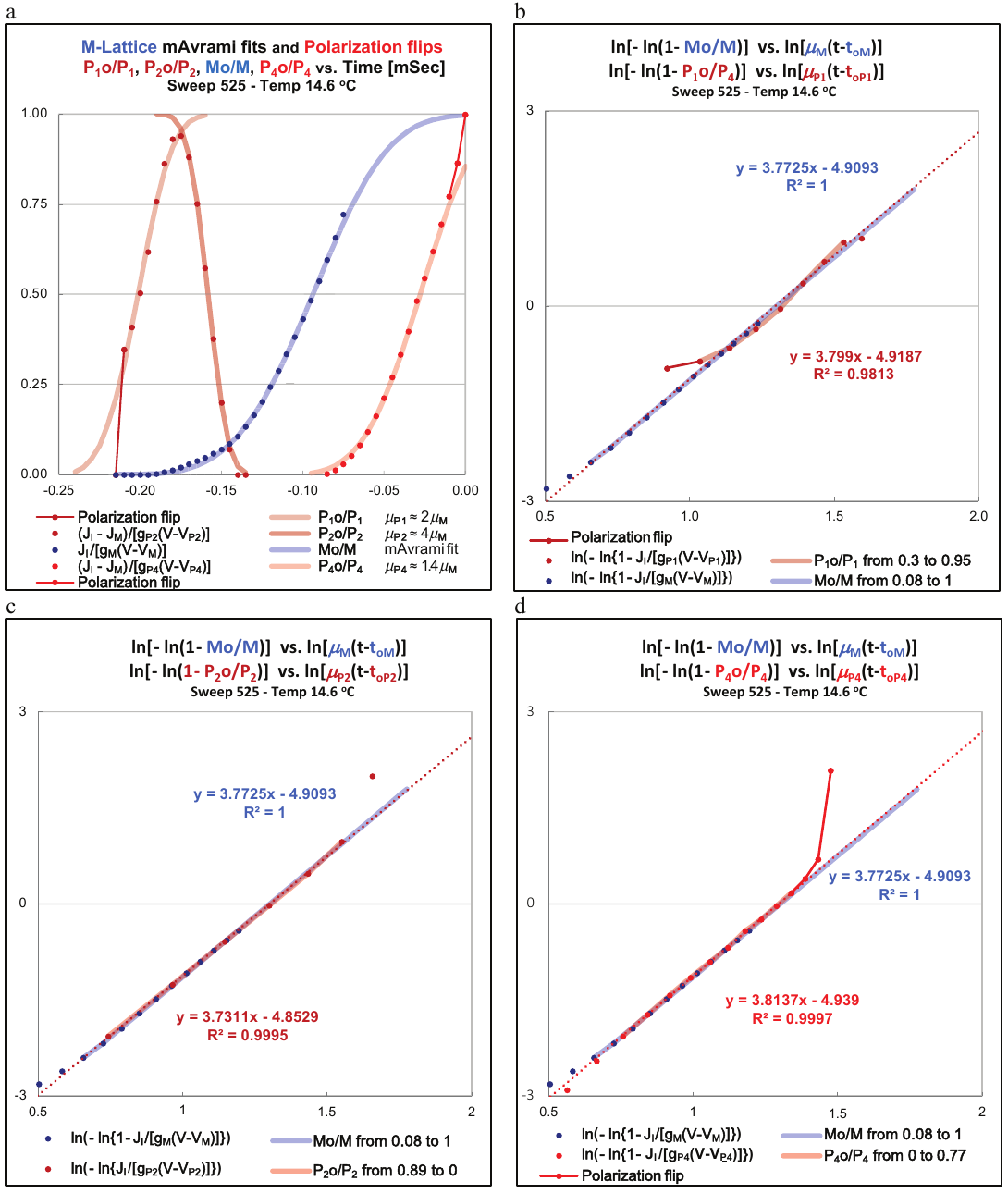}
\caption{Rising-edge mAvrami fits of the fraction of open sodium M channels, $M_o/M$, and of the completed polarization fractions $Pn_{o}/Pn$. Inception and AP-peak polarization flips are displayed. Parameters $\alpha_M$ and $\alpha_{Pn}$ from Eq.~\eqref{mAvrami} are seeded with the value of the fine-structure constant, $\alpha = 0.007297352$, and the parameters $\theta_M$ and $\theta_{Pn}$ are seeded with the value $3.78$. \textbf{Note:} $g_{P1}=g_{P2}$, $V_{P1}=V_{P2}$, and there is no segment $P3$. \textbf{(a)} The inception polarization segment consists of two concatenated portions with different time rates. The AP-peak polarization segment $P4$ is single. The polarization time rates $\mu_{P1}$ and $\mu_{P2}$ are close multiples of the sodium M-channel time rate $\mu_M$, whereas $\mu_{P4}$ is not. \textbf{(b)--(d)} $\ln \alpha = \ln(0.007297352\ldots) = -4.920243\ldots$.}
\label{fig:mAvramiLabFitSweep525L}
\end{figure*}
\begin{figure*}
\includegraphics[width=2.1\columnwidth]{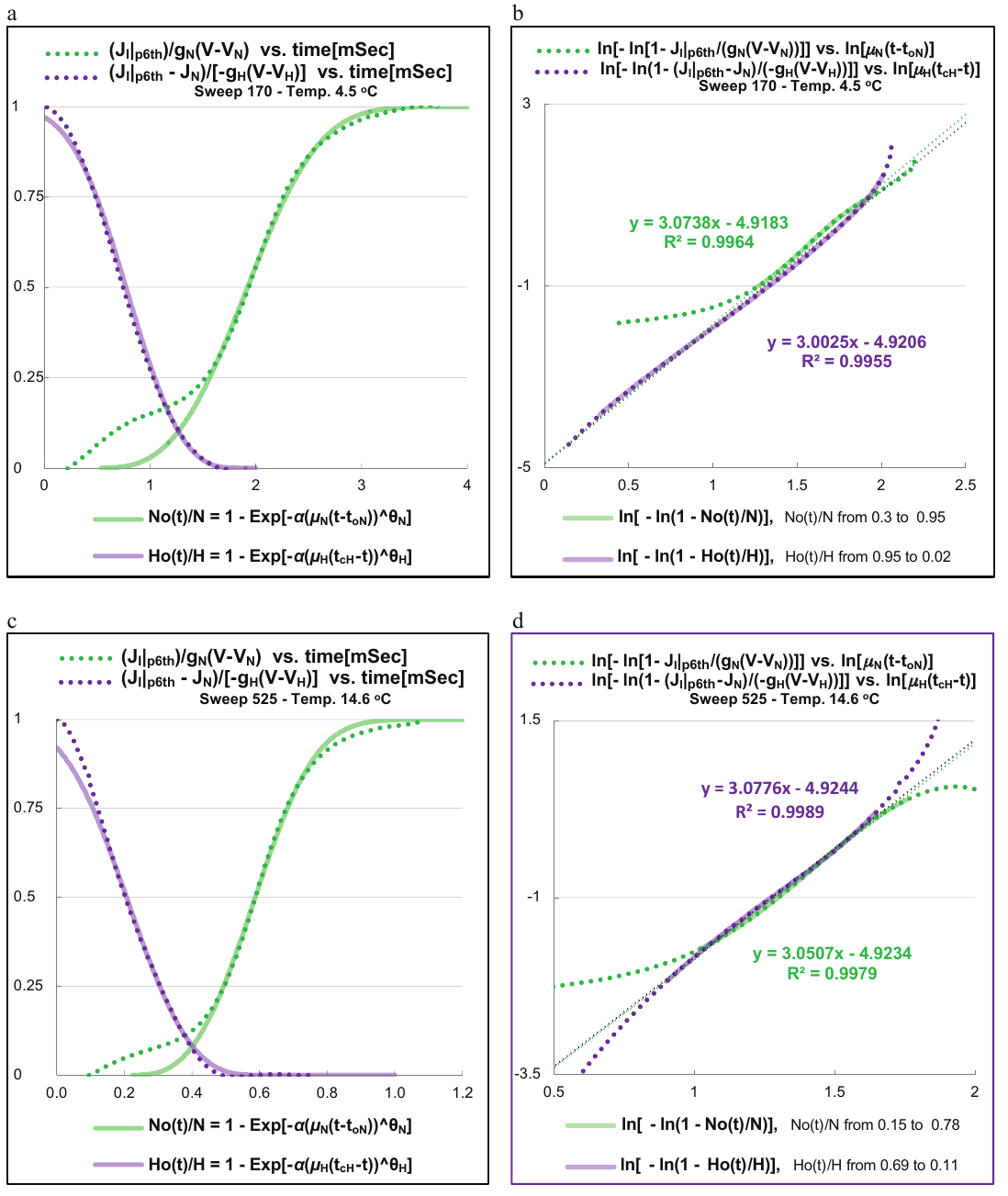}

\caption{Recovery mAvrami fits of fraction of open sodium H-channels $H_o/H$ and potassium N-channels $N_o/N$. 
Parameters $ \alpha_H$ and $\alpha_N$ from \protect(Eq.~\eqref{mAvrami})  are seeded with the value of the fine-structure constant $\alpha$ = 0.007297352 and parameters $\theta_H$, $\theta_N$ are adjusted by fitting. {\bf (a),(c)} Pnots of experimental fractions $H_o(t)/H$ and $N_o(t)/N$ vs. time[mSec] are fitted with mAvrami Eq.\eqref{mAvramib} and Eq.\protect \eqref{mAvramic} respectively. {\bf (b), (d)} Pnots of experimental $\ln[-\ln(1-H_o(t)/H)]$ vs. $\ln[\mu_H(t_{cH}-t)]$ and $\ln[-\ln(1-N_o(t)/N)]$ vs. $\ln[\mu_N(t-t_{iN})]$ are fitted with linear functions. {\bf Note:} $\ln \alpha$ = $\ln(0.007297352...)$ = $- 4.920243...$. 
}
 \label{fig:mAvramiRecoverySweep170_525L} 
\end{figure*}

\begin{figure*}
\includegraphics[width=2.09\columnwidth]{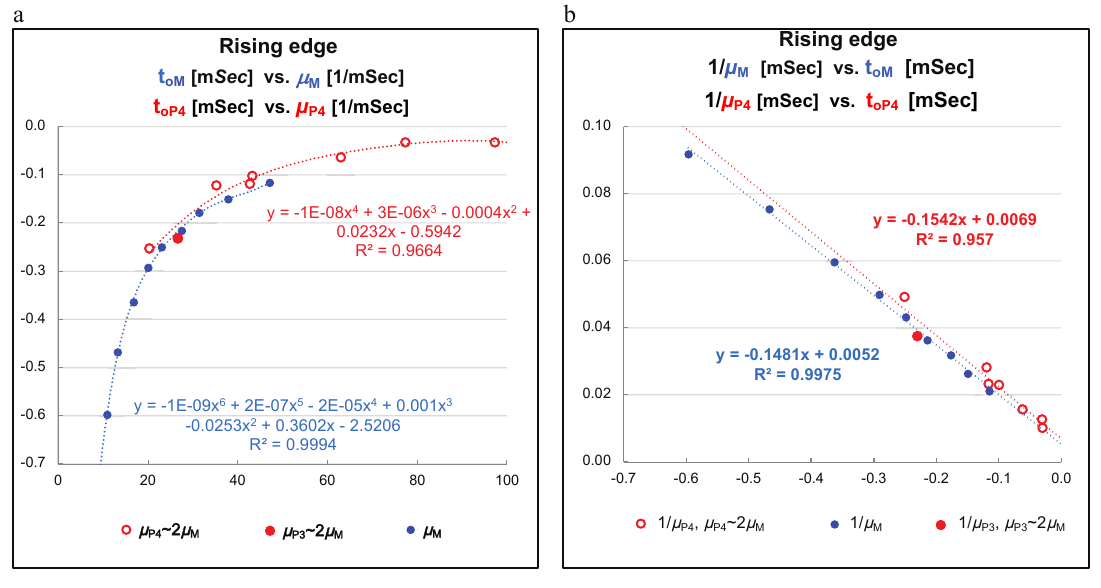}
\caption{}
\label{fig:TimeRateVsTimeL}
\end{figure*}

\clearpage

\end{document}